\ifdefined\pdfoutput\pdfoutput=1\fi
\documentclass[pdflatex,sn-mathphys-num]{sn-jnl}

\usepackage{graphicx}%
\usepackage{multirow}%
\usepackage{amsmath,amssymb,amsfonts}%
\usepackage{amsthm}%
\usepackage{mathrsfs}%
\usepackage[title]{appendix}%
\usepackage{float}%
\usepackage{xcolor}%
\usepackage{textcomp}%
\usepackage{manyfoot}%
\usepackage{booktabs}%
\usepackage{algorithm}%
\usepackage{algorithmicx}%
\usepackage{algpseudocode}%
\usepackage{listings}%
\usepackage[utf8]{inputenc}   
\ifdefined\DeclareUnicodeCharacter
  \DeclareUnicodeCharacter{229E}{\ensuremath{\boxplus}}
\fi

\providecommand{\burl}[1]{\url{#1}}

\theoremstyle{thmstyleone}%
\theoremstyle{thmstyletwo}%

\theoremstyle{thmstylethree}%

\begin{document}

\title[Cross-Community Work in Open Source]{Building Digital Societies as Ecosystems:
How Recognition and Repeat Relationships Sustain Cross-Community Work in Open Source}


\author[1,2]{\fnm{Lucia} \sur{Gomez} Tejeiro}\email{lucia.gomez@unige.ch}
\equalcont{These authors contributed equally to this work.}

\author[1]{\fnm{Thibaut} \sur{Chataing}}\email{thibaut.chataing@unige.ch}

\author[3]{\fnm{Julian} \sur{Jang-Jaccard}}\email{julian.jang-jaccard@armasuisse.ch}

\author[3]{\fnm{Alain} \sur{Mermoud}}\email{alain.mermoud@armasuisse.ch}

\affil*[1]{\orgdiv{Geneva School of Economics and Management}, \orgname{University of Geneva}, \orgaddress{\street{Bvd. Pont-d'Arve, 40}, \city{Geneva}, \postcode{1205},  \country{Switzerland}}}

\affil[2]{\orgdiv{Institute for Applied Data Science and Finance}, \orgname{Bern University of Applied Sciences}, \orgaddress{\street{Brückenstrasse 73}, \city{Bern}, \postcode{3012}, \country{Switzerland}}}

\affil[3]{\orgname{armasuisse Science and Technology}, \orgaddress{\street{EPFL Innovation Park, Bâtiment I}, \city{Lausanne}, \postcode{1015}, \country{Switzerland}}}

\author*[1]{\fnm{Thomas} \sur{Maillart}}\email{thomas.maillart@unige.ch}
\equalcont{These authors contributed equally to this work.}

\maketitle

\section*{Abstract}
\begin{abstract}

We measure cross-boundary collaboration in an open-source software (OSS) ecosystem by reconstructing the bipartite contributor--repository graph of $464$ cybersecurity-focused projects and $11{,}372$ contributor identities active over October 2001--May 2022, seeded from the curated Rawsec Cybersecurity Inventory. Louvain community detection identifies $163$ non-singleton communities; per-community contributor count scales superlinearly with repository count ($n_\mathrm{contributors}\sim {n_\mathrm{repos}}^{1.4}$), and community formation follows a logistic trajectory whose inflection falls in $2018$. Three patterns support a recognition \& repeat-relationship account of cross-boundary work. \emph{First}, cross-community work concentrates in a thin \emph{carrier layer}: only nine contributors span seven or more communities at the commit level, authoring $14\%$ of $4{,}015$ inter-community merged pull-request records; the top-$50$ cross-community contributors produce $54\%$, and half of all cross-boundary work comes from $36$ contributors ($0.34\%$ of the population). \emph{Second}, boundary friction is \emph{relationship-specific} rather than a fixed property of the boundary: under repository and author fixed effects, prior write access in the destination repository raises the odds of a cross-community pull-request being merged $2.3$-fold and cuts integration latency by $66\%$, whereas a contributor's breadth across \emph{other} communities buys nothing. \emph{Third}, community survival is cohort-structured: per-cohort residualisation hazard rises an order of magnitude between pre-2010 and 2018 cohorts, and external community reach predicts survival mainly through community size, leaving late-cohort communities under-served. The corpus closes just before large-language-model coding assistants entered mainstream use, so these measurements form a pre-AI baseline for a system now changing fast. Their joint implication is sharp: the cross-boundary work holding this ecosystem together is carried by very few people and priced by relationships that take years to accumulate -- a dependency invisible to aggregate activity metrics, and erodible without any of them moving. Whether it generalises beyond one curated cybersecurity corpus is untested; this baseline is what makes that testable.

\end{abstract}


\section{Introduction}
Open source software (OSS) has moved from economic oddity \cite{lerner_dynamics_2006} to load-bearing infrastructure: encryption libraries, security tooling, build systems, and operating-system components are now produced and maintained as OSS \cite{sharma_comprehensive_2021}, and the digital systems that depend on them increasingly mediate physical infrastructures and public services \cite{floridi_fourth_2014, helbing_cocreating_2024, wang_digital_2024}. The governance template -- bottom-up peer production \cite{benkler_coases_2002} under permissive licensing \cite{lerner_scope_2005} -- has proven robust enough to scale into the mainstream \cite{benkler_penguin_2011}, but the same expansion has made the question of \emph{ecosystem} sustainability concrete: many projects are abandoned or under-maintained \cite{coelho_identifying_2018, coelho_is_2020, calefato_will_2022}, and complex dependency networks transmit fragility across organisational boundaries \cite{ladisa_taxonomy_2022}. This raises a focused, measurable question: what holds an OSS ecosystem together when individual projects come and go? The answer matters beyond software, because OSS-like codes and processes already structure live coding for music \cite{collins_live_2003, collins_live_2011}, peer production on Wikipedia \cite{wilkinson_cooperation_2007, klein_virtuous_2015}, and the AI-driven reshaping of scientific discovery itself \cite{jumper_highly_2021, wang_scientific_2023}, while increasingly mediating physical systems from urban mobility \cite{batty_big_2013} to energy grids \cite{farhangi_path_2010} and civic platforms \cite{linders_e-government_2012}. Cybersecurity, the domain of the ecosystem under scrutiny here, is itself produced and coordinated through these same peer-production processes -- from collective vulnerability discovery and open threat-intelligence sharing to the maintenance of critical encryption and security tooling \cite{wagner_misp_2016, maillart_given_2017, gillard_efficient_2023}.

It is well established that contribution \emph{within} OSS projects is heavy-tailed: a small core absorbs most of the integration and quality-control work while a long tail contributes occasionally \cite{mockus_two_2002, lee_firm-based_2003, dabbish_social_2012}, with productive bursts that match self-organised-criticality signatures \cite{maillart_empirical_2008, sornette_how_2014}. This intra-project picture has been extensively documented and is not in dispute. What has received much less ecosystem-scale attention is the \emph{inter}-project layer: the contributors whose commits and pull requests reach across community boundaries, link otherwise loosely coupled projects, and absorb the coordination cost of doing so. Boundary-crossing work is known to be more contested in peer-produced systems -- higher review scrutiny, lower acceptance, longer deliberation \cite{kittur_he_2007, halfaker_dont_2011, yasseri_dynamics_2012, tsay_influence_2014} -- yet ecosystem-scale evidence on \emph{who} sustains it, and on what such carriers can and cannot do for the communities they bridge, remains scarce.

We ask how a cybersecurity OSS ecosystem is \emph{born}, \emph{develops}, and is held together across its internal boundaries, on the bipartite contributor--repository graph of the Rawsec Cybersecurity Inventory (October 2001--May 2022; $11{,}372$ contributor identities, $464$ repositories, $163$ non-singleton communities). Community formation traces from the pre-GitHub era through the post-2008 acceleration to a logistic inflection in $2018$: an ecosystem that has largely stopped growing by adding communities, and that must therefore be held together, if at all, by the links among those it already has.

Three findings emerge, and they compound. Cross-boundary work is carried by very few people: nine contributors span seven or more communities, and $36$ contributors account for half of all cross-community pull-request volume. Boundary-crossing contributions are measurably more costly, facing a $20.8$-percentage-point acceptance gap and two- to four-fold slower integration, but that cost is largely waived for contributors who already have a history with the destination repository, and not at all for contributors who are merely broad. And community survival is cohort-structured: late-born communities enter dormancy at an order-of-magnitude higher annual rate, while more than half of all communities currently host no active cross-community contributor. We call the resulting pattern \emph{friction with stable trust} -- boundary work is contested wherever a relationship has not yet been built, and cheap wherever one has.

That pattern bears on what such a system can and cannot be optimised for. The friction is not simply waste to be engineered away: the slow, contested, repeatedly negotiated crossing is also the process by which the relationships that make later crossings cheap are accumulated, and those relationships are held by very few people and take years to form. An intervention that removed the cost without preserving the relationship would risk removing the mechanism with it -- and, because ecosystem-wide activity is dominated by within-community work, would do so without moving any headline metric. We test no such intervention here. What we provide is the measurement that would make its effects legible, on a corpus that closes just before large-language-model coding assistants entered mainstream use.

The remainder of the paper is organised as follows. Section~\ref{sec:background} sets out the complex-adaptive-systems and peer-production foundations and motivates the breadth/depth distinction. Section~\ref{sec:methods} describes the corpus and the measurement strategy. Section~\ref{sec:results} reports the community-scale anatomy, the carrier layer, community survival, and the boundary-friction analysis. Section~\ref{sec:discussion} draws out what follows for the governance of OSS-like ecosystems.

\section{Background}
\label{sec:background}
OSS ecosystems share with biological, social, and technological complex adaptive systems a small set of organising features: decentralised control, nonlinear interaction, and emergent structure across scales \cite{holland_complex_1992, mitchell_complexity_2009}. Two ideas anchor the present study. The first is Holling's distinction between optimisation and \emph{resilience} -- the capacity of a system to absorb disturbance and reorganise rather than maintain a fixed point \cite{holling_resilience_1973}. The second is Simon's argument that hierarchical, near-decomposable structure is what lets complex systems evolve and persist \cite{simon_architecture_1962}. Both frame the question this paper makes measurable: an OSS ecosystem is workable not when its boundaries are erased but when boundary-spanning structures absorb the perturbations that an explicitly modular system would otherwise transmit.

At the network level, the structural features that shape both robustness and vulnerability are well established: heterogeneous degree distributions, community structure, and core--periphery organisation. Heavy-tailed affiliation arises naturally from multiplicative stochastic growth, in which prior advantage compounds into further attention -- the Matthew effect of cumulative advantage \cite{merton_matthew_1968, barabasi_emergence_1999, maillart_empirical_2008}. Such systems are robust to random failure but fragile under targeted disruption \cite{albert_error_2000}, a ``robust-yet-fragile'' regime formalised by the Highly Optimized Tolerance framework \cite{carlson_highly_2000} and observable as far afield as the Internet backbone \cite{stoger_bgp_2025}. Self-organised-criticality dynamics complement this picture by showing that punctuated, cascade-like activity is the norm rather than the exception in evolving networks \cite{bak_self-organized_1987, sornette_how_2014}, with hierarchical branching processes providing a generative account of how local interactions cascade into observed bursty, multi-scale activity patterns \cite{saichev_hierarchy_2013}.

OSS communities exhibit these dynamics directly. Production is bursty, contribution is heavy-tailed, and a small core handles integration and quality control while a long tail contributes occasionally \cite{mockus_two_2002, maillart_empirical_2008, sornette_how_2014}. Concentration is a productivity feature rather than a sign of weakness: super-linear scaling between team size and output was first reported across OSS projects by Sornette et al.\ \cite{sornette_how_2014}, contested by Scholtes et al., who recovered a sublinear Ringelmann effect instead \cite{scholtes_aristotle_2016}, and then confirmed by Maillart and Sornette, whose re-examination finds the two sets of results less contradictory than had been claimed \cite{maillart_aristotle_2019}; across GitHub and Wikipedia alike, productivity scales super-linearly for smaller groups and saturates for larger ones \cite{muric_collaboration_2019}. The temporal organisation of bursts -- short windows of intense, coordinated activity -- lets communities respond to emerging needs without sustaining permanent overhead \cite{maillart_computational_2024}. At the ecosystem scale, the relevant unit of decomposition is \emph{community modularity} -- the natural partition of the contributor--repository network into densely connected, weakly coupled groups recovered by network-detection methods \cite{fortunato_community_2010, de_meo_generalized_2011}. This network-level modularity lets parallel work proceed across many projects with minimal coordination, while the inter-community ties through which knowledge and trust travel define where that parallelism breaks down. Peer-production governance \cite{benkler_coases_2002} provides the matching social layer, allowing authority to accrue meritocratically through contribution rather than hierarchy. Throughout the rest of this paper, \emph{community} refers to a Louvain community of contributors and repositories on the bipartite graph -- the network-detection sense of modularity -- and not to a code module or library dependency.

What sustains an OSS ecosystem at the scale of many such communities is the comparatively under-studied question of who and what carries collaboration \emph{across} community boundaries. Sustained engagement and the presence of cross-community contributors have long been associated with knowledge recombination and adaptability in collaborative innovation systems \cite{powell_network_2005, fleming_collaborative_2007}, and a pronounced core--periphery organisation, in which a small core sustains most of the integration work, is well documented on OSS data \cite{crowston_core_2006, valverde_self-organization_2007, bird_latent_2008, sornette_how_2014} -- structure that motivates the keystone-contributor analogy to ecological systems we adopt here for its inter-community counterpart. Boundary work, however, is not free. Comparative evidence from Wikipedia treats talk-page activity and revert rates as visible traces of coordination cost, showing that cross-boundary edits face higher scrutiny and survive less often than within-group edits \cite{kittur_he_2007, yasseri_dynamics_2012}. In GitHub code review, Tsay et al. show that social signals -- prior contribution history, community membership -- shape whether a pull request is accepted, revised, or rejected \cite{tsay_influence_2014}, and Halfaker et al. link concentrated rejections of newcomer contributions to later attrition \cite{halfaker_dont_2011}. These threads motivate a structured pair of measurement strategies we adopt jointly here: a \emph{breadth} view -- how many cross-community links exist and how dense they are -- and a \emph{depth} view operationalised through four indicators -- pull-request acceptance, integration latency, review-state mix and issue-comment depth -- so that ecosystem-scale traces speak to whether boundary-spanning collaboration is contested, sustained, and costly, not just whether it occurs.

\section{Methods}
\label{sec:methods}

\subsection{Data and contributor identity}
\label{sec:methods-data}

We collected GitHub commit, pull-request, and issue activity for the 464 repositories listed under the cybersecurity domain in Rawsec's Cybersecurity Inventory (RCI) as of May 2022,\footnote{Rawsec's Cybersecurity Inventory \url{https://inventory.raw.pm/tools.html}} produced by 11{,}372 contributors over October 2001--May 2022. The RCI is a curated inventory of state-of-the-art open-source cybersecurity software: projects are listed because they cover a recognised cybersecurity capability (scanning, exploitation, forensics, threat-intelligence sharing, and so on), not because they share a software-functional dependency. There is, a priori, no expectation that one listed project depends on another in the sense of a package--dependency graph; the corpus is therefore not a software ecosystem in the dependency-network sense \cite{ladisa_taxonomy_2022}, but an \emph{ecosystem of specialisation} in contributing to open-source cybersecurity software, in which contributors and repositories interact across thematic boundaries through shared maintainers, cross-project pull-requests, and issue-level deliberation rather than through code-level coupling. Two complementary streams support the analyses. The first, collected through the GitHub REST API, retains $602{,}103$ commit records with creation/merge timestamps, user identities, and event types, of which $214{,}930$ carry a non-null \texttt{merged\_at} and so land through a \emph{merged} pull-request (\texttt{state=closed}); those records resolve to $69{,}348$ distinct merged pull-requests, roughly three commit records apiece.

\paragraph{GitHub Archive enrichment for collaboration-depth indicators.}\label{sec:gharchive-enrichment}%
The second stream, drawn from the GitHub Archive (\url{https://www.gharchive.org}) via Google BigQuery for February 2011--May 2022, adds every pull-request, review, and issue event in the same window -- including rejected and abandoned PRs (\texttt{state=closed}, \texttt{merged\_at=null}), review states (\texttt{APPROVED}, \texttt{CHANGES\_REQUESTED}, \texttt{COMMENTED}, \texttt{DISMISSED}), and issue-level activity. The two streams support both the intra- versus inter-community collaboration ratios used at the community scale and the four collaboration-depth indicators reported in Section~\ref{sec:results-depth}: pull-request acceptance rate, integration latency, review-state mix (share of merged PRs receiving at least one \texttt{CHANGES\_REQUESTED} review) and issue discussion depth. Each indicator is stratified by whether the actor's home (modal) community matches the recipient repository's community, in line with the comparative literature on cross-boundary coordination cost in peer production \cite{kittur_he_2007, halfaker_dont_2011, yasseri_dynamics_2012, tsay_influence_2014}.

The commit metadata uses a numeric \texttt{author\_uniqueID} in $86.9\%$ of rows; the GitHub Archive stream uses login strings throughout. We canonicalise contributor identity via an \texttt{actor\_id $\leftrightarrow$ actor\_login} map built from the GitHub Archive ($490{,}990$ \texttt{actor\_id}s, $507{,}395$ logins), then exclude $11$ explicit bots (logins matching \texttt{*[bot]} or on a curated automation list) and $688$ phantom git identities (numeric IDs that do not resolve to a login plus the generic placeholders \texttt{root}/\texttt{admin}/\texttt{git}/\texttt{user}), leaving $10{,}490$ canonical human contributors on which all cross-community contributor analyses are computed. The full bot list, and the rule defining the phantom identities, are given in Supplementary~\ref{sec:SI_friction_by_k}, and the sensitivity of the cross-community contributor layer to the breadth cutoff in Supplementary~\ref{sec:SI_carrier_threshold}; the technology annotation of the repository sample is shown in Supplementary~\ref{sec:SI_bipartite_views}.

\paragraph{Contributor breadth, and a non-circular variant.} A contributor's breadth $k$ is the number of distinct Louvain communities they commit to over the observation window. Because a merged cross-community pull-request itself produces a commit in the destination community, $k$ is partly a consequence of pull-request acceptance and cannot be used as a regressor for it. Wherever breadth enters a model of acceptance or integration latency we therefore use $k^{\mathrm{prior}}_{\mathrm{excl}}$, the number of distinct communities the author had committed to strictly before the focal pull-request opened, excluding the destination community. The distinction is developed, and its consequences for the boundary-friction results reported in Section~\ref{sec:results-depth}, in Supplementary~\ref{sec:SI_friction_controls}. The models in that section additionally use pull-request size (\texttt{additions}, \texttt{deletions}, \texttt{changed\_files}, \texttt{commits}), the merging account (\texttt{pr\_merged\_by\_login}), and \texttt{PushEvent} and \texttt{MemberEvent} records, all present in the same GitHub Archive extraction, to construct measures of contributor experience, technical effort and write access in the destination repository.

\subsection{Bipartite community detection}
\label{sec:methods-network}
We model the data as a weighted bipartite graph $G=(U,V,E)$ with contributor nodes $U$, repository nodes $V$, and edges of weight $\log(1+c)$ where $c$ is the per-pair commit count, in line with prior bipartite-affiliation measurement of cooperation structures in peer production \cite{klein_virtuous_2015}. Louvain community detection \cite{blondel_fast_2008} (\texttt{igraph} implementation, resolution $=0.5$) on this weighted bipartite graph yields $163$ non-singleton contributor--repository communities. Per community $c$ we compute intra- and inter-community link densities and their ratio (rescaled to $[0,1]$ for plotting), the contributor-/repository-heavy size coordinates $(|U_c|, |V_c|)$, and the inter-community pull-request (PR) share. At the contributor scale we use the participation breadth $k_i$ -- the number of distinct Louvain communities contributor $i$ touches at the commit level -- and treat the upper-tail population $k_i\geq 7$ identified by the data-driven two-regime fit of Figure~\ref{fig:cross_community}C (Supplementary~\ref{sec:SI_breadth_ccdf_ssr}) as the boundary-spanning carrier layer used throughout the Results. The bipartite graph is the natural object of inference here because community-scale structure is not recoverable from one-mode projections of $G$, which inflate degree and discard the asymmetry between occasional co-presence and sustained co-affiliation \cite{newman_scientific_2001, latapy_basic_2008}; the projection-versus-bipartite diagnostic (adjusted Rand index, ARI; normalised mutual information, NMI; purity; Jaccard) is reported in Supplementary~\ref{sec:SI_bipartite_views}. Throughout, \emph{community} denotes a Louvain community of contributors and repositories on the bipartite graph and not a code module or library dependency.

\subsection{Temporal dynamics}
\label{sec:methods-temporal}
At the ecosystem scale we compute yearly \emph{inflection scores} -- the relative change in total commit counts between consecutive years -- and Gaussian-kernel-smooth them to expose periods of acceleration, deceleration, or stasis (Figure~\ref{fig:methods_network}A). At the community scale we summarise each non-singleton Louvain community by its vector of annual commit totals and visualise the per-community trajectories as a temporal heatmap sorted by birth year (Figure~\ref{fig:anatomy}A); an exploratory unsupervised clustering of these trajectories using a Seurat-style pipeline of log-normalisation, principal component analysis (PCA), uniform manifold approximation and projection (UMAP) and Louvain clustering \cite{mcinnes_umap_2020} is reported in Supplementary~\ref{sec:SI_temporal_clustering} and used only as an exploratory device. We additionally fit a three-parameter logistic to the cumulative-births series $N_\mathrm{born}(y)$ and decompose the ecosystem-wide annual commit total $T(y)$ multiplicatively into an extensive margin (active-community count $N(y)$) and an intensive margin (mean commits per active community $\bar e(y) = T(y)/N(y)$) following the Kaya-identity decomposition tradition \cite{kaya_impact_1989, ang_lmdi_2005}; the resulting log-additive identity $\Delta\log T = \Delta\log N + \Delta\log\bar e$ and the regime-mean decomposition are reported in Supplementary~\ref{sec:SI_decomposition}.

\subsection{Community survival analysis}
\label{sec:methods-survival}
To characterise community survival we treat the residual phase as a right-censored time-to-event outcome. For each non-singleton community $c$ with annual commit series $a_c(y)$ over $y\in[b_c, Y_\mathrm{end}]$, where $b_c=\min\{y:a_c(y)>0\}$ is the birth year and $Y_\mathrm{end}=2021$ is the last full year of GitHub Archive coverage, let $\mu_c$ be the mean yearly activity over that span and $\theta_c = 0.05\,\mu_c$ a residual threshold. The last non-residual year is $y^*_c = \max\{y\in[b_c,Y_\mathrm{end}] : a_c(y)\geq\theta_c\}$; the non-residual span is $y^*_c-b_c+1$ years; communities with $y^*_c = Y_\mathrm{end}$ are right-censored. Across the $163$ non-singleton communities, $55$ enter residual by $2021$ and $108$ are censored. We summarise the cohort dependence as the per-cohort annual hazard
\begin{equation}
h(\text{cohort}) \;=\; \frac{\#\{\text{events in cohort}\}}{\sum_{c\,\in\,\text{cohort}} (y^*_c-b_c+1)},
\label{eq:cohort_hazard_methods}
\end{equation}
which normalises by years-at-risk and is therefore not mechanically driven by the geometric censoring envelope $Y_\mathrm{end}-b_c+1$. We then test whether cross-community connectivity is associated with survival in two complementary ways. First, Kaplan--Meier non-residual probability is estimated separately on tertiles of (i) the inter-community PR share and (ii) $\log_{10}(1+d^\mathrm{ext}_c)$, where $d^\mathrm{ext}_c$ is the symmetric number of other communities $c$ exchanges commits with, and the curves are compared with multi-group log-rank tests. Second, a multivariate Cox proportional-hazards model (\texttt{lifelines.CoxPHFitter}) regresses time-to-residual on the same connectivity covariates while controlling for cohort (birth year) and size ($\log_{10}$\,\#\,repositories, $\log_{10}$\,\#\,contributors). Sensitivity to the inactivity rule -- thresholds at 1\%, 2.5\%, 5\% and 10\% of mean yearly activity, a median-based variant, absolute floors of 5 and 10 commits per year, a threshold-free zero-activity definition, and variants requiring the residual state to persist for two or three consecutive years -- together with the full univariate and multivariate Cox specifications and the static-window/reverse-causation caveats, is documented in Supplementary~\ref{sec:SI_residual} and Supplementary~\ref{sec:SI_hazard_predictors}.

\subsection{Predictive forecasting of repository activity}
\label{sec:methods-forecasting}
To test whether community-level structural features carry predictive content for repository activity beyond short-run autocorrelation, we trained two Random Forest regressions \cite{breiman_random_2001} of six-month commit volume per repository (\texttt{randomForest} in R) -- one with and one without the previous six months of commits as a feature -- and ranked features by the percentage increase in mean square error (MSE) under permutation. Inputs comprised contributor-subtype indicators, cumulative commit activity and repository age, event-ratio scores, accumulated PR counts, repository-level inter-to-intra link-density summaries, community membership labels, the number of inter-community links, and clustering annotations. The full ablation comparison and prediction-error scatters are reported in Supplementary~\ref{sec:SI_forecasting}.

\section{Results}
\label{sec:results}

\begin{figure}[h!]
    \centering
    \includegraphics[width=\textwidth]{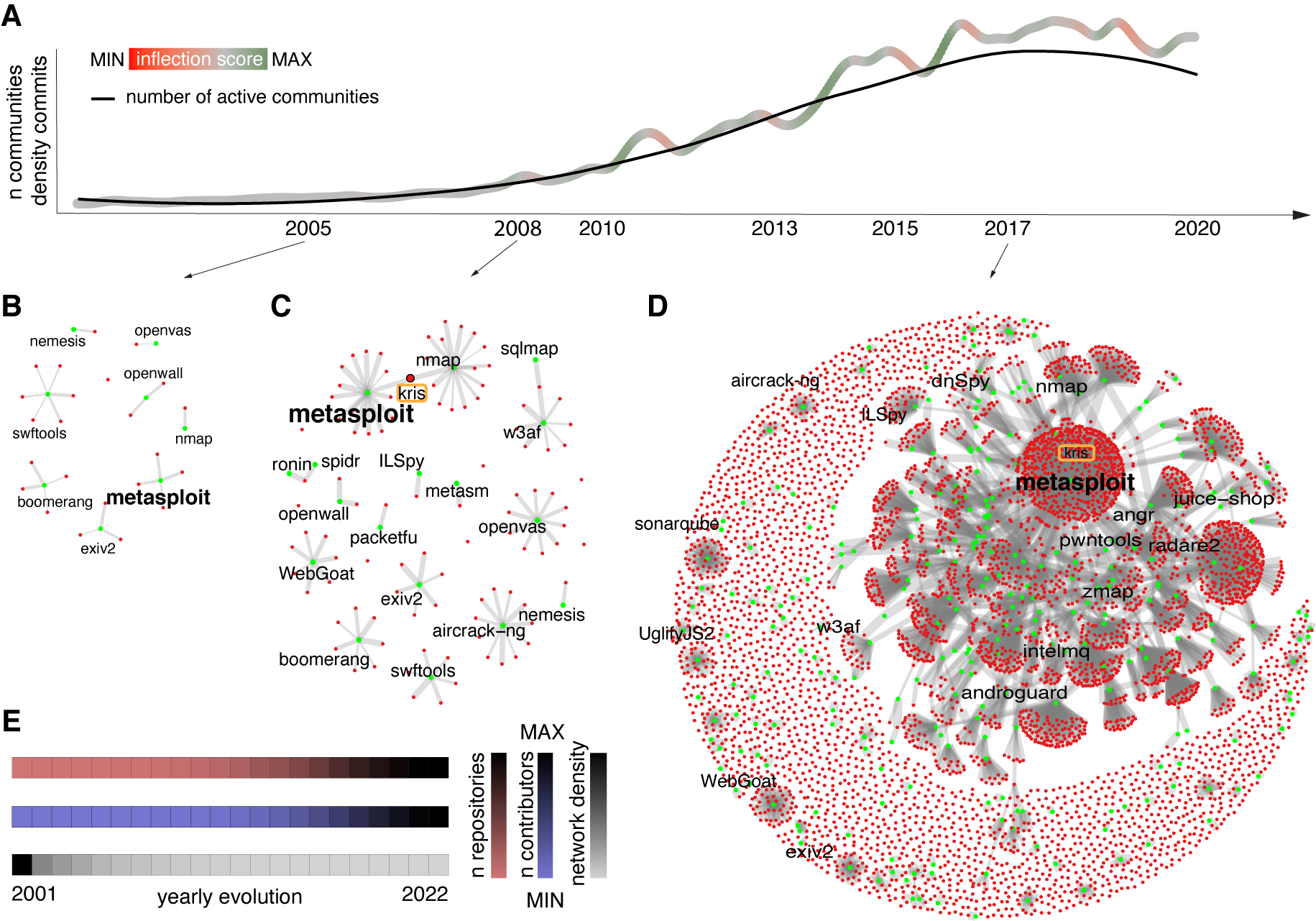}
    \caption{\footnotesize{\textbf{Bipartite contributor--repository graph: ecosystem-level activity, representative snapshots, and longitudinal metrics.} \textbf{A.}~Locally estimated scatterplot smoothing (LOESS) curve fit on the number of communities through time (black line) overlaid on a Gaussian kernel density of commit activity colour-coded by the inflection score (point-to-point differential): red (negative inflection), grey (neutral), green (positive); positive peaks correspond to surges in network-wide development activity. \textbf{B.}~Commit-activity bipartite network snapshot at the end of 2005 (contributors in red, repositories in green, grey edges weighted by commit counts). \textbf{C.}~Snapshot at the end of 2008: one contributor (\emph{kris}, highlighted in orange) marks the onset of measurable cross-repository ties, coinciding with the official launch of GitHub on April 10, 2008. \textbf{D.}~Snapshot at the end of 2017: the ecosystem has settled into a dense community-scale core--periphery layout, with \emph{metasploit} (initially linked to \emph{nmap} by \emph{kris} in 2008) standing out as a highly affiliated repository among interconnected contributor groups. \textbf{E.}~Stack of heatmaps tracking three network-level metrics over time: number of repositories (gradient: low (red) -- high (black)), number of contributors (low (blue) -- high (black)) and global network density (low (grey) -- high (black)); the per-community equivalent of the activity timeline appears in Figure~\ref{fig:anatomy}A.}}
    \label{fig:methods_network}
\end{figure}

The reconstructed bipartite network of contributor--repository interactions, aggregated from October 2001 to May 2022, comprises $11{,}372$ contributors and $464$ repositories drawn from the Rawsec Cybersecurity Inventory. Edges represent at least one commit event and are weighted by interaction frequency. Louvain community detection on this weighted bipartite graph identifies $163$ non-singleton contributor--repository communities, varying in internal cohesion and cross-community coupling. The structural transformation of the ecosystem across three representative snapshots (2005, 2008, 2017) of the bipartite contributor--repository graph and the corresponding longitudinal network-level metrics are summarised in Figure~\ref{fig:methods_network}. Figure~\ref{fig:anatomy} then characterises these $163$ communities at the community level: their temporal development across the observation window (panel~A) and the joint distribution of repositories and contributors per community in the heavy-tailed plane (panel~B).

\subsection{Community structure and cross-community coupling}
\label{sec:results-coupling}
\begin{figure}[hbt!]
    \centering
    \includegraphics[width=\textwidth]{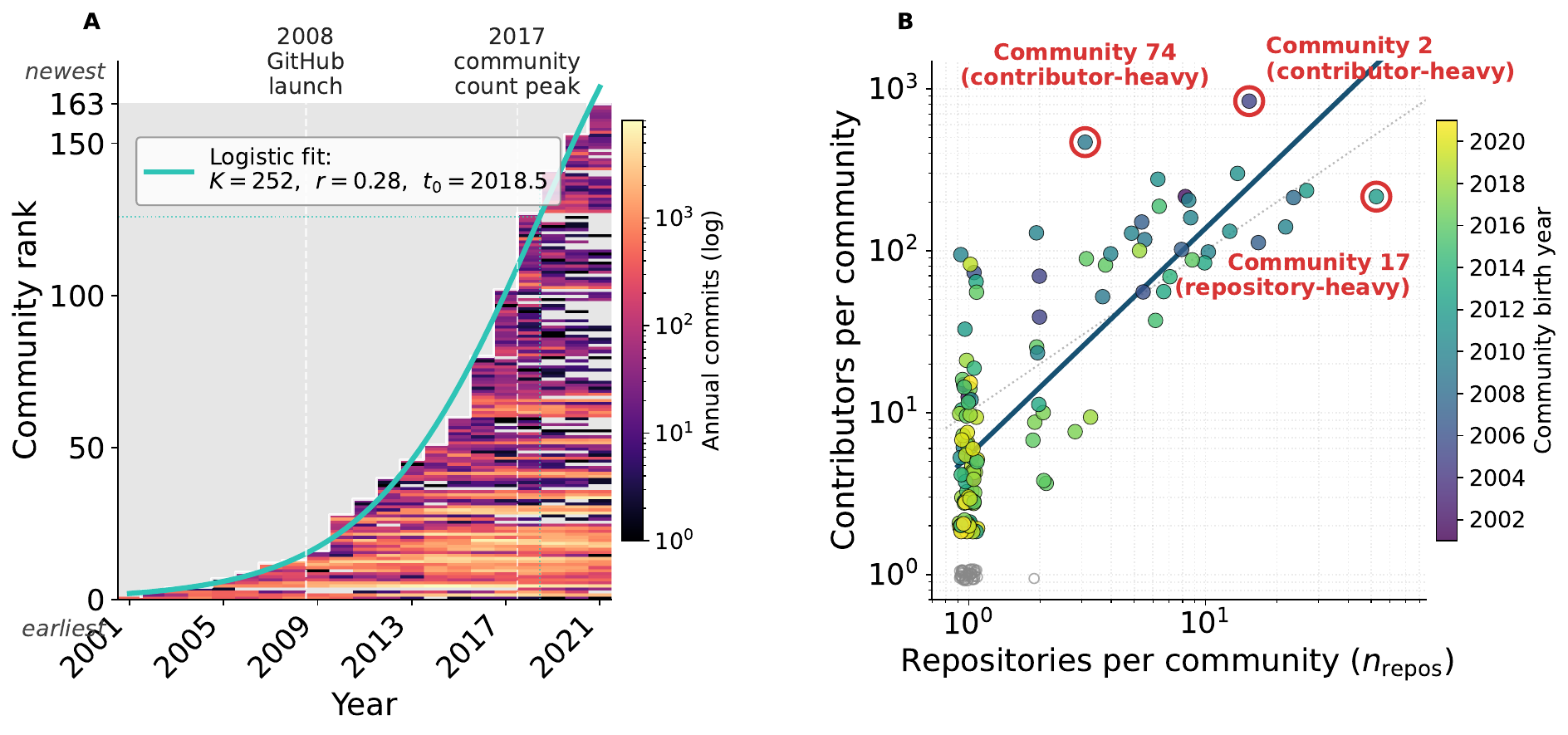}
    \caption{\footnotesize{\textbf{Anatomy of communities: temporal development and size scaling.} \textbf{A.}~Annual commits per community on a log colour scale (rows = $163$ non-singleton Louvain communities sorted by birth year, columns = the full calendar years 2001--2021, the partial year 2022 being excluded; light grey cells are pre-birth or zero-activity years). The white step traces the empirical cumulative number of communities born by each year; the cyan curve is a logistic fit (parameters in the in-panel legend and in the main text). White dashed verticals mark the GitHub launch (April 2008) and the $2017$ peak in community count; the fitted inflection year~$t_0$ and the half-capacity level $K/2$ are the faint cyan dotted lines. \textbf{B.}~Per-community joint distribution in the $(n_\mathrm{repos}, n_\mathrm{contributors})$ plane on log--log axes. Filled markers (viridis by community birth year, sharing the year axis of panel~A) are the $n=124$ communities with $n_\mathrm{contributors}>1$ that enter the OLS power-law fit (blue line; parameters in main text); the $39$ single-contributor communities are shown as small open grey markers and excluded from the fit (justification in Supplementary~\ref{sec:SI_singleton_exclusion}). The dotted diagonal marks $n_\mathrm{contributors}=10\,n_\mathrm{repos}$. Communities~2, 74 (contributor-heavy) and~17 (repository-heavy), cited in the text, are circled in red.}}
    \label{fig:anatomy}
\end{figure}

The structural analysis reveals that the OSS cybersecurity ecosystem exhibits a
highly skewed distribution of contributor–repository associations, with a small
number of communities accounting for a large share of contributors or
repositories. Quantifying this skew across the 163 non-singleton Louvain communities,
the Gini coefficients of the distributions of contributor counts and repository
counts per community are approximately 0.80 and 0.57, respectively (median
community size: 4 contributors and 1 repository; 90th percentile: 109
contributors and 7 repositories). For example, community~2 comprises 868
contributors and 14 repositories, community~74 comprises 502 contributors and 3
repositories, and community~17 comprises 235 contributors and 50 repositories,
illustrating \textbf{contributor-heavy} versus \textbf{repository-heavy} community
extremes in the community--size plane (Figure~\ref{fig:anatomy}B). The contributor-heavy communities such as $2$ and $74$ that combine many active contributors with relatively few repositories sit toward the upper range of inter-to-intra link-density ratios, while smaller and less active communities are largely localised within their own boundaries. 
Across the $124$ multi-contributor communities, contributor count scales superlinearly with repository count: an ordinary-least-squares (OLS) fit in log--log space gives ${n_{\mathrm{contributors}}} \sim {n_{\mathrm{repos}}}^{\beta}$ with $\beta = 1.4(1)$ ($R^{2} = 0.64$, $p < 10^{-3}$), so communities accrue contributors faster than they accrue repositories and the largest communities sit well above a constant-team-size line. 
These labels refer to community-scale affiliation imbalance, not to unipartite degree or eigenvector centrality among individuals; the community-scale structure reported here is not recoverable from one-mode projections of the bipartite graph, which systematically inflate degree and blur fine structure (Supplementary Figure~\ref{fig:SFig7_projection}). Commit activity is positively associated with cross-community coupling at the community scale -- a Spearman rank correlation of $\rho \approx 0.75$ ($p<10^{-3}$) between per-community total commits and the rescaled inter-to-intra link-density ratio (Supplementary Figure~\ref{fig:SFig_commits_vs_ratio}) -- consistent with prior research on modular software architecture and productivity in OSS ecosystems \cite{baldwin_architecture_2006}.
This exponent, close to $4/3$, places the community-scale measurement squarely within the superlinear team-size-to-productivity scaling first reported on individual OSS projects -- where the median exponent is likewise close to $4/3$, and the whole production is more than the sum of its parts through critical cascades of contributions~\cite{sornette_how_2014} -- a regime contested for large teams~\cite{scholtes_aristotle_2016} and subsequently confirmed~\cite{maillart_aristotle_2019}. Our result extends that within-team superlinearity by one level of aggregation: at the community scale, bigger repository portfolios attract disproportionately more contributors, as the same coordination dynamics that make larger OSS teams individually more productive also recruit them at higher rates, possibly offering diversified opportunities for joining and specialisation in a more modular community organisation \cite{von_krogh_community_2003}.

Together these features delineate a pronounced \emph{community-scale} core--periphery pattern in which high-coupling, high-activity communities may act as engines of collaboration and knowledge flow as well as potential points of systemic fragility.

\subsection{Temporal Evolution of the OSS Ecosystem}
\label{sec:results-temporal}

The temporal trajectory of community formation and commit activity in the RCI ecosystem from 2001 to 2022 reveals a sustained increase in the number of active communities through a peak in 2017, followed by a plateau. A LOESS-smoothed curve tracking the number of communities over time (Figure~\ref{fig:methods_network}A) delineates three distinct phases: a period of gradual expansion (2001--2007), a phase of rapid growth (2008--2014), and a slower expansion approaching saturation (2015--2017); the overlaid Gaussian kernel density of commit activity, colour-coded by inflection score, highlights alternating periods of acceleration (green) and contraction (red). The three bipartite-graph snapshots in Figure~\ref{fig:methods_network}B--D trace the corresponding structural transformation, from small disconnected communities before GitHub, through the emergence of the first bridging contributor in 2008, to the dense community-scale layout of 2017.

Longitudinal heatmaps tracking network-level metrics (Figure~\ref{fig:methods_network}E) show that while the number of repositories and contributors grows steadily, overall network density peaks early and then declines sharply. This decline reflects a scaling effect, whereby the addition of new nodes and links does not fully compensate for the dilution of global graph density. Structurally, this reflects a shift toward decentralised collaboration and increasing community specialisation, with growth driven more by the formation of new clusters than by the expansion of existing, tightly coupled core communities. This decentralisation--specialisation dynamic marks a transition from a highly integrated growth phase toward a more federated, domain-focused organisational structure. The same three regimes are visible at the per-community level in the temporal heatmap of Figure~\ref{fig:anatomy}A, which exposes substantial heterogeneity in when individual communities are born, peak, and decline within the observation window. We model the empirical cumulative count of community births $N(t)$ with the canonical Verhulst logistic equation \cite{verhulst_notice_1838} -- the standard model for growth bounded by a finite carrying capacity, in which the per-capita growth rate decreases linearly with current size,
\begin{equation}
\frac{\mathrm{d}N}{\mathrm{d}t} \;=\; r\,N\!\left(1 - \frac{N}{K}\right),
\qquad
N(t) \;=\; \frac{K}{1 + e^{-r\,(t - t_0)}},
\label{eq:logistic}
\end{equation}
where $K$ is the carrying capacity, $r$ the intrinsic growth rate, and $t_0$ the inflection year. Logistic trajectories are the usual form taken by the expansion of a technology domain, including in information security, where scientometric measurement across arXiv e-prints finds most technologies following a logistic-growth process even as security \emph{development} itself does not \cite{percia_david_measuring_2023}. The model is appropriate here for the same reason: the niche the corpus indexes -- canonical OSS cybersecurity capabilities (scanning, exploitation, forensics, threat-intelligence sharing, $\ldots$) -- is itself bounded: once the recognised capability space is largely filled, new communities form less frequently, and the cumulative-births trajectory necessarily saturates. Fitting Equation~\eqref{eq:logistic} to the empirical cumulative-births series over the full calendar years 2001--2021 (Figure~\ref{fig:anatomy}A, white step / cyan curve; the partial-year 2022 is excluded for consistency) returns $K = 252 \pm 27$ communities, $r = 0.28 \pm 0.02$\,yr$^{-1}$, and $t_0 = 2018.5 \pm 0.8$, quantifying the saturation visible in the cumulative trajectory.

These temporal patterns display \emph{at once} the vigour of the OSS cybersecurity ecosystem of communities and the seeds of its fragility. The vigour is visible in the trajectory itself: a roughly two-decade run from a handful of disconnected communities pre-GitHub to a dense, broadly cohesive bipartite graph by $2017$, with cumulative community births well-fit by a single logistic growth law -- a clean signature of an ecosystem whose niche of recognised cybersecurity capabilities was being rapidly populated under the institutional shift to GitHub. The fragility is built into the same trajectory: the carrying capacity $K$ is finite by construction, the births inflection has already passed ($t_0 \approx 2018.5$), and a multiplicative decomposition of ecosystem-wide annual commits into an extensive margin (the active-community count $N$) and an intensive margin (mean commits per active community $\bar e$) shows that $\bar e$ had already started to fall in $2014$, four years before the extensive margin saturated, so the headline ecosystem-wide commit total was sustained for half a decade by an inflow of new communities that masked an internal slowdown invisible to aggregate metrics [Supplementary~\ref{sec:SI_decomposition}, Table~\ref{tab:SI_decomposition_regimes}; the $2014$ peak is robust to the activity floor used to define an active community]. The cohort hazard developed in Section~\ref{sec:results-survival} sharpens the same point quantitatively: communities born after the count-peak transition are intrinsically more residualisation-prone, so the same maturation that produced the ecosystem's vigorous growth phase is also the dynamic by which it has begun to encode its own structural exposure.

The implication for the rest of this section is direct. Once the extensive margin has saturated and the per-community intensive margin has begun to thin, what sustains a matured ecosystem of $163$ non-singleton communities is no longer the inflow of new communities but whatever \emph{cross-community} structure links the existing ones to each other. The next subsection examines that structure at two scales -- community-scale heterogeneity in intra/inter pull-request activity and contributor-scale concentration of community-spanning commits -- before the survival analysis (Section~\ref{sec:results-survival}) and boundary-friction stratification (Section~\ref{sec:results-depth}) quantify the consequences.

\subsection{Cross-Community Collaboration and Structural Drivers of Activity}
\label{sec:results-cross-community}

\begin{figure}[h!]
    \centering
    \includegraphics[width=\textwidth]{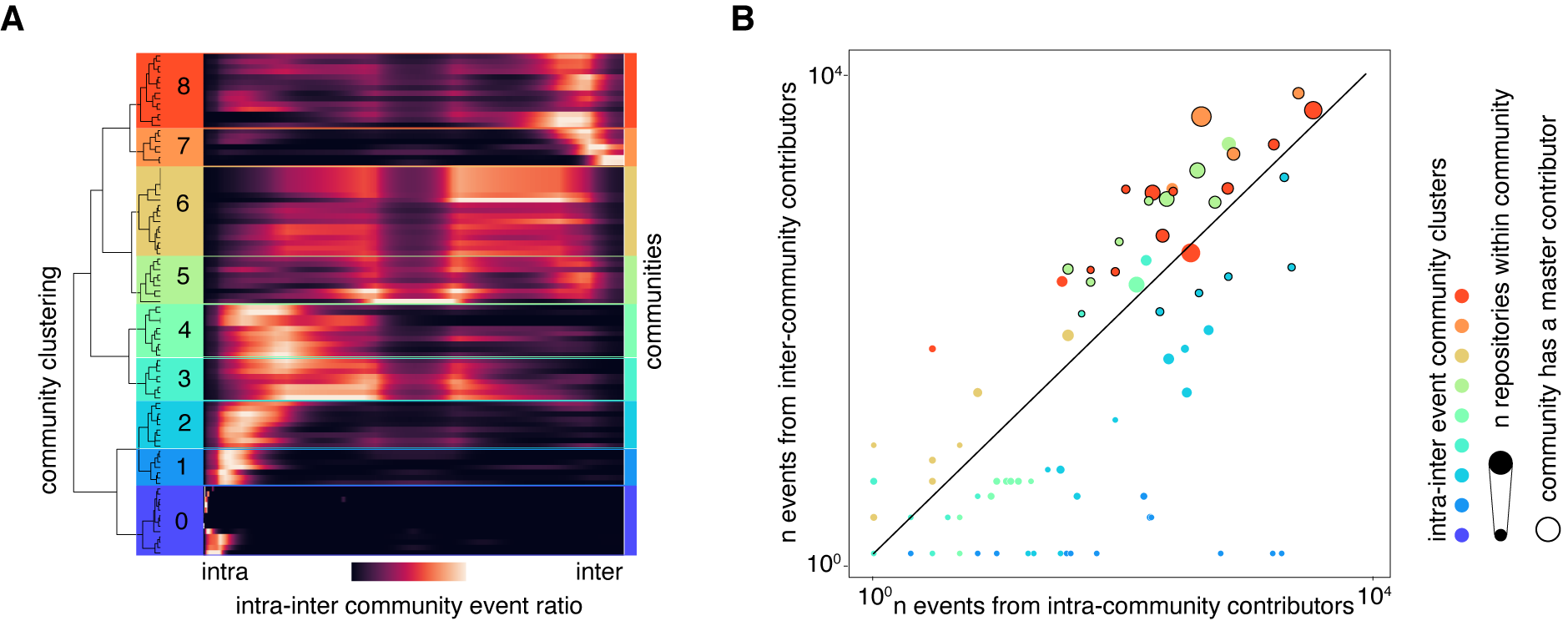}\\[6pt]
    \includegraphics[width=\textwidth]{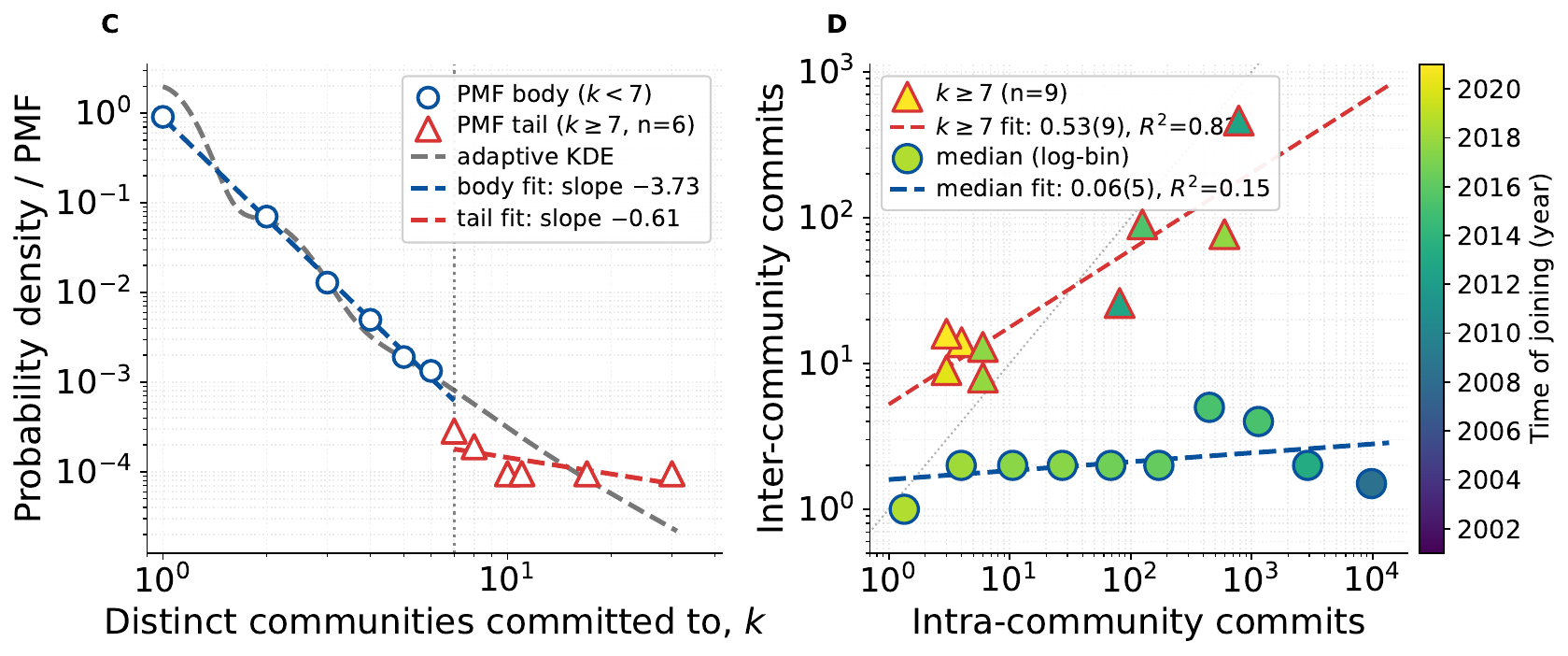}
    \caption{\footnotesize{\textbf{Cross-community collaboration at the community and contributor scales.} \textbf{A.}~Heatmap and dendrogram clustering, by intra-/inter-community pull-request event ratio, those non-singleton communities with enough pull-request activity for the ratio to be defined (rows; $137$ of the $163$ communities record any merged pull-request event at all, and the panel applies a further minimum-activity screen); the ratio runs along the $x$-axis from intra to inter and the gradient fill encodes event density within each bin; cluster colours and IDs shown in the dendrogram. \textbf{B.}~Communities ordered by within-community ($x$) and across-community ($y$) event counts; dot colour follows the panel-A clustering, size encodes constituent repositories, and contoured dots flag communities containing at least one contributor active across more than nine repositories -- the ``master contributor'' overlay retained from the earlier version of this figure, and a repository-count criterion distinct from the commit-level breadth $k$ used in panels C--D. \textbf{C.}~Empirical PMF of contributor breadth $k$ on $10{,}490$ canonical humans after bot/phantom filtering (Methods Section~\ref{sec:methods-data}): blue open circles for the body ($k<7$) and red open triangles for the tail ($k\geq 7$). The tail holds nine contributors spread over six distinct values of $k$, so the panel legend, which counts plotted points, reads $n=6$ where panel~D, which counts contributors, reads $n=9$. Adaptive Abramson KDE on $\log_{10} k$ in grey dashed. OLS body and tail power-law fits in matching colours (blue/red dashed) with breakpoint $k^{*}=7$ chosen by minimum body$+$tail SSR (slopes and SSR scan in main text and Supplementary~\ref{sec:SI_breadth_ccdf_ssr}). \textbf{D.}~Per-contributor intra- vs. inter-community commit split, with markers and colours matching the panel-C regimes. Red triangles ($n=9$): humans at $k\geq 7$ (panel-C tail). Blue circles: per-bin medians across ten log-spaced $n_\mathrm{intra}$ bins over all $966$ multi-community contributors (the nine carriers included, so this is the whole multi-community population rather than the panel-C body alone). Marker fill: year of first intra-community commit (viridis over 2001--2021, matching Figure~\ref{fig:anatomy}B). Dashed lines: OLS power-law fits in matching regime colours (exponents in main text); dotted diagonal: $n_\mathrm{inter}=n_\mathrm{intra}$.}}
    \label{fig:cross_community}
\end{figure}

Following the maturation picture above, we ask how collaboration extends across the boundaries between the $163$ non-singleton communities once the ecosystem stops being driven by the inflow of new ones. We examine that question at two scales: pull-request activity at the community level (Figure~\ref{fig:cross_community}A--B) and the underlying contributor-level commit pattern that supports it (Figure~\ref{fig:cross_community}C--D). A heatmap and dendrogram of intra- versus inter-community pull-request ratios (Figure~\ref{fig:cross_community}A) reveal strong heterogeneity in collaborative behaviour: some communities remain highly inward-focused, while others resolve a substantial portion of development issues across community lines. These more outward-oriented communities cluster together, and their collaborative capacity correlates with community size, contributor engagement, and the presence of cross-community contributors (Figure~\ref{fig:cross_community}B). These patterns reinforce prior findings that sustained engagement and cross-boundary collaboration are central to distributed innovation systems and community productivity \cite{von_krogh_community_2003, moradi-jamei_community_2022}.

The community-level breadth signal observed in Figure~\ref{fig:cross_community}A--B is supported, at the contributor scale, by a sharply heavy-tailed distribution of community participation (Figure~\ref{fig:cross_community}C): of the $10{,}490$ canonical human contributors that remain after canonicalising identity and removing $11$ bots and $688$ phantom git identities (Methods Section~\ref{sec:methods-data}), $9{,}524$ ($90.8\%$) commit only inside a single community, $966$ ($9.2\%$) cross at least one boundary at the commit level, only $43$ ($0.4\%$) span five or more distinct communities, $9$ ($0.09\%$) span seven or more, and the most-spanning contributor reaches $k=30$. The empirical probability mass function (PMF), smoothed by an adaptive Abramson kernel-density estimate (KDE) on $\log_{10} k$, has a two-regime structure with a data-driven inflection at $k^{*}=7$, selected by minimising the body+tail OLS sum of squared residuals (SSR) over candidate transitions (Supplementary~\ref{sec:SI_breadth_ccdf_ssr}). Calibrating straight lines on double-logarithmic axes recovers, for the body ($k<7$), a single power-law regime with PMF slope $\alpha = 3.73$ -- equivalently a complementary CDF (CCDF) exponent $\mu = \alpha - 1 = 2.73$; with $\mu > 2$ the body is \emph{not} heavy-tailed in the standard sense (finite mean and finite variance). The tail ($k\geq 7$, nine humans) is by contrast essentially flat in log--log space (slope $-0.61$): probability is very nearly constant across the upper tail rather than continuing the body's rapid decay. The six-fold change in apparent exponent at $k^{*}=7$ separates the bulk multi-community population from a numerically thin upper tail. We name this tail the \emph{carrier layer} -- the $k \geq 7$ canonical humans (nine individuals after bot/phantom filtering, red triangles in Figure~\ref{fig:cross_community}C--D, panel-C tail) -- and adopt this term throughout the rest of the paper to refer to the contributor population that the analyses below and in Sections~\ref{sec:results-survival}--\ref{sec:results-depth} show to repeatedly cross community boundaries and absorb cross-community integration cost on behalf of the rest of the ecosystem. Among multi-community contributors, the home community (the modal community by commit count) typically still dominates: most points in Figure~\ref{fig:cross_community}D fall below the diagonal, with a median share of inter-community commits equal to 25\% (mean 27\%). Restricted to the carrier layer, inter-community commits scale sublinearly with intra-community commits, $n_\mathrm{inter}\sim {{n_\mathrm{intra}}}^{0.53(9)}$ [$R^{2}=0.82$, $p<10^{-3}$], whereas the per-bin median across all $966$ multi-community contributors is essentially flat in $n_\mathrm{intra}$ [slope $0.06(5)$, $p=0.26$, $R^{2}=0.15$]: the carrier layer therefore sits roughly a decade above the typical contributor at any given $n_\mathrm{intra}$ and absorbs the inter-community workload disproportionately. The boundary-crossing minority therefore remains anchored in a home community while distributing a substantial part of its activity elsewhere -- a pattern whose collaboration-depth consequences are unpacked in Section~\ref{sec:results-depth}.

The value $k^{*}=7$ is the inflection selected from the shape of the breadth distribution and is used to \emph{name} this layer; no conclusion in what follows depends on it. Recomputing every carrier-layer quantity at each cutoff from $k\geq 4$ to $k\geq 10$ (Supplementary~\ref{sec:SI_carrier_threshold}) leaves the disproportion that the argument rests on intact: the layer holds between $95$ and $4$ contributors as the cutoff rises, and is over-represented in cross-community pull-request work by a factor of $51$ to $166$ at every one of them. The sublinear scaling exponent is likewise stable, at $0.36$--$0.53$ across $k\geq 5$ to $k\geq 9$ with overlapping standard errors, and significantly steeper than the flat all-contributor reference throughout. The claim can also be made without any cutoff at all: ranking canonical humans by inter-community pull-request output, half of all cross-boundary work is produced by $36$ contributors ($0.34\%$ of the population), $80\%$ by $249$ ($2.37\%$) and $90\%$ by $479$ ($4.57\%$). Treating $k$ continuously over the $966$ multi-community contributors, breadth and cross-community output are rank-correlated at Spearman $\rho=+0.59$ for pull-requests and $\rho=+0.61$ for commits (both $p<10^{-89}$) -- an association between quantities that share a data-generating step, as Section~\ref{sec:results-depth} makes explicit, rather than a behavioural gradient.

A complementary, predictive angle on the importance of community links comes from a repository-level forecasting exercise reported in Supplementary~\ref{sec:SI_forecasting}. Two random-forest regressions of six-month commit volume per repository are compared: one with the previous six months of activity as a feature (capturing auto-regressive memory), and one without. Removing the temporal-memory term reveals that structural features -- the number of community links, the inter-/intra-community pull-request ratio, repository-level link-density summaries, and lifetime -- retain substantial explanatory power. This is consistent with the cross-community contributor story above: ecosystem embedding is observable enough at the repository scale to forecast future activity even when short-run persistence is removed, providing an additional, independent line of evidence that cross-community structure carries predictive weight beyond local dynamics.

\subsection{Community survival and the cohort hazard}
\label{sec:results-survival}

\begin{figure}[h!]
    \centering
    \includegraphics[width=\textwidth]{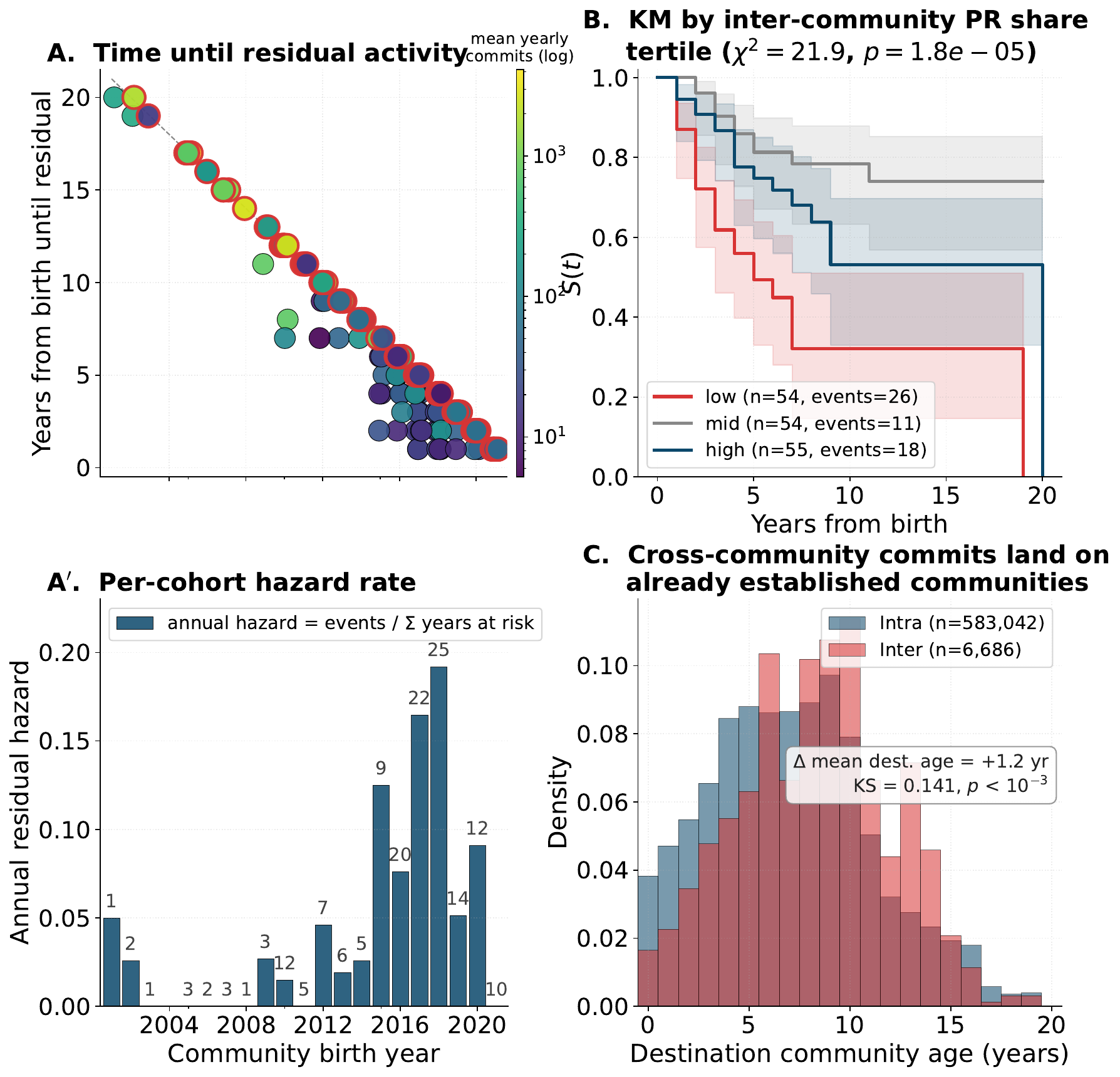}
    \caption{\footnotesize{\textbf{Community survival, per-cohort hazard, and Kaplan--Meier stratification by cross-community connectivity.} A community is in the \emph{residual phase} from the year after the last year in which its activity reached $5\%$ of its mean yearly commits, i.e.\ a dip below the threshold that later recovers does not start the phase; definitions for the residual phase and the per-cohort hazard are given in Methods Section~\ref{sec:methods-survival}. Panels~A and~A$'$ share the birth-year $x$-axis on the left; panels~B and~C each have their own $x$-axis. \textbf{A.}~Time from birth to residual, one marker per community, jittered within its birth year: black-edged markers are the $55$ communities that entered residual, red-edged the $108$ right-censored at $Y_\mathrm{end}=2021$. Marker fill is mean yearly commits since birth (logarithmic colour scale); the grey dashed line is the observation horizon $Y_\mathrm{end}-b_c+1$, which no marker can exceed. \textbf{A$'$.}~Per-cohort annual hazard rate of entering residual; bar labels are cohort sizes. The hazard rises an order of magnitude between pre-2010 cohorts and the $2015$--$2018$ cohorts (peaking at $0.192$ for the $2018$ cohort). \textbf{B.}~Kaplan--Meier non-residual probability by tertiles of inter-community pull-request share, over all $163$ communities (multi-group log-rank $\chi^{2}=21.9$, $p=1.8\times 10^{-5}$; stratum sizes and events in the legend, shaded bands $95\%$ confidence intervals). Communities with no pull-request activity take a share of $0$ rather than being dropped, as for the Cox model of Table~\ref{tab:cox_survival}. \textbf{C.}~Density of destination-community age at the time of the commit -- years elapsed since the receiving community's birth -- for intra-community commits (blue, $n=583{,}042$) against inter-community commits (red, $n=6{,}686$); a contributor's home community is their modal community by commit count. Cross-boundary work targets communities on average $1.2$ years older (two-sample Kolmogorov--Smirnov $D=0.141$, $p<10^{-3}$), that is, already-established ones. The accompanying multivariate Cox proportional-hazards model is in Table~\ref{tab:cox_survival}.}}
    \label{fig:survival}
\end{figure}

The cross-community contributor layer identified above carries boundary work, but its presence does not by itself account for whether communities persist. To characterise community survival we compute, for each non-singleton community, the time from birth to its first year below $5\%$ of its mean yearly commit activity that is never recovered (Methods Section~\ref{sec:methods}; full procedure and threshold sensitivity in Supplementary~\ref{sec:SI_residual}). Of the $163$ non-singleton communities, $55$ have entered residual by $2021$ and $108$ are right-censored at the observation horizon $Y_\mathrm{end}=2021$. Figure~\ref{fig:survival}A plots the per-community time-to-residual against birth year, with markers coloured by mean yearly commits and red edges marking censored cases. Communities born in $2001$--$2010$ sit close to the censoring envelope (most have not yet entered residual after $11$--$20$ years); the $2015$--$2018$ cohorts are sharply bimodal between very short non-residual spans and a tail of censored cases.

The cohort effect is not an artefact of shorter observation windows. Following the standard right-censored survival framework \cite{kaplan_nonparametric_1958, cox_regression_1972}, Figure~\ref{fig:survival}A$'$ reports the per-cohort annual hazard of Equation~\eqref{eq:cohort_hazard_methods}: events divided by the years each cohort was actually observed at risk, so the comparison is not driven by later cohorts having been watched for less time. Pre-2010 cohorts sit at $0$--$0.05\,\text{yr}^{-1}$; the hazard rises sharply for the $2015$--$2018$ cohorts and peaks at $0.192\,\text{yr}^{-1}$ for the $2018$ cohort ($n=25$). Communities born after the community-count peak transition ($\sim 2017$) are intrinsically more likely to drop into residual within any given year, consistent with the broader observation that abandonment and reduced maintenance are themselves cohort-structured features of the OSS landscape \cite{coelho_identifying_2018, coelho_is_2020}.

The residual threshold is not load-bearing. Recomputing the whole analysis under ten alternative inactivity rules -- thresholds at $1\%$, $2.5\%$, $5\%$ and $10\%$ of mean yearly activity, a median-based variant, absolute floors of $5$ and $10$ commits per year, a threshold-free zero-activity definition, and variants requiring the residual state to persist for two or three consecutive years -- leaves the cohort pattern intact in every case, with the pre-2010 to 2015--2018 hazard ratio ranging from $9.8\times$ to $23.8\times$ against $10.5\times$ at the reported threshold (Supplementary~\ref{sec:SI_residual}). The protective association of the inter-community pull-request share reported below is likewise stable in direction (hazard ratio, HR $0.05$--$0.29$) and significant under seven of the ten rules, while the external-degree association, already non-significant in the multivariate baseline, remains so under most of them.

The cross-community contributor layer is associated with longer survival at the community scale. Figure~\ref{fig:survival}B shows Kaplan--Meier non-residual probability stratified by tertiles of inter-community pull-request share: the low-share tertile sustains a $0.32$ ten-year non-residual probability against $0.53$ for the high-share tertile, with a multi-group log-rank statistic of $\chi^{2}=21.9$ ($p=1.8\times 10^{-5}$). The separation is driven by the low tertile rather than being monotone across the three: the middle tertile in fact sits highest, at $0.78$, so what the stratification establishes is that communities with little or no cross-boundary pull-request activity fare markedly worse, not that survival increases steadily with that share. A multivariate Cox proportional-hazards model controlling for cohort and size confirms the protective association of the inter-community pull-request share. We report each predictor's effect as a \emph{hazard ratio} [HR$\,=\exp(\beta)$, the multiplicative change in the per-year residualisation hazard for a one-unit increase in the predictor: HR${<}1$ is protective, HR${>}1$ accelerates the event], accompanied by its $95\%$ Wald \emph{confidence interval} [CI, the 2.5\%--97.5\% range of the hazard ratio under the asymptotic normality of the maximum-likelihood estimate]. For the inter-community pull-request share we obtain HR${\,=\,}0.11$, $95\%$ CI $[0.02,\,0.61]$, $p=0.012$, with model concordance $C=0.843$. That figure is the fitted per-unit coefficient, and a unit of a share is the whole $0$--$1$ range; across the $163$ communities the share itself has median $0.043$ and interquartile range $0$--$0.167$, so the interpretable reading is per interquartile range: a community one interquartile range (IQR) higher in cross-community pull-request share has roughly a $30\%$ lower instantaneous hazard (HR${\,=\,}0.69$ per IQR, $0.66$ per standard deviation). The full model is reported in Table~\ref{tab:cox_survival} and discussed in Supplementary~\ref{sec:SI_hazard_predictors}.

Stratifying the same survival outcome by tertiles of the symmetric inter-community degree $\log_{10}(1+d^{\mathrm{ext}}_c)$ -- the count of \emph{distinct} other communities a focal community exchanges commits with, irrespective of pull-request volume -- gives a sharper univariate separation still than the pull-request share of Figure~\ref{fig:survival}B (curves not shown; the underlying model is reported in Supplementary~\ref{sec:SI_hazard_predictors}): the high-degree tertile sustains a $\sim 0.95$ ten-year non-residual probability against $\sim 0.10$ in the lowest tertile (multi-group log-rank $\chi^{2}=49.1$, $p=2.2\times 10^{-11}$). In the multivariate Cox model, however, this effect attenuates substantially (HR$=0.19$, $95\%$ CI $[0.02,\,2.09]$, $p=0.176$): much of the apparent univariate protection from external reach is mediated by community size, which is itself protective and strongly correlated with the number of external communities a community manages to engage. The cross-community degree therefore reads as a partial proxy for community resourcedness rather than as an independent causal mechanism, but the qualitative ordering -- communities with broader cross-community reach are dramatically more likely to remain non-residual -- is unambiguous in the data, and is consistent with the carrier-layer reading: communities that succeed in being engaged by carrier-layer contributors are also the ones that survive longest. We read this as an end-of-window association rather than a causal-direction test: communities that survived longer accumulated more cross-community pull-requests in part because they were around to receive them. The recognition effect documented at the contributor scale (Section~\ref{sec:results-depth}, Supplementary~\ref{sec:SI_friction_controls}) concerns how cross-community work is integrated \emph{once it occurs}, and operates through a contributor's prior relationship with a particular destination rather than through membership of the carrier layer as such. The cohort hazard documents that survival depends on more than the availability of such relationships: late-cohort communities are over-represented among the residualisation events. Established relationships therefore make cross-boundary work cheap where they exist, but cannot, by themselves, equalise cohort-level access to that work; the carrier-layer-and-cohort tension is the substrate of the broader optimisation/resilience discussion we return to in Section~\ref{sec:discussion}.

\begin{table}[h]
\centering
\footnotesize
\caption{Multivariate Cox proportional-hazards model for community residualisation on $n=163$ non-singleton Louvain communities ($55$ events, $108$ censored at $Y_\mathrm{end}=2021$; concordance $C=0.843$). Hazard ratios with $95\%$ Wald confidence intervals; the cross-community connectivity covariates are tested while controlling for cohort and size. Hazard ratios are per unit of the listed predictor on its own scale (years for birth year, $\log_{10}$ units for log-transformed predictors); for share-valued predictors a unit is the full $0$--$1$ range, which for the inter-community PR share is far wider than the observed interquartile range of $0$--$0.167$ (per IQR, HR${\,=\,}0.69$). \textbf{Bold rows} mark predictors significant at $p<0.05$.}
\label{tab:cox_survival}
\begin{tabular}{lrrrr}
\toprule
predictor & HR & $95\%$ CI lower & $95\%$ CI upper & $p$ \\
\midrule
\textbf{Birth year}                                  & $\mathbf{1.18}$ & $\mathbf{1.03}$ & $\mathbf{1.35}$ & $\mathbf{0.021}$ \\
$\log_{10}$ \# repositories                  & $0.21$ & $0.01$ & $3.87$ & $0.292$ \\
$\log_{10}$ \# contributors                  & $0.39$ & $0.15$ & $1.02$ & $0.055$ \\
\textbf{Inter-community PR share}                    & $\mathbf{0.11}$ & $\mathbf{0.02}$ & $\mathbf{0.61}$ & $\mathbf{0.012}$ \\
$\log_{10}$ \# external communities reached  & $0.19$ & $0.02$ & $2.09$ & $0.176$ \\
Share of contributors with $k\geq 2$        & $4.72$ & $0.39$ & $56.6$ & $0.221$ \\
\bottomrule
\end{tabular}
\end{table}

\subsection{Boundary Friction and Collaboration Depth}
\label{sec:results-depth}

\begin{figure}[h!]
    \centering
    \includegraphics[width=\textwidth]{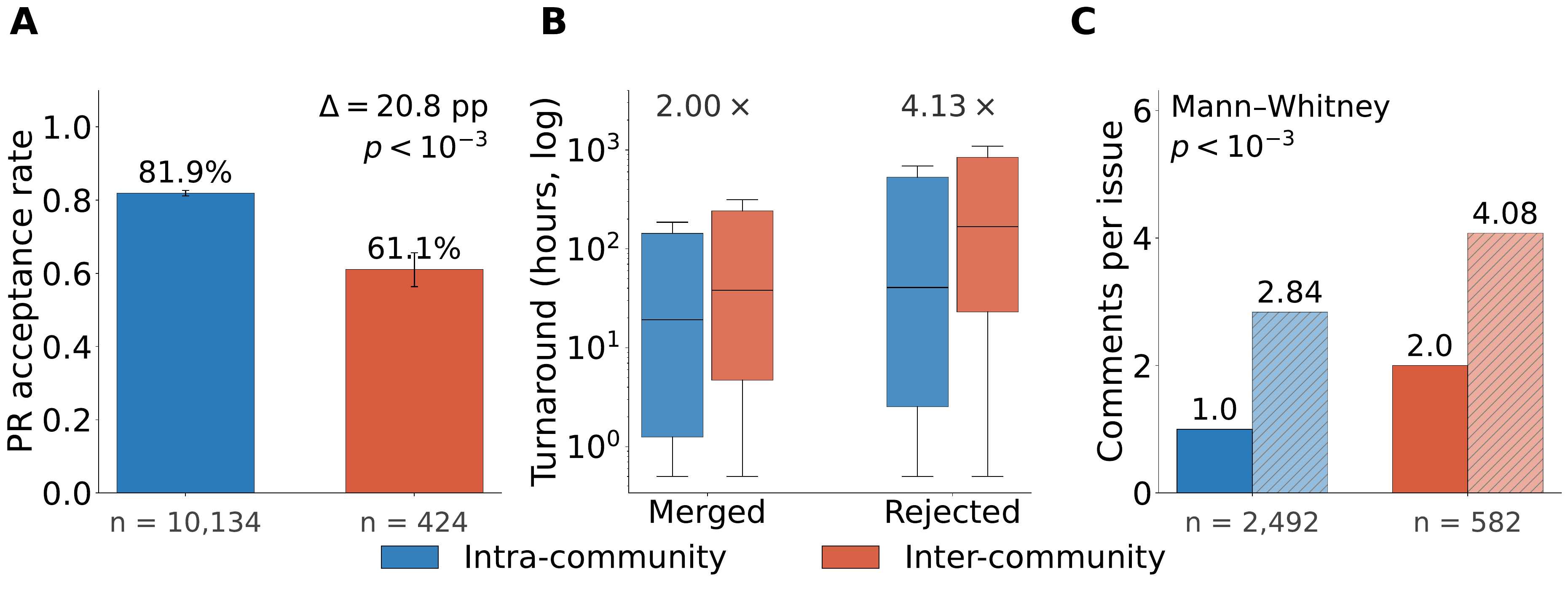}
    \caption{\footnotesize{\textbf{Boundary friction at the pull-request and issue level.} Each panel compares an intra-community to an inter-community slice of the same depth indicator (colour legend below); per-bar sample sizes are annotated beneath the axis in panels A and C. \textbf{A.}~Pull-request acceptance rate, per (repository, PR number) pair on the terminal state, with Wilson $95\%$ confidence intervals; the gap is $20.8$ pp. \textbf{B.}~Turnaround distribution split by outcome, counted in pull-request-linked commit records rather than distinct pull requests (merged: $n=249{,}061$ intra, $5{,}693$ inter; rejected: $n=11{,}309$ intra, $883$ inter); box heights show the IQR around the median and the whiskers are drawn at a fixed fraction of the IQR for legibility on log axes rather than as a percentile statistic; $\times$-labels are the inter-to-intra median ratio. \textbf{C.}~Median (solid) and mean (hatched) comments per issue, by the issue author's home community relative to the recipient repository. The smaller secondary CHANGES\_REQUESTED signal is reported in the main text and stratified by author breadth $k$ in Supplementary~\ref{sec:SI_friction_by_k}.}}
    \label{fig:Fig6_depth}
\end{figure}

Cross-community coupling metrics reported above capture the \emph{breadth} of inter-community collaboration -- how much of it takes place and where. A complementary question, motivated in Background Section~\ref{sec:background}, concerns its \emph{depth}: how contested, sustained, and costly cross-community work is in practice. The primary commit$+$PR dataset preserves only merged pull requests; to probe depth directly we draw on the second GitHub Archive event stream described in Methods Section~\ref{sec:methods-data} and derive four depth indicators -- pull-request acceptance rate, integration latency, review-state mix and issue discussion depth after a boundary crossing. Each indicator is computed both globally and stratified by whether the submitter's home community matches the recipient repository's community, yielding a principled intra- versus inter-community comparison that is summarised in Figure~\ref{fig:Fig6_depth}.

At the \emph{pull-request} level, boundary-crossing contributions are markedly more costly on every depth indicator (Figure~\ref{fig:Fig6_depth}). The submitter's home community resolves for $10{,}558$ of the $83{,}075$ closed pull requests ($12.7\%$; the remainder are authored by contributors with no commit history in the corpus and cannot be classified). On that subset, intra-community submissions are accepted at $81.9\%$ [$n=10{,}134$] whereas inter-community submissions are accepted at $61.1\%$ [$n=424$], a $20.8$ percentage-point gap that is highly significant ($\chi^{2} = 113.9$, $p=1.4\times10^{-26}$; panel A). Integration latency amplifies the same pattern: the median time between a pull request's creation and its terminal state is $19.2$ hours within-community versus $38.3$ hours across-community for merged PRs (ratio $2.0$), and widens to $40.6$ hours versus $168.0$ hours for rejected PRs (ratio $4.1$; $p<10^{-3}$; panel B). A smaller secondary signal echoes the same pattern: among the merged pull-requests carrying at least one review event -- $17.6\%$ of post-2017 merged pull-requests, rising to $27.4\%$ if any review signal, including inline review comments, is counted -- $13.0\%$ of intra-community PRs receive at least one \texttt{CHANGES\_REQUESTED} review [$n=6{,}062$] against $8.3\%$ for inter-community PRs [$n=351$] (Fisher exact-test odds ratio $1.66$, $p = 0.008$), so inter-community contributions are not merely slower but are handled more \emph{binarily}, with less iterative revision and a higher share of outright rejection \cite{tsay_influence_2014}. At the \emph{issue} level, the same boundary accrues coordination cost in deliberation rather than code review: of the $164{,}125$ issues in the corpus, $3{,}074$ can be assigned an intra/inter status by the author's home community, and on those inter-community issues attract a median of $2$ comments against $1$ for intra-community issues [$n=582$ versus $n=2{,}492$; means $4.08$ versus $2.84$; Mann--Whitney $p=1.3\times10^{-29}$; panel C], paralleling Kittur et al.'s observation that cross-boundary work in peer-produced systems is reflected in talk-page volume \cite{kittur_he_2007}.

Whether a contributor can close that gap, and by what means, is the substantive question. Stratifying the same indicators by the author's commit-level breadth $k$ appears to show that they can (Supplementary~\ref{sec:SI_friction_by_k}), but the stratification cannot be taken at face value: $k$ counts the communities an author commits to over the whole window, and a merged cross-community pull-request itself produces a commit in the destination community, so $k$ is in part an outcome of the acceptance it is invoked to explain. Among the $746$ inter-community pull-requests submitted by authors with no prior commit in the destination community, $67.0\%$ of the merged requests are followed by that author appearing as a committer there, against $16.9\%$ of the rejected ones. Recomputing breadth strictly before the request, and excluding the destination community, removes the gradient: across $51{,}709$ closed pull-requests [$2{,}409$ inter-community, $6{,}447$ canonical human authors, $354$ repositories], the interaction between breadth and boundary-crossing is statistically indistinguishable from zero in every specification, including one with author fixed effects absorbing all time-invariant contributor quality (Supplementary~\ref{sec:SI_friction_controls}). What does survive controls for experience, prior activity volume, prior acceptance elsewhere, pull-request size, and repository and year fixed effects is a recognition effect tied to the contributor's prior relationship with the \emph{specific} destination: prior write access in the target repository is associated with an \emph{odds ratio} of $2.26$ on acceptance [OR$\,=\exp(\beta)$, the multiplicative change in the odds of a merge for a one-unit increase in the predictor; against the $68.8\%$ baseline acceptance of inter-community requests by non-insiders this corresponds to a predicted $83\%$] and a $66\%$ reduction in integration latency [$p<10^{-14}$ and $p<10^{-19}$ respectively], and prior pull-requests to the same repository with an OR of $1.70$ per decade [$p=0.0006$]. Boundary-crossing contributions are integrated cheaply where a relationship with the destination already exists, not where the contributor is broadly spread across the ecosystem.

The cross-community work that bridges these costs is concentrated in a very thin layer of contributors -- the same minority isolated at the contributor scale in Figure~\ref{fig:cross_community}C--D. Of the same $10{,}490$ canonical human contributors, $880$ ($8.4\%$) ever author a merged pull request that crosses a community boundary; the top $10$ produce $37\%$ of all $4{,}015$ inter-community merged pull-request records, the top $50$ produce $54\%$, and the top $250$ produce $80\%$. A record here is a commit landing through a merged pull request, so the $4{,}015$ records correspond to $2{,}529$ distinct inter-community pull requests, and the concentration shares quoted above are computed on the record base (Supplementary~\ref{sec:SI_carrier_threshold}). Without the filter, the two Dependabot accounts together would account for $\sim 24\%$ of the unfiltered $5{,}612$ inter-community pull-request volume (\texttt{dependabot[bot]} alone $18\%$). Commit-level breadth and pull-request-level cross-boundary activity are two views of the same underlying events rather than two independent measurements: a merged inter-community pull-request lands a commit outside its author's home community, so every contributor holding one is multi-community by construction. The non-trivial question is how strongly the two scale together \emph{within} that population, and there the association is moderate rather than near-perfect: across the $966$ multi-community contributors, breadth is rank-correlated with inter-community pull-request output at Spearman $\rho = +0.59$ ($p<10^{-89}$). A correlation computed over all $10{,}490$ canonical humans would reach $\rho \approx 0.95$, but that figure reflects only the $90.8\%$ of contributors sitting at $k=1$ with zero cross-boundary output and should not be quoted (Supplementary~\ref{sec:SI_carrier_threshold}). Within the 255 contributor--repository pairs with three or more cross-community merged pull requests, a fixed-effects regression of $\log$-turnaround on within-pair sequence index yields a small but significant negative slope ($\beta=-0.0010$, $p=0.005$); we report this only as a within-pair effect, since the marginal medians by sequence index in fact \emph{rise} with sequence due to selection (pairs that recur tend to involve repositories with longer baseline turnaround).

This carrier layer is also unevenly distributed across the ecosystem and visibly fragile. Of the 163 non-singleton communities, $64$ ($39\%$) have never had a single human contributor with cross-community pull-request experience; in another $21$ communities ($13\%$) cross-community contributors are present in the historical record but \emph{all} have gone dormant in the final twelve months of the observation window (no commit since 2021-05-31). Together, $85$ communities ($52\%$) currently lack any active cross-community contributor at all, and only $78$ ($48\%$) retain at least one. That shortfall is not spread evenly over the ecosystem's history but concentrated in the later cohorts: $14\%$ of communities born in $2001$--$2010$ lack an active cross-community contributor, against $35\%$ of the $2011$--$2014$ cohort, $68\%$ of $2015$--$2018$ and $58\%$ of $2019$--$2021$ ($\chi^{2}=27.5$, $3$ d.f., $p=4.7\times10^{-6}$; Cram\'er's $V=0.41$). The gradient is, however, largely a size effect: median community size falls from $99$ contributors in the earliest band to $2$ in the latest, and once $\log_{10}$ community size is controlled the association with birth year disappears (odds ratio $1.01$ per year, $p=0.81$; likelihood-ratio $\chi^{2}=0.06$ against size alone). Late-born communities are under-served because they are small, not because of their vintage as such. The depth indicators in Figure~\ref{fig:Fig6_depth} therefore portray a coherent pattern we call \emph{friction with stable trust}: cross-community work is measurably more contested on every dimension, yet is carried by a thin, identifiable layer of contributors whose presence in a community co-varies with its cross-boundary capacity. Modular ecosystems are workable not because friction is absent, but because a small set of recurring contributors absorbs it; the nearly half of communities that currently lack any active cross-community contributor are correspondingly more exposed to fragmentation when boundary-crossing collaboration is needed.

\section{Discussion}
\label{sec:discussion}

Our analyses jointly support a \emph{recognition \& repeat-relationship account} of sustained cross-boundary work in modular socio-technical ecosystems. Because the design is observational throughout, we keep three registers distinct in what follows. \emph{Measured} quantities are counts, shares, rates, latencies and review-state mixes read directly off the corpus. \emph{Associated} quantities are model estimates -- hazard ratios, odds ratios, log-rank separations, rank correlations -- which describe covariation under stated controls and carry no directional claim. \emph{Interpretation} is the recognition \& repeat-relationship account itself, which we offer as a reading consistent with the first two registers rather than as a mechanism the design can demonstrate. Boundary-crossing pull-requests are accepted 20.8 percentage points less often than within-community ones (61.1\% vs.\ 81.9\%), take two to four times longer to integrate, and attract a more binary review pattern -- gaps that sit in the range the peer-production literature associates with active coordination effort at group boundaries rather than technical noise \cite{kittur_he_2007, halfaker_dont_2011, yasseri_dynamics_2012, tsay_influence_2014}. The friction is not, however, a permanent property of the boundary. Controlling for contributor experience, prior activity volume, prior acceptance elsewhere, pull-request size, and repository, year and author fixed effects, boundary-crossing requests are integrated at close to within-community cost wherever the contributor already has a relationship with the destination: prior write access in the target repository is associated with $2.3$-fold odds of acceptance and a $66\%$ shorter integration latency, and prior pull-requests to the same repository with a further $1.7$-fold per decade (Supplementary~\ref{sec:SI_friction_controls}). We read this as a recognition account in a specific sense -- what is recognised is a history with a particular destination, not a general standing across the ecosystem. A contributor's breadth over \emph{other} communities carries no acceptance or latency premium at the boundary once breadth is measured before the request and the destination community is excluded from it; an apparent breadth gradient in the raw data arises instead because a merged cross-community pull-request itself creates a commit in the destination community, and so inflates the very breadth measure invoked to explain it.

Two features of this carrier of inter-community collaboration, established in Section~\ref{sec:results-depth}, matter for what follows. It is numerically thin -- fewer than one contributor in ten ($8.4\%$) ever authors a boundary-crossing merged pull-request, and within that minority a few dozen individuals produce the majority of the volume, the same individuals the commit-level breadth measure isolates. And it is unevenly distributed: rather more than half of the $163$ non-singleton communities currently host no active cross-community contributor at all, either because they never had one or because the ones they had have gone dormant. Read against Benkler's account of commons-based peer production \cite{benkler_penguin_2011}, this pattern suggests that the health of an OSS ecosystem's inter-community links does not rest on friction being absent but on whether a small subset of contributors actively spans community boundaries -- extending to the inter-community scale the small integrating core that earlier OSS literature documents \emph{within} projects \cite{crowston_core_2006, valverde_self-organization_2007, bird_latent_2008, sornette_how_2014}, and playing the brokerage role through which recombination travels \cite{fleming_collaborative_2007}. The static-Cox analysis in Supplementary~\ref{sec:SI_hazard_predictors} establishes this as an end-of-window association, not as a causal-direction test: the recognition account offered here concerns how cross-boundary work is integrated \emph{once it occurs}, and the coverage figure is a fragility signal at the carrier of inter-community collaboration rather than a demonstrated structural risk to community survival. The same reading extends to the second cross-community survival signal (Figure~\ref{fig:survival}C): the symmetric inter-community degree $d^{\mathrm{ext}}_c$ separates the ten-year non-residual probabilities of the highest and lowest tertiles by an order of magnitude, but the multivariate Cox attenuates the effect substantially because external reach is mediated by community size. The cross-community degree therefore reads, like the inter-community PR share, as a partial proxy for community resourcedness rather than as an independent causal mechanism for survival -- a reading the inactivity-rule sensitivity supports, since its multivariate effect survives under only three of ten alternative definitions while the inter-community PR share survives under seven (Supplementary~\ref{sec:SI_residual}); what both panels agree on is the qualitative ordering -- communities that succeed in being engaged across many boundaries are dramatically more likely to remain non-residual, the same population on which the carrier layer's repeat cross-boundary submissions land.

\paragraph{What we can and cannot optimise for.}
The observation window closes at May~2022, immediately before the mainstream diffusion of large-language-model (LLM) assisted development tools \cite{moradi_dakhel_github_2023, chen_evaluating_2025}. The depth indicators reported here therefore characterise \emph{human-coordinated} boundary-crossing in OSS at ecosystem scale, before this technology became part of the day-to-day workflow. What follows in this paragraph is forward-looking and rests on a single curated corpus: we set out what an AI-mediated ecosystem could plausibly change and how it would be measured, not what it has changed, and every expectation stated below is conditional and untested here. Each indicator has a plausible mechanism by which AI-assisted authoring and review could shift it: acceptance might rise if assistants raise baseline code quality or fall if they enable lower-effort contributions; latency might compress as AI summarisation triages faster; the review-state mix might tilt toward iterative revision as suggestions become cheap to apply; and the recognition effect itself might attenuate if AI lowers the cost of establishing credibility with an unfamiliar destination. We therefore present the corpus and the depth-indicator panel as a measurable \emph{pre-AI baseline}: replicating the same panel on a 2023--2026 OSS corpus would let claims about ``the impact of AI on open collaboration'' rest on side-by-side ecosystem-scale evidence rather than on rhetoric. Three AI-driven shifts deserve specific attention in such a follow-up. First, on \emph{trust-building}: the controlled analysis of Section~\ref{sec:results-depth} attributes the inter-community pull-request acceptance gap to a recognition cost that contributors without a prior history at the destination pay on each submission, and AI-mediated review and contextual summarisation could in principle compress that cost by surfacing the contribution's provenance and prior code in a form that a busy maintainer can absorb cheaply, attenuating the gap between one-off and repeat cross-boundary submitters; whether this happens in practice depends on whether AI suggestions are read as a credible trust signal or as another low-effort artefact, an open question on which the existing usability and security evidence is mixed \cite{vaithilingam_expectation_2022, pearce_asleep_2022}. Second, on \emph{cohort-level access to cross-community attention}: late-cohort communities in our corpus are sharply over-represented among residualisation events (Section~\ref{sec:results-survival}) and the communities currently without an active cross-community contributor are themselves concentrated in those later cohorts (Section~\ref{sec:results-depth}), so the carrier-layer supply we document does not, on its own, reach late-born communities at anything like the rate it reaches early ones -- a shortfall that tracks community size rather than vintage, and that the recognition account, resting as it does on accumulated destination-specific history, gives small and recent communities no obvious way to close -- though we do not directly test the contributor-side routing that would explain the gap. Cognitive substrates that could plausibly underlie an asymmetric routing of contributor attention toward already-visible destinations include the \emph{availability heuristic} of Tversky and Kahneman \cite{tversky_availability_1973} and a \emph{status-quo bias} \cite{samuelson_status_1988}; AI-driven recommenders trained on activity traces of the kind we report here would, if blindly applied, be well placed to amplify any such asymmetry \cite{felzmann_transparency_2019, sambasivan_re-imagining_2021}, and counteracting it would require explicitly fairness-aware routing of contributor attention towards under-engaged communities -- a measurable design objective rather than a hope. Third, on \emph{alleviating the cohort hazard}: the survival-side counterpart to the under-served-late-cohort pattern is the per-cohort residualisation hazard documented in Section~\ref{sec:results-survival}, which rises an order of magnitude between pre-2010 cohorts and the $2018$ cohort. AI tools that lower the operational entry cost for small or late-cohort communities -- automated dependency-update pull-requests, AI-assisted documentation, generative authoring of boilerplate code -- could in principle reduce that hazard by substituting for some of the maintenance labour the carrier layer is currently asked to provide \cite{moradi_dakhel_github_2023, chen_evaluating_2025}. There are limits, however, to what such an optimisation can deliver. The mechanisms our analyses surface -- recognition (Section~\ref{sec:results-depth}) and the cohort-structured fragility of the carrier layer (Section~\ref{sec:results-survival}) -- are not failure modes that an efficiency-oriented optimiser can simply remove, because the same dynamics that produce the friction also produce the destination-specific histories on which cheap integration rests; the Highly Optimised Tolerance literature warns explicitly that systems optimised for expected workloads become acutely fragile to unanticipated shocks \cite{may_will_1972, holling_resilience_1973, carlson_highly_2000} -- how that fragility would unfold in an AI-mediated OSS ecosystem, and what governance or platform mechanisms could pre-empt it, remains an open question at this stage of the technology's diffusion. AI-assisted productivity gains in OSS ecosystems are therefore best read as a question of what \emph{can} be optimised away (boilerplate, dependency hygiene, documentation drift) versus what \emph{should not} be (the human social fabric of trust accumulation that the recognition account rests on, and the bot/human identification boundary on which our identity reconciliation depends -- given the thinness of the carrier layer, nine humans at $k\geq 7$, even small misclassifications between human and AI agents would qualitatively shift the picture). The depth-indicator panel of this paper provides the measurement substrate to tell those two apart. A deeper question, however, hangs over the next decade of OSS production. The technology is moving fast enough that an increasing share of code on public platforms is now authored or co-authored by large language models \cite{li_competition_2022, moradi_dakhel_github_2023, chen_evaluating_2025}, and it is genuinely uncertain how much of the contributor distribution itself will remain human in any meaningful sense by the end of this decade. The carrier layer the present paper isolates -- a thin tail of repeat cross-boundary humans on whom the cross-community fabric of a $464$-repository ecosystem currently rests -- is precisely the kind of human-mediated structure that an aggressive AI-substitution strategy could displace: directly, by routing cross-community pull-requests through LLM agents instead of repeat human contributors; and indirectly, by eroding the trust-accumulation process on which the destination-specific recognition effect of Section~\ref{sec:results-depth} rests once authorship signals stop reliably indexing a human history of repeated cross-boundary engagement. Whether such a substitution is welfare-improving, and on what terms, is not a question optimisation alone can answer. \emph{What we ultimately cannot optimise for is our own place in the digital societies these ecosystems prefigure}: the carrier layer is not just a piece of infrastructure to be made more efficient, it is part of how a community of humans recognises its own collective work, and that recognition is what an OSS-style commons ultimately rests on \cite{benkler_penguin_2011}.

\paragraph{Digital societies and the scope of transfer.}
Cybersecurity OSS is part of the digital infrastructure on which many societal functions rely, and OSS-like codes and processes already structure other domains of digital production: in live coding for music and performance, creators modify executable code in real time, often projecting it publicly and iterating through rapid, auditable edits that can be shared, remixed, and reused across a community \cite{collins_live_2003, collins_live_2011}; on Wikipedia, distributed editors recursively co-produce article quality through bipartite contributor--article cooperation patterns whose dynamics resemble those of OSS \cite{wilkinson_cooperation_2007, klein_virtuous_2015}. In parallel, the broader epistemic environment in which open collaboration takes place is being reshaped by dramatic AI-driven advances in science -- from near-experimental-accuracy protein structure prediction \cite{jumper_highly_2021} and scaled materials discovery \cite{merchant_scaling_2023} to a generalised reframing of scientific discovery itself \cite{wang_scientific_2023}, with code-generation systems now matching competitive-programming performance \cite{li_competition_2022} -- which makes trustworthy validation, provenance, and review at the boundaries of communities the binding constraint, not authoring throughput. A plausible vision -- not tested here as a welfare claim -- is that robust digital societies will need institutions and infrastructures that preserve community-level autonomy \emph{and} workable cross-boundary coordination under uncertainty; OSS ecosystems are historically where those tensions first became observable at scale through public traces. This systems perspective has broader implications: as digital infrastructures increasingly shape the design, coordination, and adaptation of physical systems -- urban mobility \cite{batty_big_2013}, energy grids \cite{farhangi_path_2010}, and civic platforms \cite{linders_e-government_2012, mahajan_participatory_2022} -- the architecture of open, adaptive software ecosystems becomes directly relevant to the governance of digital societies \cite{pohle_digital_2020, helbing_cocreating_2024, wang_digital_2024}. Two things should be distinguished when asking how far this travels. The \emph{measurement strategy} -- carrier-layer identification, friction stratification with a destination-excluding breadth measure, and survival analysis (Sections~\ref{sec:results-cross-community}--\ref{sec:results-depth}) -- carries over directly wherever an analogous time-evolving bipartite interaction graph can be reconstructed, whether for citizen-science platforms, federated open-data ecosystems or multi-team enterprise software. Whether the \emph{empirical pattern} recurs there -- a thin carrier layer, a boundary cost waived by prior relationships with the destination, a cohort-structured hazard -- is untested by us and is offered as a hypothesis, not a finding. We treat parallels to urban, mobility, or energy systems as hypothesis-generating, not as findings of this paper.

One such parallel is worth stating explicitly, because it follows from the temporal decomposition rather than from the carrier layer. Until about $2014$ the ecosystem's aggregate output grew on both margins at once; between $2014$ and $2018$ per-community intensity was already falling while continued community births kept the total rising; after $2018$ the extensive margin saturated as well and the total turned down. What then remained available to sustain aggregate activity was the coupling among the communities that already existed (Section~\ref{sec:results-temporal}). The structure of that transition -- a saturating extensive margin, a declining intensive margin, and an aggregate that conceals the second behind the first for several years -- is not specific to software. It is the generic situation of any system whose expansion has been carried by an inflow of new units and whose inflow then stops, which is the arithmetic that urban and regional systems face under demographic stagnation: value creation and attractiveness have to come increasingly from exchange among existing units rather than from their multiplication \cite{batty_big_2013, helbing_cocreating_2024}. We test nothing demographic here and make no claim about cities; the corpus supplies a worked instance in which such a transition is measurable end-to-end, and with it one transferable warning. For roughly four years the ecosystem-wide commit total kept rising while per-community activity was already declining, so an observer optimising on the aggregate alone would have recognised the change of regime four years late. That is a concrete instance of the distinction this special issue asks about: the quantity that was easiest to measure and optimise was precisely the one that concealed the transition, and the structure that mattered afterwards -- cross-community coupling -- is the harder one to see.

\paragraph{Limitations.}
An open-source ecosystem has no natural boundary. Repositories are bound together by contributors, dependencies, forks, shared standards and shared threat models in an open-ended web, and no membership criterion closes; any empirical delineation is therefore a choice of \emph{seed}, and what an analysis describes is the collaboration structure induced by that seed rather than a population that exists independently of it. The Rawsec Cybersecurity Inventory is best read in those terms -- a human-curated list of security tooling whose closure under contributor co-participation is what our bipartite graph represents. The $464$ repositories are not a random draw, not an exhaustive enumeration and not a sample frame, and the $163$ communities are communities \emph{of that seed}; community counts, coverage shares and cohort compositions are correspondingly seed-relative. The corpus is also single-domain and GitHub-only, so GitLab, self-hosted and fork-level activity are invisible to it.

What the seed selects for can be measured rather than asserted. Using the star and fork traces in the same GitHub Archive extraction, the seeded repositories are a high-visibility set (median $974$ unique stargazers, first quartile $343$), and within the corpus visibility is associated with the behaviour we analyse: across the $375$ repositories carrying both measures, stargazers correlate with the cross-community pull-request \emph{share} at Spearman $\rho=+0.18$ ($p=0.0004$), and the probability that a repository has no boundary-crossing pull-request at all falls monotonically from $56.4\%$ in the least visible quartile to $0\%$ in the most visible one. Because inclusion in a curated inventory plausibly tracks visibility, the corpus over-represents precisely the repositories whose cross-boundary behaviour we measure, and an ecosystem-wide rate of cross-community collaboration would very likely be \emph{lower} than the levels reported here. The exposure is uneven across our results. Levels and coverage shares -- the share of communities without an active cross-community contributor, the per-cohort hazard, the community and contributor counts -- are seed-relative and should be read as such. Internal contrasts, in which the seed is held fixed on both sides of the comparison, are considerably less exposed: the intra- versus inter-community friction gap, the review-state mix, and the controlled per-pull-request results, which are identified within repositories and, in the strongest specification, within authors.

Two measurement caveats bear repeating. The $k\geq 7$ threshold is a single cut through a continuous affiliation distribution, so every quantity resting on it is also reported at nearby cutoffs (Supplementary~\ref{sec:SI_carrier_threshold}); and commit-derived breadth is not exogenous to acceptance -- a merged cross-community pull-request produces a commit in the destination community -- a circularity affecting any affiliation measure built from accepted contributions, which is why breadth enters the friction models in a time-resolved, destination-excluding form (Methods Section~\ref{sec:methods-data}, Supplementary~\ref{sec:SI_friction_controls}). The survival results are end-of-window associations: the year-resolved specification returns a null on both connectivity terms it carries and has no counterpart for the pull-request share of Table~\ref{tab:cox_survival}, so we do not claim that cross-community connectivity protects communities (Supplementary~\ref{sec:SI_time_varying_cox}). Sentiment-level analysis of review comments, organisational-affiliation proxies and off-platform coordination remain out of scope; the most informative extension would expand the seed by snowballing along contributor overlap, testing whether the carrier-layer statistics are properties of the ecosystem or of the seed. Beyond that, future work could combine ethnography or surveys with network traces and examine how embedded coding assistants change collaboration patterns on platforms, including fairness and transparency risks \cite{binns_fairness_2018, felzmann_transparency_2019, sambasivan_re-imagining_2021}.

\section{Conclusion}
\label{sec:conclusion}

This paper measures the architecture of cross-boundary collaboration in a longitudinal OSS corpus seeded from a curated cybersecurity inventory. At the ecosystem scale, per-community contributor count scales superlinearly with repository count ($n_\mathrm{contributors}\sim {n_\mathrm{repos}}^{1.4}$) and community formation follows a logistic trajectory whose inflection falls at $t_0 = 2018.5 \pm 0.8$, with the fitted carrying capacity ($K = 252 \pm 27$) still well above the $163$ communities observed. Within that maturing structure we isolate a thin \emph{carrier layer} -- nine canonical humans at $k\geq 7$ -- that carries a disproportionate share of cross-community work, but is unevenly distributed across the $163$ non-singleton communities and cannot, by itself, equalise the cohort-level hazard of residualisation. What makes a boundary crossing cheap to integrate, once controls for experience, effort and repository and author heterogeneity are in place, is not breadth as such but a prior relationship with the destination. Late-cohort communities are over-represented among the residualisation events, leaving them under-served on cross-community attention. OSS now underpins the digital systems on which public services, critical infrastructures and the reliability of cybersecurity itself depend, and the way it sustains work across its own internal boundaries is therefore worth treating as a source of design intuition rather than as a special case. The measurement template developed here -- carrier-layer identification, boundary-friction stratification, and survival analysis -- transfers to any time-evolving bipartite interaction system in which boundary-crossing carries an integration cost, and the mechanism it isolates -- that durable relationships, rather than breadth as such, are what make crossings cheap -- is one that cities and critical-infrastructure networks could deliberately cultivate.

Because the corpus closes just before the mainstream diffusion of large-language-model coding assistants, the boundary-friction gap, the carrier-layer thinness, and the cohort hazard together constitute a measurable pre-AI baseline against which AI-mediated OSS ecosystems can be compared. What such a comparison would put at stake -- set out in Section~\ref{sec:discussion} -- is the distinction between what OSS production can afford to optimise away and what it cannot: the carrier layer is not only infrastructure to be made more efficient, but part of how a community of humans recognises its own collective work.

\section*{Declarations}

\backmatter


\bmhead{Data and Code Availability}
The datasets generated and analyzed during the current study, as well as the code scripts used, are available in the repository \url{https://doi.org/10.17605/OSF.IO/5RWEK}.

\bmhead{Competing interests}
Not applicable.

\bmhead{Funding}
Thomas Maillart (T.M.), Lucia Gomez Tejeiro (L.G.T.) and Thibaut Chataing (T.C.) acknowledge financial support by armasuisse Science and Technology (S+T).

\bmhead{Author contributions}
T.M. and L.G.T conceptualized the research. L.G.T, T.M. and T.C. performed the analyses, T.M., L.G.T wrote the manuscript. Julian Jang-Jaccard (J.J.J.) and Alain Mermoud (A.M.) revised and improved the manuscript.

\bmhead{Use of large language models (LLMs) and AI assistants}
During the preparation of this work, the authors used large-language-model assistants (Claude and GitHub Copilot) to support code refactoring and figure-generation scripts, and to copy-edit prose for clarity. After using these tools, the authors reviewed and edited the output as needed and take full responsibility for the content of the publication. LLMs do not satisfy authorship criteria and are not listed as authors; all research questions, study design, analytical decisions, interpretation of results, and conclusions are the authors' own. This disclosure follows the Springer Nature editorial policy on the use of generative AI in scientific writing.





\clearpage
\section*{Supplementary Material}
\setcounter{section}{0}
\renewcommand{\thesection}{S.\arabic{section}}
\renewcommand{\thesubsection}{\thesection.\arabic{subsection}}
\setcounter{figure}{0}
\renewcommand{\thefigure}{S\arabic{figure}}
\setcounter{table}{0}
\renewcommand{\thetable}{S\arabic{table}}
\renewcommand{\theHsection}{SM.\arabic{section}}
\renewcommand{\theHsubsection}{SM.\arabic{section}.\arabic{subsection}}
\renewcommand{\theHfigure}{SM.\arabic{figure}}
\renewcommand{\theHtable}{SM.\arabic{table}}

\noindent\textbf{Contents}
\smallskip

{\footnotesize
\begin{tabular}{@{}p{1.2cm}p{12cm}r@{}}
\ref{sec:SI_bipartite_views} & Bipartite-graph representation and aggregate ecosystem views & p.~\pageref{sec:SI_bipartite_views} \\
\ref{sec:SI_anatomy} & Community anatomy: temporal trajectories and size scaling & p.~\pageref{sec:SI_anatomy} \\
\hspace{0.8em}\ref{sec:SI_singleton_exclusion} & Why single-contributor communities are excluded from the size-scaling fit & p.~\pageref{sec:SI_singleton_exclusion} \\
\ref{sec:SI_temporal_evolution} & Temporal evolution of ecosystem-wide commit activity & p.~\pageref{sec:SI_temporal_evolution} \\
\hspace{0.8em}\ref{sec:SI_decomposition} & Multiplicative decomposition of ecosystem-wide annual commits & p.~\pageref{sec:SI_decomposition} \\
\ref{sec:SI_community_scale} & Community-scale cross-community coupling & p.~\pageref{sec:SI_community_scale} \\
\ref{sec:SI_contributor_scale} & Contributor-scale cross-community structure & p.~\pageref{sec:SI_contributor_scale} \\
\hspace{0.8em}\ref{sec:SI_breadth_ccdf_ssr} & Diagnostic views of the contributor-breadth distribution & p.~\pageref{sec:SI_breadth_ccdf_ssr} \\
\hspace{0.8em}\ref{sec:SI_carrier_threshold} & Sensitivity of the carrier layer to the breadth cutoff & p.~\pageref{sec:SI_carrier_threshold} \\
\ref{sec:SI_forecasting} & Repository activity forecasting & p.~\pageref{sec:SI_forecasting} \\
\ref{sec:SI_survival} & Community survival: residual phase and hazard predictors & p.~\pageref{sec:SI_survival} \\
\hspace{0.8em}\ref{sec:SI_residual} & Per-community residual phase: time and hazard & p.~\pageref{sec:SI_residual} \\
\hspace{0.8em}\ref{sec:SI_hazard_predictors} & Predictors of residualisation hazard (static-covariate Cox) & p.~\pageref{sec:SI_hazard_predictors} \\
\hspace{0.8em}\ref{sec:SI_time_varying_cox} & Time-varying incoming connectivity: a null & p.~\pageref{sec:SI_time_varying_cox} \\
\ref{sec:SI_friction} & Boundary friction and collaboration depth & p.~\pageref{sec:SI_friction} \\
\hspace{0.8em}\ref{sec:SI_friction_by_k} & How boundary friction varies with contributor breadth $k$ & p.~\pageref{sec:SI_friction_by_k} \\
\hspace{0.8em}\ref{sec:SI_friction_controls} & Do the friction gradients survive controls? & p.~\pageref{sec:SI_friction_controls} \\
\ref{sec:SI_legacy} & Legacy analyses retained for traceability & p.~\pageref{sec:SI_legacy} \\
\end{tabular}
}

\medskip

\section{Bipartite-graph representation and aggregate ecosystem views}\label{sec:SI_bipartite_views}

This section assembles the figures that establish the data substrate of the study: the cybersecurity OSS ecosystem viewed as a single bipartite contributor--repository graph, the longitudinal community-level metrics that motivate the per-community treatment in Figure~\ref{fig:anatomy}A, and the comparison of the bipartite representation against its unipartite projections that motivates working with the bipartite graph directly throughout Methods and Results.

\begin{figure}[H]
    \centering
    \includegraphics[width=0.9\textwidth]{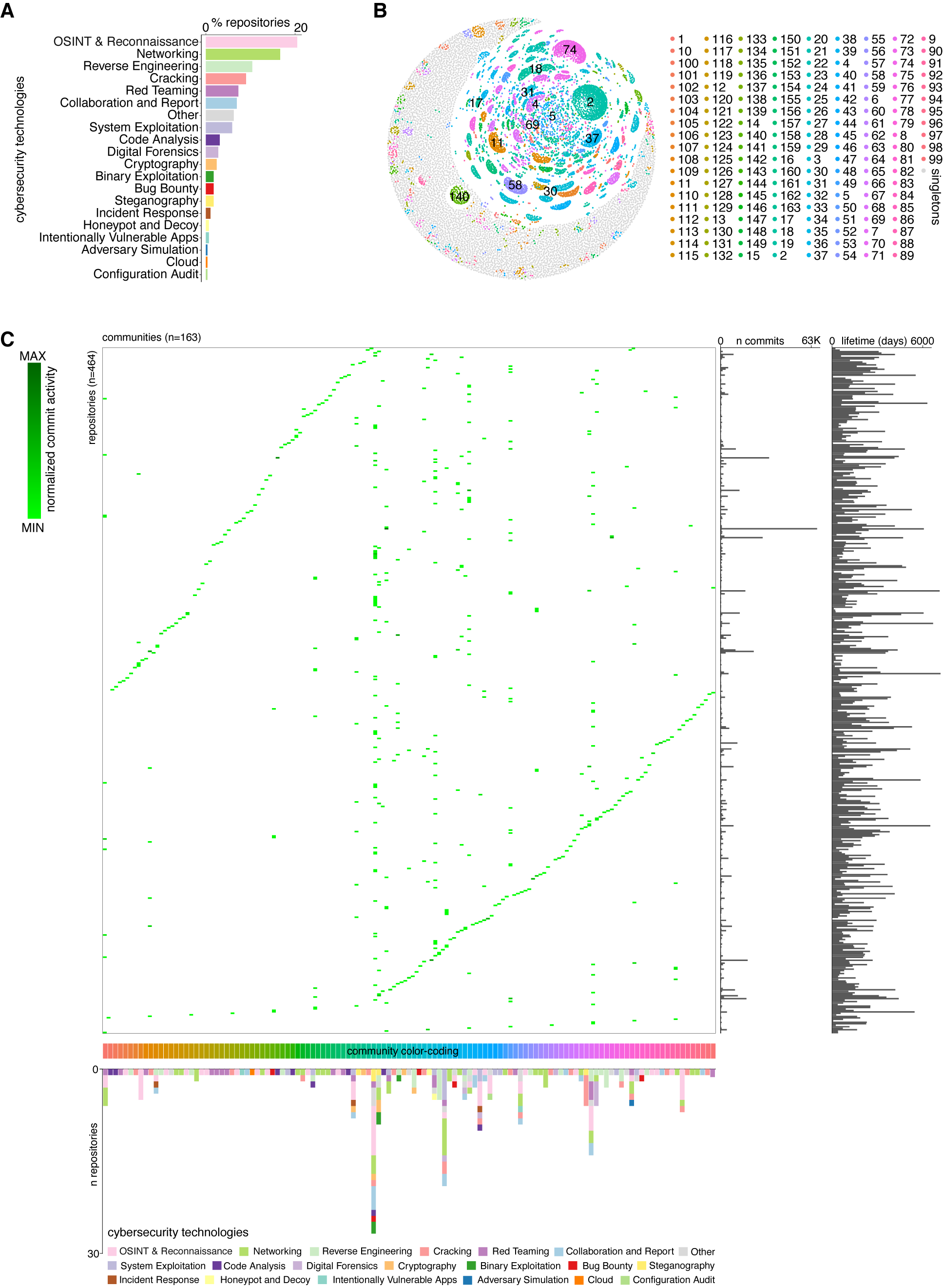}
    \caption[Activity patterns and representation of OSS technologies and communities]{\footnotesize{{\bf Activity patterns and representation of OSS Technologies and Communities under study. } {\bf A.} Barplot illustrating the representation of cybersecurity technologies in the OSS ecosystem under study: percentage of repositories by technology subtype. {\bf B.} Scatter plot depicting coordinates of Bipartite network built on commit relationships between contributors and repositories [historical aggregate] color-coded by identified communities. Numerical ID and color-coding associated with each community is detailed in legend. Bigger communities are highlighted in network by displaying their IDs. {\bf C.} Heatmap illustrating the normalized commit activity of each repository and  associated community: number of commits divided by repository lifetime [color gradient: light-green (relative low normalized activity) to dark-green (relative high normalized activity)]. Y-axis joint barplots indicate counts of number of commits and repository lifetime. X-axis joint barplot indicates composition of technologies for each community [technology color-coding as in A., community color-coding as in B.].}}
    \label{fig:SFig1_gitpaper}
\end{figure}

\begin{figure}[H]
    \centering
    \includegraphics[width=\textwidth]{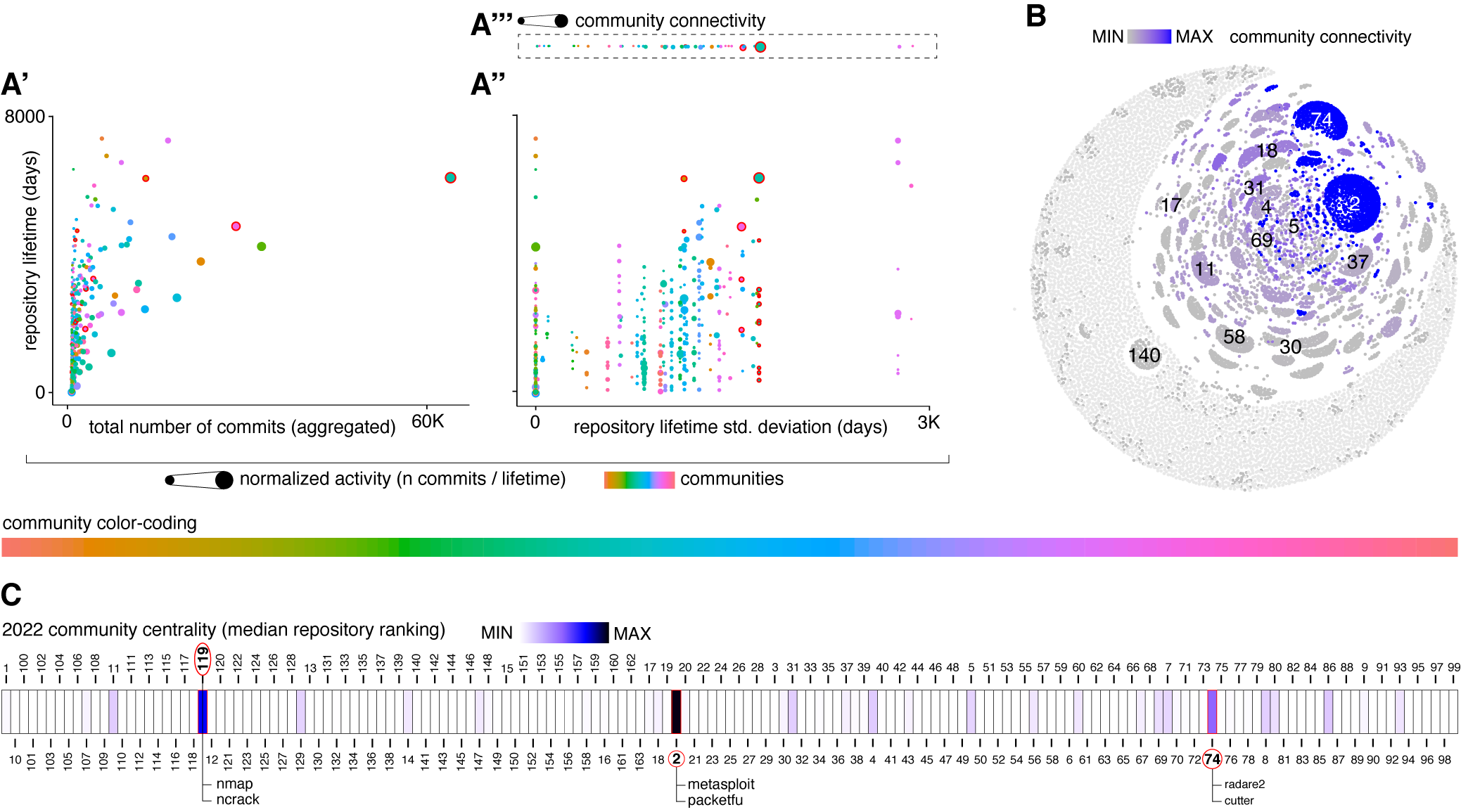}
    \caption[Community lifetime, connectivity and centrality]{\footnotesize{{\bf Community Lifetime, Connectivity and Centrality. }{\bf A'.} Scatter plot depicting repositories ordered by their total number of commits through time (y-axis) and their lifetime in days (y-axis; days of activity). Color-coding groups repositories by community belonging [rainbow discrete palette from orange to pink]. Dot size indicates repository normalized activity (total number of commits divided by lifetime) {\bf A''.} Scatter plot depicting repositories ordered by their lifetime standard deviation [variation in days between commit activity registry] (x-axis) and their lifetime in days (y-axis; days of activity). Color-coding groups repositories by community belonging [rainbow discrete palette from orange to pink]. Dot size indicates repository normalized activity (total number of commits divided by lifetime) {\bf A'''.} Scatter plot depicting repositories ordered by their lifetime standard deviation [as in A'' x-axis] and sized by their inter community connectivity. In {\bf A'.}, {\bf A''.} and {\bf A'''.}, red-contoured dots correspond to repositories whose community is among the top-3 centrality in 2022 network. {\bf B.} Bipartite network scatter plot color-coded by community inter connectivity [gradient scale: grey (low inter connectivity) to blue (high inter connectivity)]. ID highlighted communities: biggest in terms of contributors and / or highest in terms of commit activity. {\bf C.} Heatmap color-coded by centrality score of communities as of 2022 aggregated network [gradient scale: grey (low centrality) to blue (high centrality)]. Red-contoured fields correspond to repositories whose community is among the top-3 centrality. For those, top 2 centrality repositories are indicated.}}
    \label{fig:SFig2_gitpaper}
\end{figure}

\begin{figure}[H]
    \centering
    \includegraphics[width=0.95\textwidth]{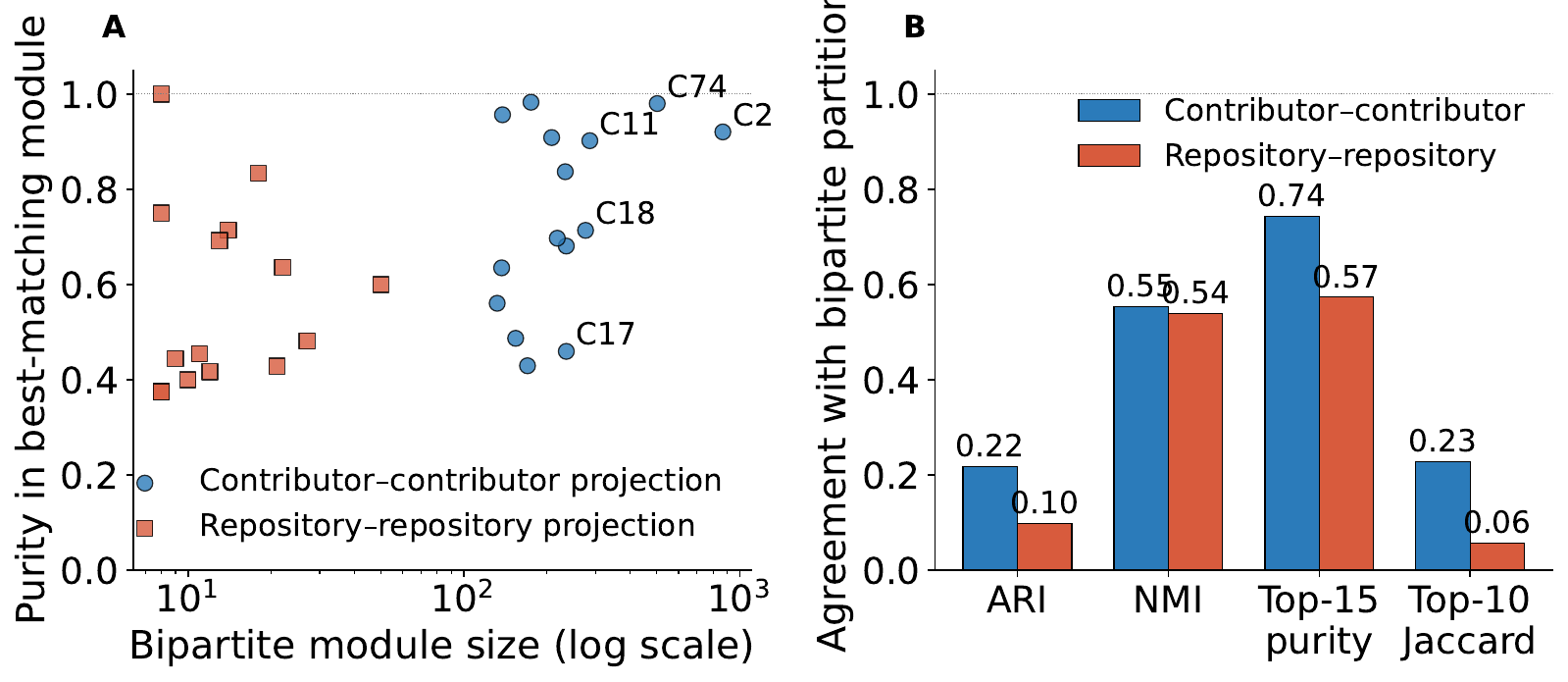}
    \caption{\footnotesize{{\bf Information content of the bipartite representation.} The community structure reported in Results on the bipartite graph $G=(U,V,E)$ is not recoverable from unipartite projections of~$G$. \textbf{A.}~Per-community purity versus community size for the 15 largest bipartite communities -- ranked by contributor count for the contributor--contributor series and by repository count for the repository--repository series, so the two sets only partly overlap -- under the contributor--contributor projection [edge weight $=$ number of shared repositories; blue circles] and the repository--repository projection [edge weight $=$ number of shared contributors; orange squares]: large bipartite communities are partially preserved on the contributor side (mean purity~$0.74$) but substantially collapsed on the repository side (mean purity~$0.57$). \textbf{B.}~Aggregate agreement with the bipartite partition [adjusted Rand index, normalised mutual information, mean top-15 purity, mean top-10 Jaccard] for the contributor--contributor and repository--repository projections. The combined evidence quantifies the information loss that motivates using the bipartite representation directly as the object of inference \cite{newman_scientific_2001,latapy_basic_2008}, rather than a one-mode projection of it.}}
    \label{fig:SFig7_projection}
\end{figure}

\section{Community anatomy: temporal trajectories and size scaling}\label{sec:SI_anatomy}

This section supports Figure~\ref{fig:anatomy}B of the main text by justifying the singleton exclusion in the size-scaling fit.

\subsection{Why single-contributor communities are excluded from the size-scaling fit}\label{sec:SI_singleton_exclusion}

The OLS power-law fit reported alongside Figure~\ref{fig:anatomy}B uses the $124$ non-singleton Louvain communities for which $n_\mathrm{contributors} > 1$, and excludes the $39$ communities for which $n_\mathrm{contributors} = 1$. The exclusion is principled rather than aesthetic. The quantity that the fit characterises is how many contributors a community accrues as it accrues repositories, and that scaling is undefined for a community whose contributor count is fixed at one regardless of $n_\mathrm{repos}$: $38$ of the $39$ excluded points sit at the corner $(n_\mathrm{repos}, n_\mathrm{contributors}) = (1, 1)$ and the remaining one at $(2, 1)$, so the excluded subset encodes no within-community contributor-pair collaboration and is a degenerate strip rather than the lower tail of a distribution. We retain the singletons in the panel as small open grey markers for visual completeness only.

\section{Temporal evolution of ecosystem-wide commit activity}\label{sec:SI_temporal_evolution}

This section quantifies the per-community temporal heterogeneity visible in Figure~\ref{fig:anatomy}A through a multiplicative (Kaya-style) decomposition of ecosystem-wide annual commits into an extensive and an intensive margin, The exploratory unsupervised clustering of the per-community trajectories, which does not enter the main analytical narrative, is reported with the other legacy analyses in Section~\ref{sec:SI_temporal_clustering}.

\subsection{Multiplicative decomposition of ecosystem-wide annual commits}\label{sec:SI_decomposition}

\begin{figure}[H]
    \centering
    \includegraphics[width=\textwidth]{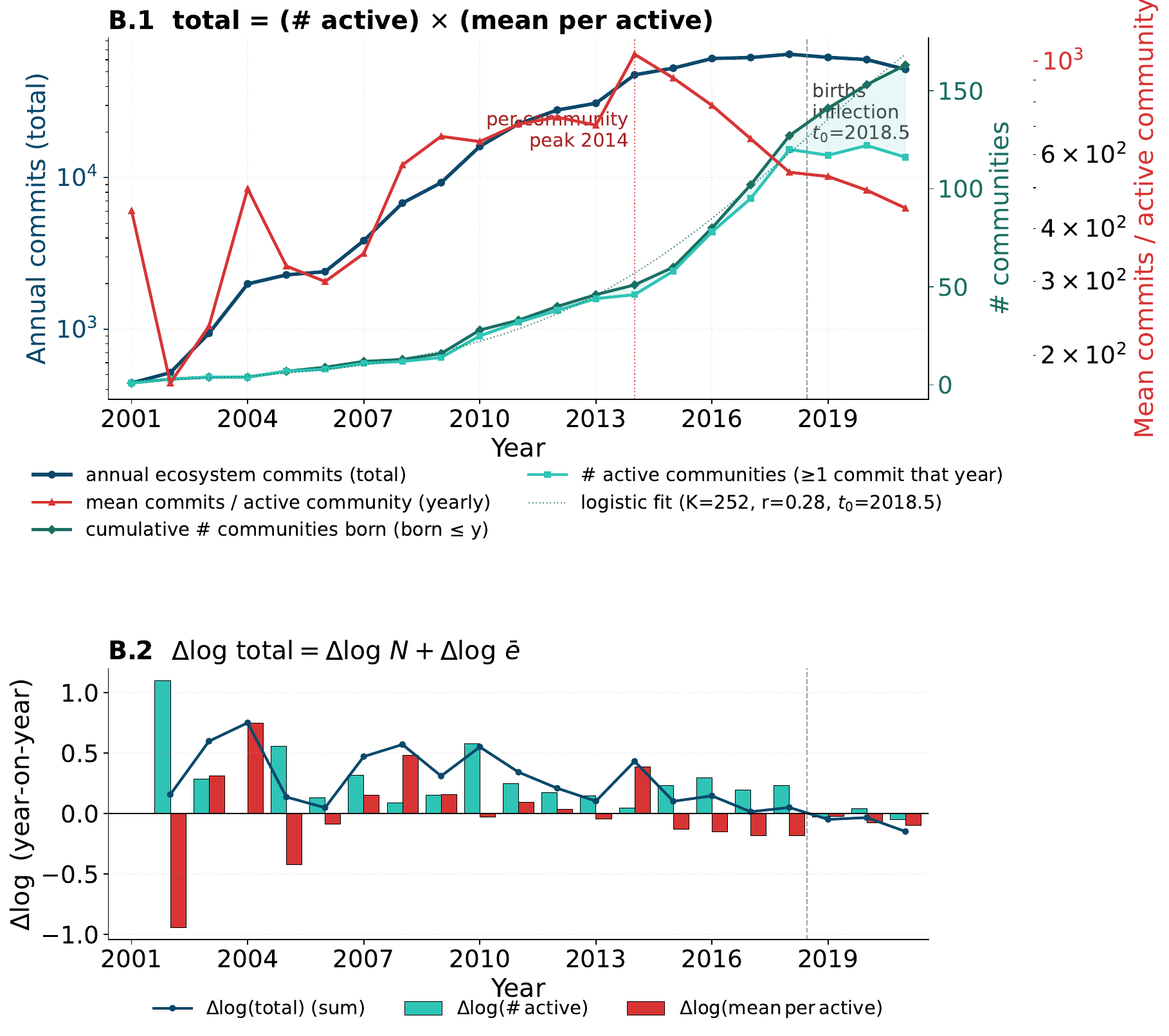}
    \caption{\footnotesize{{\bf Multiplicative decomposition of ecosystem-wide annual commits.} Aligned $x$-axis (2001--2021) for both panels. \textbf{B.1.}~Annual ecosystem commits $T(y)$ [blue, left log axis], mean commits per active community $\bar e(y) = T(y)/N(y)$ [red, right log axis], and the cumulative number of communities born $N_\mathrm{born}(y)$ [dark teal] and active $N(y)$ [light teal] on a shared linear axis, with a logistic fit ($K=252$, $r=0.28$, $t_0=2018.5$) to $N_\mathrm{born}(y)$. The light-teal fill marks the dormant gap [born but inactive that year]. Two inflections are annotated: the per-community peak $\approx$2014 [vertical dotted red] and the births inflection $t_0 \approx 2018.5$ [vertical dashed grey]. \textbf{B.2.}~Year-on-year log decomposition $\Delta\log T = \Delta\log N + \Delta\log \bar e$. Green bars: contribution from the change in the active-community count. Red bars: contribution from the change in mean commits per active community. Blue line: their algebraic sum, equal to the year-on-year growth rate of total commits. Pre-$t_0$ growth is carried by green bars; post-$t_0$ decline is carried by red bars. Regime means are reported in Table~\ref{tab:SI_decomposition_regimes}.}}
    \label{fig:SFig_decomposition}
\end{figure}

\begin{figure}[H]
    \centering
    \includegraphics[width=0.75\textwidth]{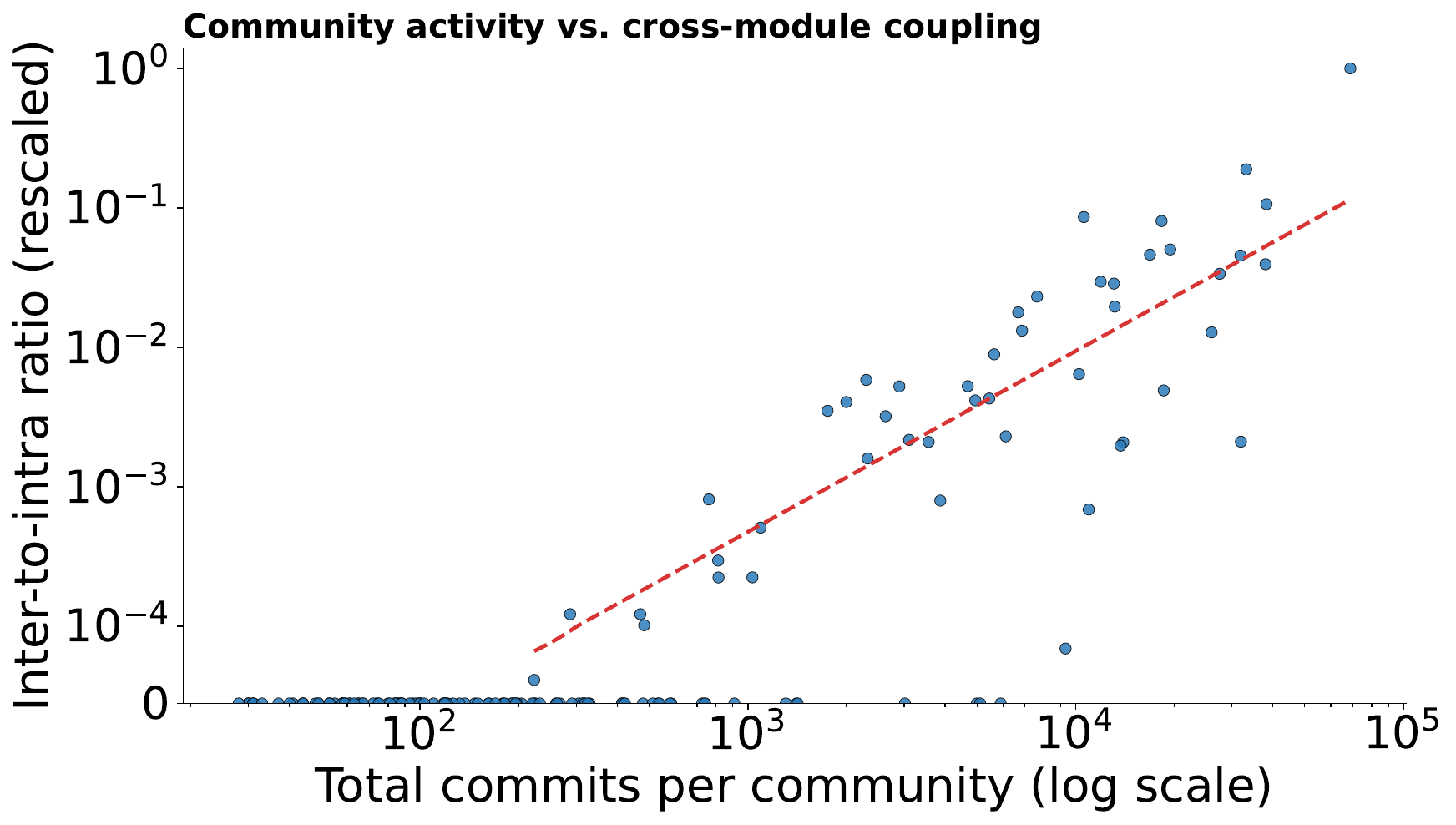}
    \caption{\footnotesize{{\bf Community-level activity versus cross-community coupling.} Per-community total commits [$x$-axis, log scale] versus rescaled inter-to-intra link-density ratio [$y$-axis], across the 163 non-singleton Louvain communities. The $y$-axis is \emph{symmetric-log} (linear below $10^{-4}$, logarithmic above): $116$ of the $163$ communities have a ratio of exactly zero, so a plain logarithmic axis would drop $71\%$ of the sample, while a linear one compresses the remaining four and a half orders of magnitude into the bottom of the panel. Total commits and the ratio are positively rank-correlated over all $163$ communities (Spearman $\rho \approx 0.75$, $p < 10^{-3}$): communities with higher total commit volumes consistently exhibit higher cross-community coupling. The dashed red line is an ordinary-least-squares power-law fit on the $47$ communities with a strictly positive ratio -- the only ones a log--log fit can use -- with slope $1.30 \pm 0.14$ and $R^2 = 0.65$. The slope exceeding one indicates that cross-community coupling grows \emph{faster} than proportionally with community size among the communities that have any, which is consistent with the size dependence reported for the survival analysis; it is estimated on a selected subsample and is descriptive only. The same association is unpacked at finer granularity in the main text via the depth-of-collaboration analysis [Section~\ref{sec:results-depth}, Figure~\ref{fig:Fig6_depth}].}}
    \label{fig:SFig_commits_vs_ratio}
\end{figure}

The qualitative observation in Figure~\ref{fig:anatomy}A -- ``more communities, less steady contributions'' -- has a clean quantitative analogue obtained by decomposing the ecosystem-wide annual commit total into an extensive and an intensive margin. Let $T(y) = \sum_c a_c(y)$ be total annual commits, $N(y) = \#\{c : a_c(y) > 0\}$ the number of active communities in year $y$, and $\bar e(y) = T(y)/N(y)$ the mean commits per active community. By construction $T(y) = N(y)\,\bar e(y)$, so taking logs and first differences yields a Kaya-style identity \cite{kaya_impact_1989, ang_lmdi_2005}
\[
\Delta\log T(y) \;=\; \Delta\log N(y) \;+\; \Delta\log \bar e(y),
\]
i.e.\ each year's growth rate of total commits is exactly the algebraic sum of the growth rate of the active-community count and the growth rate of mean per-community commits. We additionally fit a three-parameter logistic to the cumulative-births series $N_\mathrm{born}(y) = \#\{c : b_c \leq y\}$ via \texttt{scipy.optimize.curve\_fit}, recovering a carrying capacity $K$, an intrinsic growth rate $r$, and an inflection year $t_0$. We split the time series at $\lfloor t_0 \rfloor$ and report the cohort-mean values of $\Delta\log N$, $\Delta\log\bar e$, and $\Delta\log T$ over each window. Mean per active community is plotted on its own logarithmic twin axis in Figure~\ref{fig:SFig_decomposition}B.1 [rather than co-plotted on the same logarithmic axis as $T(y)$] so that its $\sim 1$-decade dynamic range is not visually compressed against the $\sim 4$-decade dynamic range of the total.

The logistic fit to cumulative births returns $K = 251.9$ communities, $r = 0.276\,\text{yr}^{-1}$, and $t_0 = 2018.45$, consistent with the values reported on the same full-calendar-year window in the main text (Figure~\ref{fig:anatomy}A). Mean commits per active community peaks earlier, in 2014, at $\bar e = 1{,}038$ commits per community per year, then declines monotonically. The two regime means are summarised in Table~\ref{tab:SI_decomposition_regimes}. Pre-$t_0$, the active-community count contributes $96\%$ of total growth [$0.282/0.294$], while per-community intensity is essentially flat [$+0.012\,\text{yr}^{-1}$]; post-$t_0$, the per-community decline contributes $85\%$ of the total drop [$-0.065/-0.077$], while the active-community count is roughly stationary [this attribution holds at the activity floor used here and reverses under stricter floors; see the sensitivity at the end of this subsection]. Two distinct inflections are therefore visible: a per-community level inflection near 2014 [the peak of $\bar e$], and a births-rate inflection at $t_0 \approx 2018.5$ [the inflection of the logistic fit to $N_\mathrm{born}$]. Between 2014 and 2018, $\bar e$ is already declining but $N_\mathrm{born}$ is still increasing -- the dormant gap between $N_\mathrm{born}$ and $N_\mathrm{active}$ [light-teal fill in Figure~\ref{fig:SFig_decomposition}B.1] widens visibly after 2018 [gap $7$ in 2018, then $24$, $31$ and $47$ in 2019--2021], indicating that some of the new births contribute zero commits in a given year and so $N_\mathrm{active}$ lags $N_\mathrm{born}$.

\begin{table}[h]
\centering
\footnotesize
\caption{Regime-mean year-on-year log-decomposition of ecosystem-wide annual commits, split at the logistic-births inflection $\lfloor t_0 \rfloor = 2018$.}
\label{tab:SI_decomposition_regimes}
\begin{tabular}{lrrr}
\toprule
period & mean $\Delta\log N$ & mean $\Delta\log \bar e$ & mean $\Delta\log T$ \\
\midrule
2002--2018 [pre-$t_0$] & $+0.282\,\text{yr}^{-1}$ & $+0.012\,\text{yr}^{-1}$ & $+0.294\,\text{yr}^{-1}$ \\
2019--2021 [post-$t_0$] & $-0.011\,\text{yr}^{-1}$ & $-0.065\,\text{yr}^{-1}$ & $-0.077\,\text{yr}^{-1}$ \\
\bottomrule
\end{tabular}
\end{table}

The decomposition recasts the ``more communities, less steady'' pattern as a two-regime story. Until 2014, the ecosystem grows because new communities are being added at high rate while established communities sustain their per-year intensity; both terms are positive but $\Delta\log N$ dominates. Between 2014 and 2018, per-community intensity has already begun to fall, but births are still numerous enough that the total keeps rising and the aggregate masks an internal shift. After 2018, the logistic-saturated births can no longer offset the persistent per-community decline, and the ecosystem total turns negative, driven almost entirely by the intensive margin. The four-year lag between the per-community peak and the births inflection is the key signature: a manager reading only $T(y)$ would have missed the per-community decline for $\sim$4 years, since the births inflow was hiding it. In a maturing open-source ecosystem, the first sign of regime change is in the intensive margin ($\bar e$), not the extensive one ($N$). \paragraph{Sensitivity to the activity floor.} A potential concern is that $N(y)$ is defined above as ``$\geq 1$ commit in year $y$'', a binary activity flag giving equal weight to a community with one commit and one with thousands, so that an influx of nominally-active communities could depress $\bar e$ and manufacture the 2014 peak. Re-running the decomposition with $N^{(k)}(y) = \#\{c : a_c(y) \geq k\}$ and $\bar e^{(k)}(y) = [\sum_{c\,:\,a_c(y)\geq k} a_c(y)]/N^{(k)}(y)$ for $k \in \{1, 5, 50\}$ shows that it does not. The peak of $\bar e^{(k)}$ remains at 2014 for $k=1$ [the specification used above, $\bar e = 1{,}038$] and for $k=5$ [$\bar e = 1{,}061$], and moves by a single year, to 2015, at $k=50$ [$\bar e = 1{,}252$]. Under every floor the intensive-margin peak precedes the births inflection $t_0 = 2018.5$ by three to four years, so the four-year lag reported above is a property of established-community behaviour rather than an artefact of counting marginal communities.

Two quantities are, however, floor-dependent and should be read accordingly. The first is the severity of the decline: $\bar e$ falls $57\%$ from peak to 2021 at $k=1$ but only $27\%$ at $k=50$, so part of the headline thinning does come from marginal communities. The second, and more consequential, is the post-$t_0$ attribution, which reverses: the intensive margin accounts for $85\%$ of the post-2018 drop at $k=1$, $44\%$ at $k=5$, and only $17\%$ at $k=50$, where the extensive margin instead carries $83\%$. The total decline is essentially identical across floors [$\Delta\log T = -0.077$, $-0.077$, $-0.078$ respectively]; only its decomposition moves. Among substantial communities, then, the post-2018 slowdown consists largely of communities falling below the activity floor rather than of surviving communities thinning out -- the same population dynamic that the per-cohort residualisation hazard in Section~\ref{sec:results-survival} quantifies.

\section{Community-scale cross-community coupling}\label{sec:SI_community_scale}

This section supports Figure~\ref{fig:cross_community}A--B of the main text by reporting the community-level activity-versus-coupling association; the exploratory clustering of communities along an event-ratio axis is in Section~\ref{sec:SI_legacy_umap}.

\section{Contributor-scale cross-community structure}\label{sec:SI_contributor_scale}

This section supports Figure~\ref{fig:cross_community}C--D of the main text by reporting the diagnostic views (CCDF and SSR scan) that justify the two-regime fit on the contributor-breadth distribution. The complementary contributor-typology view used in earlier work is in Section~\ref{sec:SI_contributor_types}.

\subsection{Diagnostic views of the contributor-breadth distribution}\label{sec:SI_breadth_ccdf_ssr}

\begin{figure}[H]
    \centering
    \includegraphics[width=\textwidth]{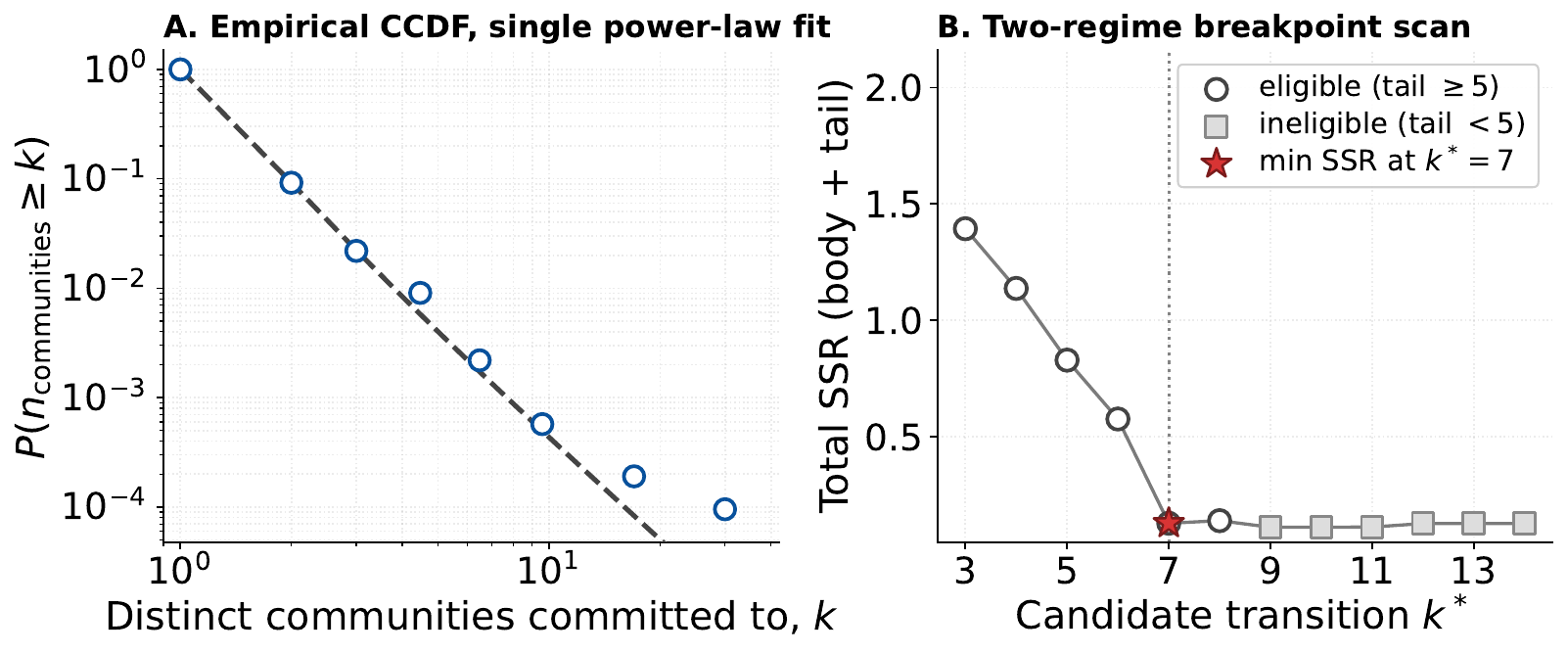}
    \caption{\footnotesize{{\bf Diagnostic views supporting the two-regime fit in Figure~\ref{fig:cross_community}C.} \textbf{A.}~Empirical CCDF of contributor breadth $k$ on log--log axes [open blue circles, $n = 10{,}490$ canonical human contributors after bot/phantom removal], log-binned for legibility into ten bins of equal width on $\log_{10} k$, of which eight are non-empty and plotted: each circle is plotted at the geometric mean of the observed integer $k$ values in its bin, with $y$ equal to the empirical CCDF evaluated at the smallest of those $k$ values [preserving the $P(X \geq k)$ semantic]. The dashed grey line is a single discrete power-law fit on the multi-community subset $k \geq 2$ [Hurwitz-zeta MLE]. The single-fit CCDF exponent is $\mu = \alpha - 1 = 3.00$ [95\% bootstrap CI $[2.80,\,3.22]$, $B = 1{,}000$]. The nine humans at $k \geq 7$ lie above the fitted CCDF, so a single straight line is insufficient to describe the upper tail. \textbf{B.}~Sum of squared residuals [body+tail OLS in log--log space on the empirical PMF] as a function of the candidate transition $k^*$. Open circles are eligible candidates [tail has $\geq 5$ unique $k$ values]; grey squares are ineligible [tail $<5$ unique $k$]. The red star marks the min-SSR breakpoint $k^* = 7$ used in the main text. The sharp drop from $0.58$ at $k^* = 6$ to $0.13$ at $k^* = 7$ is the elbow that defines the data-driven inflection.}}
    \label{fig:SFig_breadth_ccdf_ssr}
\end{figure}

The main-text Figure~\ref{fig:cross_community}C shows the empirical PMF of contributor breadth $k$ together with an adaptive Abramson KDE and a two-regime OLS fit whose breakpoint $k^* = 7$ is chosen by minimising body+tail SSR. Figure~\ref{fig:SFig_breadth_ccdf_ssr} reports the two diagnostic views that motivate that choice. \textbf{Panel A} plots the empirical CCDF of $k$ on log--log axes together with a single discrete power-law fit to the multi-community subset ($k \geq 2$), estimated by the Hurwitz-zeta maximum-likelihood estimator (MLE) of \cite{clauset2009power}. The MLE returns a CCDF exponent $\mu = \alpha - 1 = 3.00$ [95\% non-parametric bootstrap CI $[2.80,\,3.22]$, $B = 1{,}000$ resamples]. The fit captures the body of the distribution but underestimates the upper tail: the nine humans at $k \geq 7$ sit visibly above the predicted CCDF, indicating that a single power law is too restrictive a model for the full range of $k$. \textbf{Panel B} reports the SSR scan that selects the two-regime breakpoint used in the main text. For each candidate $k^* \in [3, 14]$ we fit a power law to the body ($k < k^*$) and to the tail ($k \geq k^*$) by OLS on $\log_{10} k$ versus $\log_{10}\text{PMF}$, summing the body and tail squared residuals. Open circles mark candidates with at least five unique tail $k$ values [eligible for selection]; grey squares mark candidates with fewer unique tail points [shown for completeness; an OLS fit on a tail of two or three points achieves a trivially low SSR by interpolation, so these are not eligible]. The total SSR drops sharply from $0.58$ at $k^* = 6$ to $0.13$ at $k^* = 7$ -- the elbow that defines the data-driven breakpoint -- and stays in the range $0.11$--$0.14$ for all $k^* \geq 7$, including the ineligible ones: every candidate that places at least one of the high-$k$ humans inside the tail yields essentially the same residual, so $k^{*}=7$ is both the minimum-SSR candidate and the earliest point of that plateau, with $k^{*}=8$ a near-tie at $0.14$. What the resulting layer does and does not depend on that choice is quantified separately in Supplementary~\ref{sec:SI_carrier_threshold}, which recomputes every carrier-layer quantity at each nearby cutoff and restates the thinness claim without any threshold.

\subsection{Sensitivity of the carrier layer to the breadth cutoff}\label{sec:SI_carrier_threshold}

The carrier layer is named at $k^{*}=7$, the inflection selected by minimising the body$+$tail sum of squared residuals on the empirical breadth distribution (Supplementary~\ref{sec:SI_breadth_ccdf_ssr}). Because the resulting group contains only nine contributors, this subsection reports every carrier-layer quantity at each nearby cutoff, and then restates the same claim in a form that uses no cutoff at all. The base population is the $10{,}490$ canonical human contributors remaining after identity canonicalisation and the removal of $11$ bots and $688$ phantom git identities (Methods Section~\ref{sec:methods-data}), producing $4{,}015$ inter-community merged pull-request records and $5{,}364$ inter-community commits across $163$ non-singleton communities.

\paragraph{Threshold sweep.} Table~\ref{tab:SI_carrier_cutoffs} recomputes the layer at each cutoff. The absolute share of cross-community work falls as the group shrinks, as it arithmetically must; the \emph{over-representation} ratio -- the layer's share of inter-community pull-request records divided by its share of the contributor population -- does not, remaining between $51$ and $166$ throughout. Whichever cutoff is chosen, a group two to three orders of magnitude smaller than the ecosystem carries a double-digit percentage of its cross-boundary work. The cutoffs $k\geq 9$ and $k\geq 10$ coincide because no contributor has $k$ exactly $9$. Shares computed on distinct pull requests rather than on pull-request records differ by at most $3.8$ percentage points and are reported alongside.

\begin{table}[h]
\centering
\footnotesize
\caption{Carrier layer recomputed at each breadth cutoff. Columns are, in order: layer size; its share of the $10{,}490$-contributor population; its share of inter-community pull-request records and of distinct inter-community pull-requests; its share of inter-community commits; the over-representation ratio [share of inter-community pull-request records divided by share of the population]; and the number of the $163$ communities the layer reaches. The focal cutoff used in the main text is shown in bold.}
\label{tab:SI_carrier_cutoffs}
\setlength{\tabcolsep}{5pt}
\begin{tabular}{lrrrrrrr}
\toprule
cutoff & $n$ & \% pop. & \% PR rec. & \% distinct PRs & \% commits & over-repr. & comm.\ reached \\
\midrule
$k\geq 4$ & $95$ & $0.91$ & $46.3$ & $44.2$ & $39.7$ & $51\times$ & $67$ \\
$k\geq 5$ & $43$ & $0.41$ & $39.5$ & $35.7$ & $34.0$ & $96\times$ & $59$ \\
$k\geq 6$ & $23$ & $0.22$ & $18.1$ & $18.1$ & $16.3$ & $83\times$ & $52$ \\
$\mathbf{k\geq 7}$ & $\mathbf{9}$ & $\mathbf{0.086}$ & $\mathbf{14.3}$ & $\mathbf{13.2}$ & $\mathbf{13.4}$ & $\mathbf{166\times}$ & $\mathbf{47}$ \\
$k\geq 8$ & $6$ & $0.057$ & $4.6$ & $6.8$ & $4.1$ & $81\times$ & $44$ \\
$k\geq 9$ & $4$ & $0.038$ & $4.2$ & $6.1$ & $3.7$ & $109\times$ & $41$ \\
$k\geq 10$ & $4$ & $0.038$ & $4.2$ & $6.1$ & $3.7$ & $109\times$ & $41$ \\
\bottomrule
\end{tabular}
\end{table}

\paragraph{Sublinear scaling inside the layer.} The main text reports that within the carrier layer inter-community commits scale sublinearly with intra-community commits, $n_\mathrm{inter}\sim n_\mathrm{intra}^{0.53(9)}$ [$R^{2}=0.82$], against an essentially flat per-bin median across all $966$ multi-community contributors [slope $0.06(5)$, $p=0.26$]. Refitting inside each candidate layer gives $b = 0.22(5)$ at $k\geq 4$ [$n=95$, $R^{2}=0.20$], $0.42(7)$ at $k\geq 5$ [$n=43$], $0.49(10)$ at $k\geq 6$ [$n=23$], $0.53(9)$ at $k\geq 7$ [$n=9$], $0.41(7)$ at $k\geq 8$ [$n=6$] and $0.36(9)$ at $k\geq 9$ [$n=4$]. Every candidate layer is significantly sublinear and significantly steeper than the flat all-contributor reference, and the exponents at $k\geq 5$ through $k\geq 8$ have overlapping standard errors. The one value that drifts is $k\geq 4$, where the layer has grown to $95$ contributors and begins to merge into the general multi-community population -- which is an argument for placing the boundary above $4$ rather than against $k^{*}=7$.

\paragraph{Statements that use no cutoff.} The thinness claim does not require a threshold. Ranking canonical humans by inter-community pull-request output, the top $1$ contributor accounts for $16.6\%$ of all cross-boundary records, the top $10$ for $36.8\%$, the top $50$ for $54.3\%$, the top $250$ for $80.1\%$ and the top $500$ for $90.5\%$. Equivalently, half of all cross-community pull-request work is produced by $36$ contributors [$0.34\%$ of the population], $80\%$ by $249$ [$2.37\%$] and $90\%$ by $479$ [$4.57\%$].

Treating $k$ as a continuous variable over the $966$ multi-community contributors -- the population for which $k$ varies meaningfully -- breadth is rank-correlated with inter-community pull-request records at Spearman $\rho=+0.587$ [$p=2\times10^{-90}$], with inter-community commits at $\rho=+0.609$ [$p=3\times10^{-99}$], and with the share of a contributor's commits that are inter-community at $\rho=+0.167$ [$p=2\times10^{-7}$]; a continuous fit gives $n_\mathrm{inter}\sim k^{1.83(9)}$ [$R^{2}=0.32$]. Two caveats attach to these figures and we state both. Computed over all $10{,}490$ contributors the same correlations reach $\rho=0.95$--$0.999$, but that is an artefact of $90.8\%$ of contributors sitting at $k=1$ with exactly zero cross-community output, so only the $k\geq 2$ population should be quoted. More importantly, $k$ and cross-community output share a data-generating step -- a merged cross-community pull-request creates a commit in the destination community (Supplementary~\ref{sec:SI_friction_controls}) -- so these are descriptive associations between quantities that are not independent by construction, not evidence of a behavioural gradient.

\paragraph{A convergence worth recording, and its limits.} The scaling exponent is highest at the focal cutoff, and the rise towards it is monotone. Because the sweep above fits nested subsets [$k\geq 8 \subset k\geq 7 \subset k\geq 6$], whose exponents cannot be compared as independent estimates, we also refit on disjoint bands with each contributor counted once: $b = 0.16(1)$ at $k=2$ [$n=736$], $0.18(2)$ at $k=3$--$4$ [$n=187$], $0.35(8)$ at $k=5$--$6$ [$n=34$], $0.68(13)$ at $k=7$--$8$ [$n=5$], falling back to $0.36(9)$ at $k\geq 9$ [$n=4$]. Two criteria that share no information therefore point to the same region: the breakpoint selected from the \emph{marginal} shape of the breadth distribution, in which no cross-community outcome enters the criterion, and the peak of a \emph{bivariate} scaling relation between intra- and inter-community commits.

We record this as a convergence and not as a justification. A bootstrap of the focal exponent [$n=9$, $5{,}000$ resamples] gives a $95\%$ interval of $[0.29,\,0.71]$, and the probability that it exceeds the neighbouring point estimates is only $0.65$ against $k\geq 6$ and $0.81$ against $k\geq 8$; a smooth interaction model fitted to all $966$ multi-community contributors finds no significant variation of the elasticity with $k$ [interaction coefficient $+0.101$, $p=0.59$]. The accompanying $R^{2}$ rises monotonically with the cutoff simply because $n$ falls from $95$ to $4$, and carries no information about fit quality. The conclusions of the paper rest on the cutoff-independent results above, not on the coincidence of these two criteria.

\section{Repository activity forecasting}\label{sec:SI_forecasting}

We complement the cross-community contributor analyses of Section~\ref{sec:results-depth} with a repository-level forecasting exercise. Two random-forest regressions of six-month commit volume per repository are trained: \emph{Model~1} includes the previous six months of commit activity as a feature [capturing auto-regressive memory in addition to all structural features]; \emph{Model~2} excludes the recent-history feature, which isolates the structural contribution of community membership and cross-community exposure from short-run persistence. Variable importance is reported as percentage increase in mean squared error under feature permutation; prediction error is the per-repository residual on the holdout. The headline observation -- that ecosystem embedding [number of community links, inter-/intra-community PR ratio, repository-level link-density summaries, lifetime] retains substantial explanatory power once recent history is removed -- complements the contributor-scale story: the structural substrate identified at the contributor scale is observable enough at the repository scale to forecast future activity.

\begin{figure}[H]
    \centering
    \includegraphics[width=\textwidth]{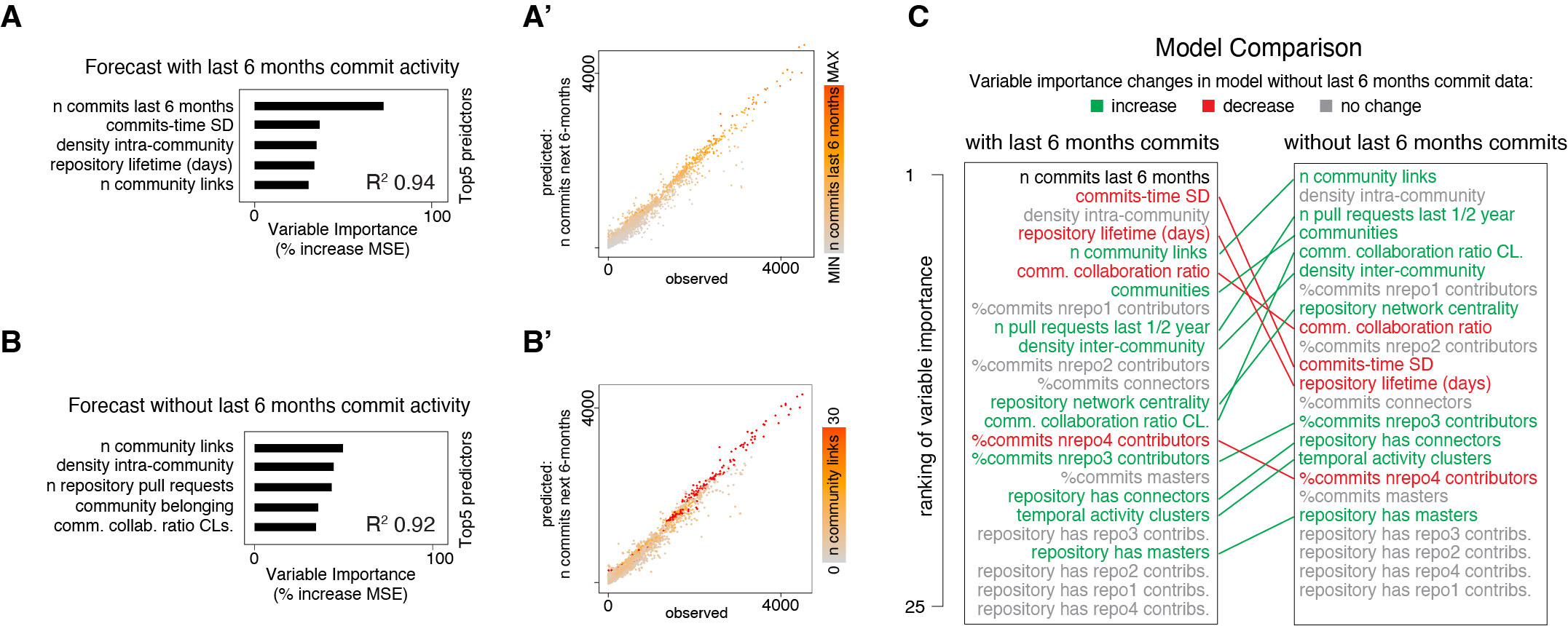}
    \caption{\footnotesize{{\bf Repository six-month commit-volume forecasting under structural-vs-temporal ablation.} \textbf{A.}~Variable-importance ranking for Model~1 [with the previous six months of commit activity included]: ranking is by percentage increase in MSE under feature permutation, which is also the quantity on the $x$-axis. \textbf{A$'$.}~Per-repository observed-vs-predicted scatter for Model~1; colour encodes the previous six months of commits [the dominant predictor]. \textbf{B.}~Variable importance for Model~2 [recent-history feature removed]: the number of community links and the inter-/intra-community PR ratio rise to the top of the ranking. \textbf{B$'$.}~Observed-vs-predicted scatter for Model~2; colour encodes the number of community links per repository [the dominant predictor in this model]. \textbf{C.}~Cross-model variable-importance comparison: variables are listed in order of importance per model, and the connecting links indicate whether each variable's importance dropped (red), increased (green), or held (grey) when the recent-history feature was removed. The persistence of community-link and inter-/intra-community-ratio importance under the ablation is the principal forecast-side evidence for the ecosystem-embedding claim made in Section~\ref{sec:results-depth}.}}
    \label{fig:SI_forecasting_rf}
\end{figure}

\begin{figure}[H]
    \centering
    \includegraphics[width=0.84\textwidth]{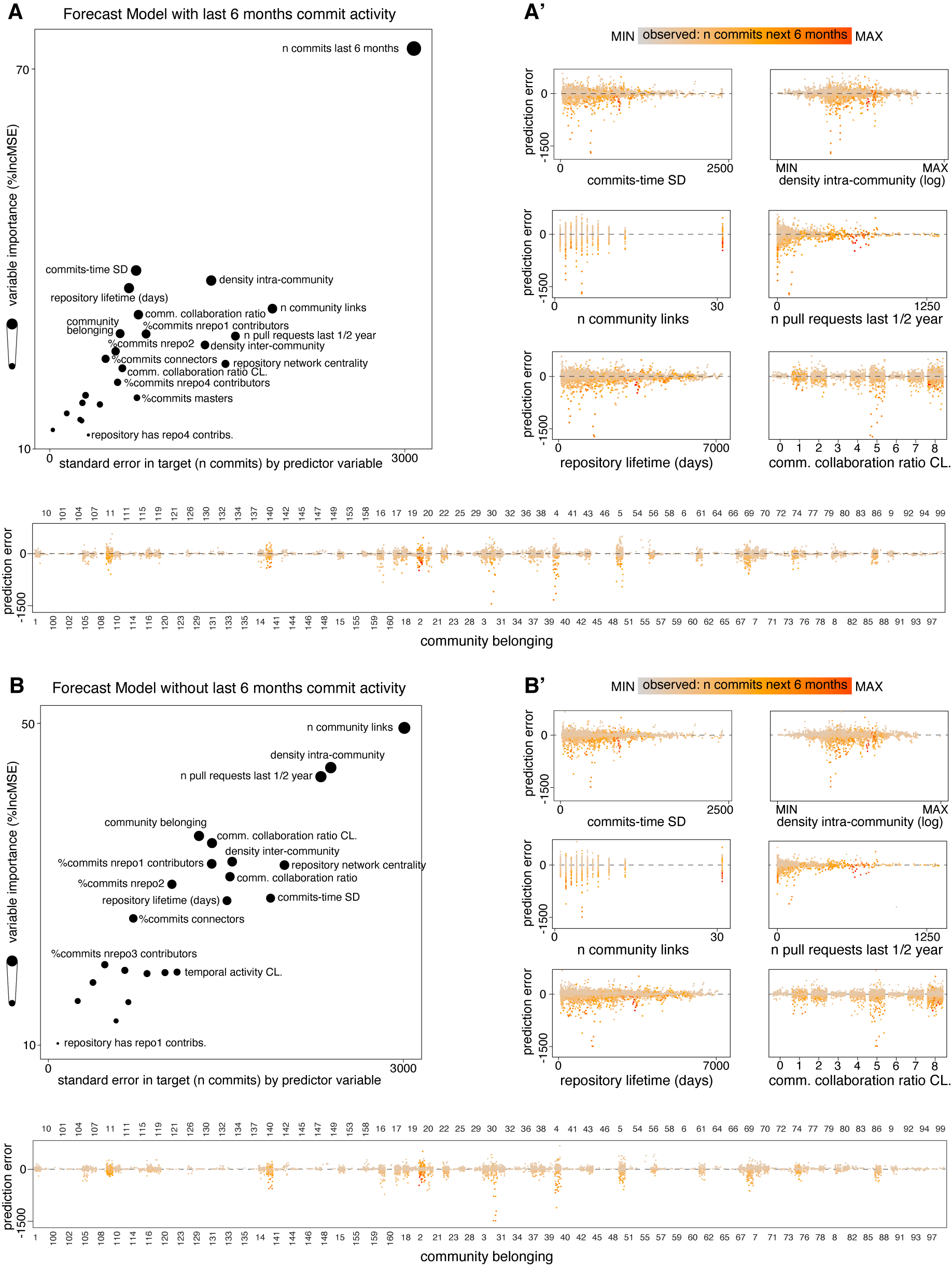}
    \caption[Forecast modeling performance comparison]{\footnotesize{{\bf Forecast modeling performance comparison.} {\bf A.}~Variable importance for Model~1 [with the previous six months of activity], with each variable's percentage increase in MSE under permutation on the $y$-axis and dot size, and the standard error of the predicted commit count on the $x$-axis. {\bf A$'$.}~Top predictive variables for both models with their per-repository prediction error on the $y$-axis, coloured by the observed next-6-month commit count. {\bf B.}~Variable importance for Model~2 [recent history excluded], same axes as panel~A. {\bf B$'$.}~Top predictive variables for both models with prediction error coloured by observed next-6-month commit count.}}
    \label{fig:SFig5_gitpaper}
\end{figure}

\section{Community survival: residual phase and hazard predictors}\label{sec:SI_survival}

This section supports the community-survival analysis in the main text (Figure~\ref{fig:survival}, Table~\ref{tab:cox_survival}). We first define the residual-phase outcome, the per-cohort annual hazard, and report the cohort-structured time-to-residual pattern; we then test whether community-level features beyond cohort and size predict residualisation, with full Cox proportional-hazards diagnostics.

\subsection{Per-community residual phase: time and hazard}\label{sec:SI_residual}

Figure~\ref{fig:anatomy}A shows that the earliest cohorts of communities sustain multi-year contribution streaks, while later cohorts appear less steady. We quantify that observation here as a residual-phase analysis. For each non-singleton Louvain community $c$, let $a_c(y)$ be the number of commits to $c$ in calendar year $y$, taken from \texttt{commit\_pr\_timedata.csv} with author/repository community labels from the bipartite Louvain partition (\texttt{out/BiNet/g\_full.csv}). The observation horizon is set to the last full year of GitHub Archive coverage in our window, $Y_\mathrm{end} = 2021$; the partial 2022 window is excluded from rate-based statistics. We define the community's birth year as $b_c = \min\{y : a_c(y) > 0\}$, its mean yearly activity as
\[
\mu_c \;=\; \frac{1}{Y_\mathrm{end} - b_c + 1}\sum_{y = b_c}^{Y_\mathrm{end}} a_c(y),
\]
and a residual threshold $\theta_c = 0.05\,\mu_c$. The last non-residual year is $y^*_c = \max\{y \in [b_c, Y_\mathrm{end}] : a_c(y) \geq \theta_c\}$, and the non-residual span [Figure~\ref{fig:survival}A, $y$-axis] is $y^*_c - b_c + 1$ years. A community is right-censored if $y^*_c = Y_\mathrm{end}$ -- the observation window terminates while the community is still above threshold. The per-cohort annual hazard rate [Figure~\ref{fig:survival}A$'$] is computed as
\[
h(\text{cohort}) \;=\; \frac{\#\{\text{events in cohort}\}}{\sum_{c\,\in\,\text{cohort}} (y^*_c - b_c + 1)},
\]
i.e.\ the empirical mean annual transition probability into residual conditional on still being non-residual. The hazard formulation is preferred over a cumulative-event share or a mean time-to-residual because it normalises by the time the cohort has actually been observed at risk; raw event shares would bias early cohorts upward [longer observation] and late cohorts downward [truncated by the censoring envelope $Y_\mathrm{end} - b_c + 1$, shown as the dashed line in Figure~\ref{fig:survival}A].

Across the 163 non-singleton communities, 55 have entered the residual phase by 2021 and 108 are right-censored. The median time-to-residual is 5 years overall, but the distribution is strongly cohort-structured. Communities born in 2001--2010 sit close to the censoring envelope, indicating that they are still active above 5\% of their long-run mean after 11--20 years; communities born in 2011--2014 show a mixed pattern split between censored and 3--7-year non-residual spans; communities born in 2015--2021 are sharply bimodal between very short [$\leq 3$-year] non-residual spans and a tail of censored cases reflecting their truncated observation windows. The colour scale [mean yearly commits] shows that the surviving [censored] communities are systematically higher-volume; small communities are over-represented among events. The per-cohort hazard [Figure~\ref{fig:survival}A$'$] makes the cohort effect explicit: pre-2010 cohorts sit at $0$--$0.05\,\text{yr}^{-1}$, then the hazard rises sharply for the 2015--2018 cohorts, peaking at $0.192\,\text{yr}^{-1}$ for the 2018 cohort [$n=25$]. The 2019--2021 cohorts have lower observed hazards, but those are small-sample [cohort sizes $n=14, 12, 10$] and bounded above by the limited time available since their birth.

Two distinct mechanisms compound in this cohort effect. The first is a geometric ceiling: a community born in 2018 cannot, by construction, accumulate more than 4 years of non-residual life by 2021, and the censoring envelope in Figure~\ref{fig:survival}A makes this constraint explicit. The hazard formulation in Figure~\ref{fig:survival}A$'$ normalises by years-at-risk and is therefore not mechanically driven by this. The second is genuine cohort heterogeneity: even after the time-at-risk normalisation, the hazard rises by roughly an order of magnitude between the pre-2010 cohorts [$0$--$0.05\,\text{yr}^{-1}$] and the 2015--2018 cohorts [$0.076$--$0.192\,\text{yr}^{-1}$, the four cohorts being $0.125$, $0.076$, $0.165$ and $0.192$], so communities born after the community-count peak transition [Figure~\ref{fig:anatomy}A, $\sim$2017] are intrinsically more likely to drop into residual within any given year. The mean-yearly-commits colour mapping suggests a complementary point: late cohorts that do survive tend to be smaller, while the highest-volume late entrants are also the ones most likely to remain censored, so volume is a precondition for sustained activity but not a sufficient one for the early cohorts [which can persist on small but continuous activity]. The 5\%-of-mean-yearly threshold is conservative for communities with long dormant tails, so the whole analysis is recomputed below under ten alternative inactivity rules.

\paragraph{Sensitivity to the inactivity rule.} We re-ran the residual-phase outcome, the per-cohort hazard, both Kaplan--Meier stratifications and the multivariate Cox model under ten rules: thresholds at $1\%$, $2.5\%$, $5\%$ [the value used above] and $10\%$ of mean yearly activity; a variant using $5\%$ of the median over active years; absolute floors of $5$ and $10$ commits per year, which do not reference a community's own scale; a threshold-free zero-activity definition [non-residual in any year with at least one commit]; and two variants requiring the residual state to have persisted for two or three consecutive years before an event is scored. Table~\ref{tab:SI_residual_rules} reports the outcome. The $5\%$ row reproduces the values used throughout this section by construction.

\begin{table}[h]
\centering
\footnotesize
\caption{Community-survival results under ten inactivity rules. Hazards are per-cohort annual rates; the ratio column is the 2015--2018 hazard divided by the 2001--2010 hazard, the contrast carrying the order-of-magnitude claim. Log-rank statistics are multi-group tests on tertiles of the inter-community pull-request share and of the external-community degree. Cox hazard ratios are from the multivariate model of Table~\ref{tab:SI_hazard_predictors}. The rule used in the main text is shown in bold.}
\label{tab:SI_residual_rules}
\setlength{\tabcolsep}{4pt}
\begin{tabular}{lrrrrrrrr}
\toprule
rule & events & cens. & $h_{2001\text{--}10}$ & $h_{2015\text{--}18}$ & ratio & $\chi^2$ share & $\chi^2$ degree & HR birth yr \\
\midrule
$1\%$ of mean            & $49$ & $114$ & $0.010$ & $0.119$ & $11.9\times$ & $22.5$ & $42.6$ & $1.19^{*}$ \\
$2.5\%$ of mean          & $51$ & $112$ & $0.010$ & $0.123$ & $12.2\times$ & $23.5$ & $45.2$ & $1.20^{*}$ \\
$\mathbf{5\%}$ \textbf{of mean} & $\mathbf{55}$ & $\mathbf{108}$ & $\mathbf{0.013}$ & $\mathbf{0.134}$ & $\mathbf{10.5\times}$ & $\mathbf{21.9}$ & $\mathbf{49.1}$ & $\mathbf{1.18^{*}}$ \\
$10\%$ of mean           & $65$ & $98$  & $0.013$ & $0.159$ & $12.1\times$ & $20.0$ & $52.6$ & $1.14^{*}$ \\
$5\%$ of median          & $52$ & $111$ & $0.013$ & $0.125$ & $9.8\times$  & $23.1$ & $46.6$ & $1.17^{*}$ \\
floor $5$ commits/yr     & $69$ & $94$  & $0.013$ & $0.183$ & $14.4\times$ & $20.4$ & $73.3$ & $1.15^{*}$ \\
floor $10$ commits/yr    & $77$ & $86$  & $0.013$ & $0.216$ & $16.9\times$ & $20.0$ & $80.3$ & $1.11^{*}$ \\
zero activity            & $47$ & $116$ & $0.005$ & $0.119$ & $23.8\times$ & $23.3$ & $42.7$ & $1.17$ \\
persist $\geq 2$ yr      & $31$ & $132$ & $0.008$ & $0.082$ & $10.7\times$ & $9.8$  & $26.8$ & $1.03$ \\
persist $\geq 3$ yr      & $20$ & $143$ & $0.005$ & $0.056$ & $10.9\times$ & $9.6$  & $23.0$ & $0.89$ \\
\bottomrule
\end{tabular}

\vspace{3pt}
{\footnotesize $^{*}$ $p<0.05$. Every log-rank test is significant under every rule [$\chi^2$ share $9.6$--$23.5$, all $p<0.01$; $\chi^2$ degree $23.0$--$80.3$, all $p\leq1.1\times10^{-5}$].}
\end{table}

Three readings follow. \emph{First, the cohort pattern is robust.} The order-of-magnitude rise in per-cohort hazard between the pre-2010 and 2015--2018 cohorts holds under all ten rules, with a ratio between $9.8\times$ and $23.8\times$. The reported threshold sits at the conservative end of that range: the threshold-free zero-activity definition gives $23.8\times$ and the absolute floors $14$--$17\times$. Nothing in this result is an artefact of the $5\%$ choice.

\emph{Second, the inter-community pull-request share association is robust in direction and significant under seven of ten rules.} Its hazard ratio stays protective in a narrow band, $0.047$--$0.291$, throughout. The three exceptions -- the $10$-commit floor [$p=0.062$] and the two persistence variants [$p=0.077$, $p=0.086$] -- are marginal, identical in sign, and occur in the specifications with the fewest events, so we read them as loss of power rather than reversal.

\emph{Third, the external-degree association is not robust in the multivariate model}, reaching significance under only three of ten rules. This confirms rather than revises what is reported above: the effect is already non-significant at the baseline [HR $=0.19$, $p=0.176$], and its univariate separation is among the most robust results in the table. The sensitivity therefore supports treating external reach as a partial proxy for community resourcedness rather than as an independent predictor.

One artefact deserves explicit statement. Under the two persistence rules the multivariate birth-year coefficient loses significance [HR $1.03$, $p=0.74$; HR $0.89$, $p=0.22$] even though the raw cohort hazard ratio is unchanged [$10.7\times$, $10.9\times$]. This is mechanical: requiring two or three consecutive sub-threshold years before scoring an event disproportionately censors \emph{late} cohorts, since a community born in 2018 has too few years before the 2021 horizon to accumulate a persistent residual run. The rule removes precisely the events that identify the cohort effect in the regression, while the hazard, being normalised by years-at-risk, is unaffected. This is a further reason for the main-text claim to rest on the hazard rather than on the Cox birth-year term.

Two implementation notes. The median-based rule is computed over \emph{active} years: fourteen of the $163$ communities have a median of zero across all years since birth, and using the all-years median would set $\theta_c = 0$ and censor them by construction. A ``two consecutive sub-threshold years'' rule implemented as a smoothing of the above-threshold indicator would be a no-op under this outcome definition, because the time-to-event is the \emph{last} year above threshold and therefore already encodes ``never recovers''; the persistence variants are consequently implemented as a censoring rule, in which an event requires the residual state to have been observed for at least two or three years by the horizon.

The visual companions to this section -- the time-to-residual scatter and the per-cohort hazard bars -- are reported in the main text as panels~A and~A$'$ of Figure~\ref{fig:survival}.

\subsection{Predictors of residualisation hazard (static-covariate Cox)}\label{sec:SI_hazard_predictors}

The headline numbers from this section are reported in the main text as Figure~\ref{fig:survival} and Table~\ref{tab:cox_survival}. This subsection retains the full Cox proportional-hazards diagnostic [univariate and multivariate model fits, two Kaplan--Meier stratifications, and the static-window caveats] for readers who want the underlying detail.

Section~\ref{sec:SI_residual} established that the per-cohort hazard rate of entering residual rises by an order of magnitude between pre-2010 cohorts and the 2015--2018 cohorts. A natural follow-up is whether community-level features beyond birth year predict residualisation, and in particular whether \emph{insufficient cross-community connectivity} -- the structural feature emphasised throughout this paper -- is associated with higher hazard once cohort and size are controlled for. We test the hypothesis with a Cox proportional-hazards (PH) model on the same 163 communities and the same outcome [time-to-residual through 2021, with right-censoring] used in Section~\ref{sec:SI_residual}.

\paragraph{Cox PH and Kaplan--Meier in a nutshell.} Both are tools for time-to-event data with right-censoring. The Kaplan--Meier (KM) estimator is non-parametric: it estimates the survival function $S(t) = \mathbb{P}(\text{community still non-residual at age } t)$ by walking through event times and updating $S$ at each by the conditional surviving fraction. Stratified KM splits the cohort into groups and renders one curve per stratum; a multi-group log-rank test compares the curves under a no-difference null. Cox PH is the regression workhorse:
\[
h(t \mid \mathbf{x}) \;=\; h_0(t)\,\exp\!\big(\beta_1 x_1 + \beta_2 x_2 + \cdots\big),
\]
where the baseline hazard $h_0(t)$ is left unspecified and only the multiplicative effect of the covariates is estimated. The hazard ratio $\mathrm{HR} = \exp(\beta)$ is the per-unit multiplier of the per-year residualisation risk: $\mathrm{HR} > 1$ accelerates the event, $\mathrm{HR} < 1$ is protective. The proportional-hazards assumption requires that the ratio $h(t \mid \mathbf{x}_a) / h(t \mid \mathbf{x}_b)$ does not change with $t$. Concordance $C$ is a discrimination measure: the probability that for a random pair where one community failed earlier, the model assigned it a higher hazard [$C=0.5$ chance, $C=1$ perfect].

\paragraph{Per-community predictors.} For each community $c$ we assemble three classes of covariates from the end-of-window snapshot. \emph{Cohort and size} [controls]: birth year $b_c$; $\log_{10}$ number of repositories from \texttt{df\_commrepo.csv}; $\log_{10}$ number of distinct contributors. \emph{Cross-community PR connectivity}: the inter-community PR share
\[
\text{inter-share}_c \;=\; \frac{n^\mathrm{inter}_c}{n^\mathrm{intra}_c + n^\mathrm{inter}_c},
\]
where $n^\mathrm{intra}_c$ [$n^\mathrm{inter}_c$] is the number of pull-requests opened by an author whose home community is [is not] the repository's community, taken from \texttt{out/revi\-sion/figure\_cross\_com\-mu\-ni\-ty/per\_com\-mu\-ni\-ty\_pr.csv}. \emph{Cross-community contributor structure}: the share of $c$'s contributors who have committed to $\geq 2$ Louvain communities (\texttt{cross\_share\_k2}); and the network-degree measure $\log_{10}(1 + d^\mathrm{ext}_c)$, where $d^\mathrm{ext}_c$ is the number of \emph{other} communities that either send commits into $c$ or receive commits from $c$'s home contributors -- the symmetric inter-community degree of $c$ in the cross-commit graph.

\paragraph{Modelling.} We fit Cox PH models with \texttt{lifelines.CoxPHFitter}, taking time-to-residual [years from birth] as the duration and $\text{event} = 1 - \text{censored}$ as the indicator. Univariate models report a single-predictor HR, 95\% Wald CI, $p$-value and concordance index. The multivariate model uses cohort + size + a parsimonious set of connectivity covariates. We also fit Kaplan--Meier survival curves stratified by tertiles of two key cross-community measures, with multi-group log-rank tests.

\paragraph{Results.} Univariately, every cohort and size variable is highly significant: birth year [HR$=1.38$ per year, $p<10^{-3}$, $C=0.73$]; $\log_{10}$\,\#\,contributors [HR$=0.17$, $p<10^{-3}$, $C=0.80$]; $\log_{10}$\,\#\,repositories [HR$=0.011$, $p<10^{-3}$, $C=0.68$]; $\log_{10}$\,total commits [HR$=0.23$, $p<10^{-3}$, $C=0.79$]. Among the cross-community measures, the strongest univariate predictor is the external-community degree $\log_{10}(1 + d^\mathrm{ext}_c)$ [HR$=0.072$, 95\% CI $[0.030,\,0.174]$, $p<10^{-3}$, $C=0.77$]: a one-decade increase in the number of other communities reached cuts the residualisation hazard by a factor of $\sim 14$. The number of inter-community PRs is similarly protective [$\log_{10}(1+n^\mathrm{inter}_c)$: HR$=0.15$, $p<10^{-3}$]. The inter-community PR \emph{share} is not significant univariately [HR$=0.67$, $p=0.62$], reflecting the fact that very small communities can have a high inter-share by accident of having a single boundary-crossing PR; only after size is controlled for does the share become an interpretable predictor.

\begin{figure}[H]
    \centering
    \includegraphics[width=\textwidth]{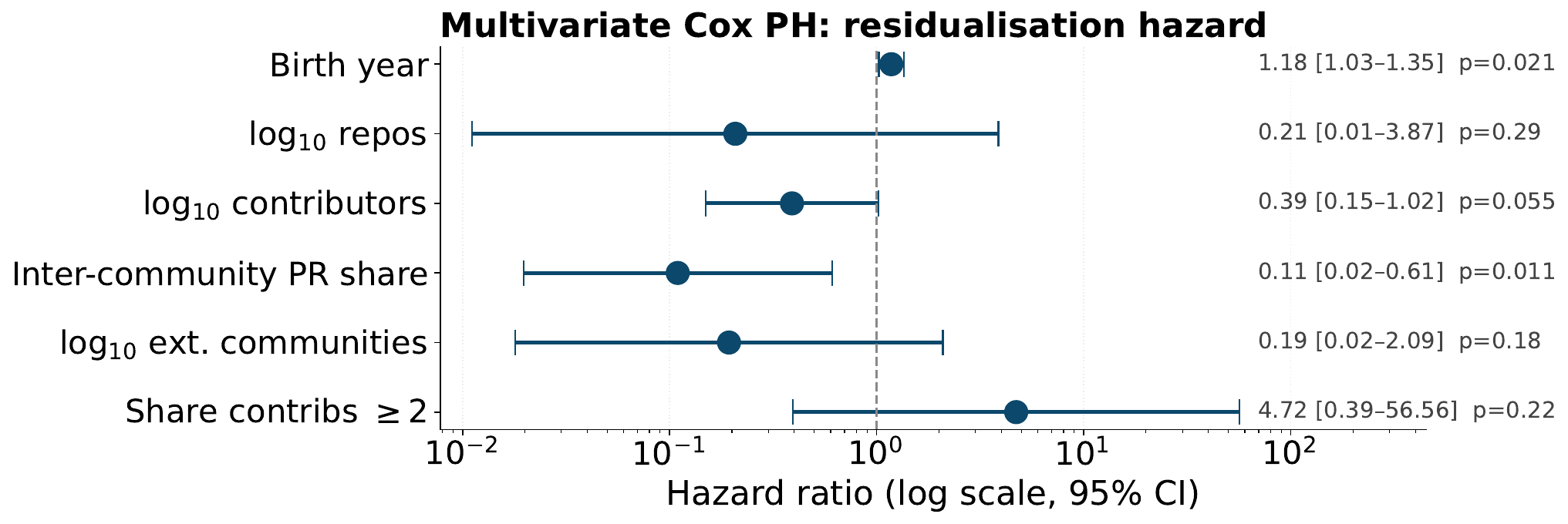}
    \caption{\footnotesize{{\bf Predictors of community residualisation hazard: multivariate Cox proportional-hazards forest plot.} $n=163$ communities, $55$ events; concordance $C=0.843$. Hazard ratios on a logarithmic axis with $95\%$ Wald confidence intervals; the dashed vertical line is HR$=1$. After controlling for cohort [birth year] and size [$\log_{10}$ \#\,repositories, $\log_{10}$ \#\,contributors], the inter-community PR share is significantly protective [HR$=0.11$, $95\%$ CI $[0.02,\,0.61]$, $p=0.012$]; birth year retains an independent unfavourable effect [HR$=1.18$ per year, $p=0.021$]; $\log_{10}$ \#\,contributors is borderline [$p=0.055$]. The same numbers are reported as Table~\ref{tab:cox_survival} in the main text; the corresponding Kaplan--Meier stratification by inter-community PR share is panel~B of Figure~\ref{fig:survival}.}}
    \label{fig:SFig_hazard_predictors}
\end{figure}

The multivariate model [Table~\ref{tab:SI_hazard_predictors}, Figure~\ref{fig:SFig_hazard_predictors}] achieves concordance $C = 0.843$. After controlling for cohort and size, three observations bear directly on the hypothesis. First, birth year remains an independent predictor [HR$=1.18$, $p=0.021$]: the cohort effect is real and not fully explained by community size or connectivity. Second, $\log_{10}$\,\#\,contributors retains a borderline-significant protective signal [HR$=0.39$, $p=0.055$]; $\log_{10}$\,\#\,repositories is no longer significant [$p=0.29$] because contributor count and repo count are highly collinear. Third -- the test -- the inter-community PR share is significantly protective once size is controlled for: HR$=0.11$ [95\% CI $[0.02,\,0.61]$, $p=0.012$]. That hazard ratio is per unit of the covariate, and a unit of a share is the full $0$--$100\%$ range, which is well outside the range over which the share actually varies: across the $163$ communities its median is $0.043$ and its interquartile range $0$--$0.167$ [standard deviation $0.188$; $26\%$ of communities have a share of exactly zero and the maximum is $1.0$], so a full-unit contrast is an extrapolation of roughly six interquartile ranges. Read on the scale of the observed variation, the same fit gives HR $=0.69$ per interquartile range [$\exp(0.167\ln 0.11)$] and HR $=0.66$ per standard deviation: a community one interquartile range higher in cross-community PR share has roughly a $30\%$ lower instantaneous hazard than a same-cohort, same-size community lower on the same measure.

\begin{table}[h]
\centering
\footnotesize
\caption{Multivariate Cox proportional-hazards model of community residualisation [n$=$163, events$=$55, concordance $C=0.843$]. HRs are per unit of the listed predictor on the original scale (years for birth year, $\log_{10}$ units for log-transformed predictors, fraction for shares).}
\label{tab:SI_hazard_predictors}
\begin{tabular}{lrrrr}
\toprule
predictor & HR & 95\% CI lower & 95\% CI upper & $p$ \\
\midrule
Birth year                          & $1.18$ & $1.03$ & $1.35$ & $0.021$ \\
$\log_{10}$ \# repositories          & $0.21$ & $0.01$ & $3.87$ & $0.292$ \\
$\log_{10}$ \# contributors          & $0.39$ & $0.15$ & $1.02$ & $0.055$ \\
Inter-community PR share            & $0.11$ & $0.02$ & $0.61$ & $0.012$ \\
$\log_{10}$ \# external communities reached & $0.19$ & $0.02$ & $2.09$ & $0.176$ \\
Share of contributors $\geq 2$ communities  & $4.72$ & $0.39$ & $56.6$ & $0.221$ \\
\bottomrule
\end{tabular}
\end{table}

Of the two Kaplan--Meier stratifications corresponding to these cross-community measures, the one by inter-community PR share is panel~B of Figure~\ref{fig:survival}; the external-degree stratification is reported numerically below and in the main text, but is not plotted. Stratifying by inter-community PR share gives a multi-group log-rank statistic of $\chi^2 = 21.9$, $p = 1.8 \times 10^{-5}$: the high-share tertile sustains a $\sim 0.55$ probability of remaining non-residual at 10 years from birth versus $\sim 0.30$ for the low-share tertile. Stratifying by external-community degree gives a sharper separation [$\chi^2 = 49.1$, $p = 2.2 \times 10^{-11}$]: communities in the highest tertile of external reach have a $\sim 0.95$ ten-year non-residual probability against $\sim 0.10$ for the lowest tertile [in the multivariate model the same predictor attenuates to HR$=0.19$, $p=0.18$, partly mediated by community size].

\paragraph{Caveats.} All predictors here are end-of-window summaries computed on the aggregate corpus, and three caveats apply. First, the inter-community PR share is itself correlated with birth year [Spearman $\rho = -0.21$, $p = 0.006$ on all 163 communities; $\rho = -0.32$, $p = 0.005$ restricted to those with $\geq 50$ PRs]: late cohorts have less time to accumulate PR history and many have no PRs at all [40\% of the 2021 cohort], inflating the cohort signal. Second, the proportional-hazards assumption is not formally tested here [Schoenfeld residuals]; the strong cohort effect makes it plausible that hazards are not strictly proportional across cohorts. Third, the static-covariate framing creates an obvious reverse-causation concern: a community already approaching residual may attract fewer external PRs because it is already perceived as inactive, so the protective association observed at end of window may partly be an effect of survival rather than its cause. The static results reported here should therefore be read as descriptions of end-of-window association rather than as causal-direction tests. Fourth, the inactivity-rule sensitivity in Supplementary~\ref{sec:SI_residual} bears directly on the external-degree covariate: its multivariate effect reaches significance under only three of the ten rules tested, so the attenuation reported above is a stable feature of the data rather than an artefact of the $5\%$ threshold, and external reach is best read as a partial proxy for community resourcedness.

\subsection{Time-varying incoming connectivity: a null}\label{sec:SI_time_varying_cox}

The Cox models in Supplementary~\ref{sec:SI_hazard_predictors} use end-of-window connectivity covariates, which conflates two readings: that early connectivity is associated with later survival, and that already-surviving communities simply had longer to accumulate cross-community ties. A specification in which connectivity is recomputed year by year separates the two for whichever connectivity measures it carries. The model reported here carries two, both read off the commit stream and both \emph{incoming}: the recent cross-community commit inflow into a community, and the number of distinct external communities that have sent commits into it. Both return nulls, and those nulls are the clearest available statement of why this paper stops at association.

\paragraph{Construction.} For each community $c$ we build one interval per year of life, $[t,\,t+1)$ with $t = y - b_c$, at risk from $t=0$ to $T_c = y^*_c - b_c + 1$, the time-to-residual used throughout Supplementary~\ref{sec:SI_residual}. The event indicator is $1$ only on the final interval, and only for communities that are not censored. Connectivity covariates are evaluated at the \emph{end} of each interval's calendar year using only commits dated on or before that year, so no future information enters: $\log_{10}(1+n^{\mathrm{inter}}_{[y-2,\,y]})$, the recent cross-community commit inflow over a three-year trailing window; and $\log_{10}(1+d^{\mathrm{ext}}_{[b_c,\,y]})$, the number of distinct external communities that have sent commits into $c$ up to that year. Birth year and the two size covariates remain static. The model is fitted with \texttt{lifelines.CoxTimeVaryingFitter}.

The inter-community pull-request share of Supplementary~\ref{sec:SI_hazard_predictors} has no counterpart among these covariates. It is a share of pull-requests, whereas the year-resolved connectivity series available on the commit stream are counts of incoming commits; the nearest share that can be built from that stream, cross-community commits over total commits, differs from it in both numerator and denominator and lives on a different scale entirely [median $0.0009$ against $0.043$], so it is not a year-resolved rendering of the same quantity and is not fitted here.

\paragraph{Result.} Neither time-varying connectivity covariate is associated with the residualisation hazard once cohort and size are controlled: incoming cross-community commit flow gives HR $=1.64$ [$p=0.66$] and incoming community breadth HR $=1.08$ [$p=0.95$]. Among the covariates this model shares with the static one, what survives is what survives there: birth year [HR $=1.23$ per year, $p=0.003$] and $\log_{10}$ \#\,contributors [HR $=0.28$, $p=0.001$], with $\log_{10}$ \#\,repositories marginal [HR $=0.089$, $p=0.087$].

\paragraph{Reading.} What this specification establishes is a null on two incoming measures: neither the cross-community commit volume a community receives over the preceding three years nor the number of distinct communities it has received commits from is associated with its residualisation hazard once cohort and size are controlled. Two readings of that null remain open and we cannot separate them with these data: either incoming cross-community activity is not protective and the end-of-window association reflects survivorship -- communities that lasted longer had more time to accumulate cross-community ties -- or the effect operates on a timescale, or through a channel, that annual trailing-window counts of incoming commits do not capture. What the specification does not do is test the inter-community pull-request share of Table~\ref{tab:cox_survival}: that covariate has no year-resolved counterpart among the series fitted here, so the association reported for it is neither confirmed nor refuted by this model and stands untested against a time-varying alternative. Either way, the honest summary is that this paper measures an association at end of window and does not establish a direction. The nulls reported here sit alongside the inactivity-rule sensitivity of Supplementary~\ref{sec:SI_residual}, where the external-degree covariate [the static, symmetric counterpart of the incoming-breadth measure] reaches multivariate significance under only three of ten definitions, and together they are the reason the Discussion frames the carrier layer as covarying with community persistence rather than as sustaining it.

\begin{figure}[H]
    \centering
    \includegraphics[width=\textwidth]{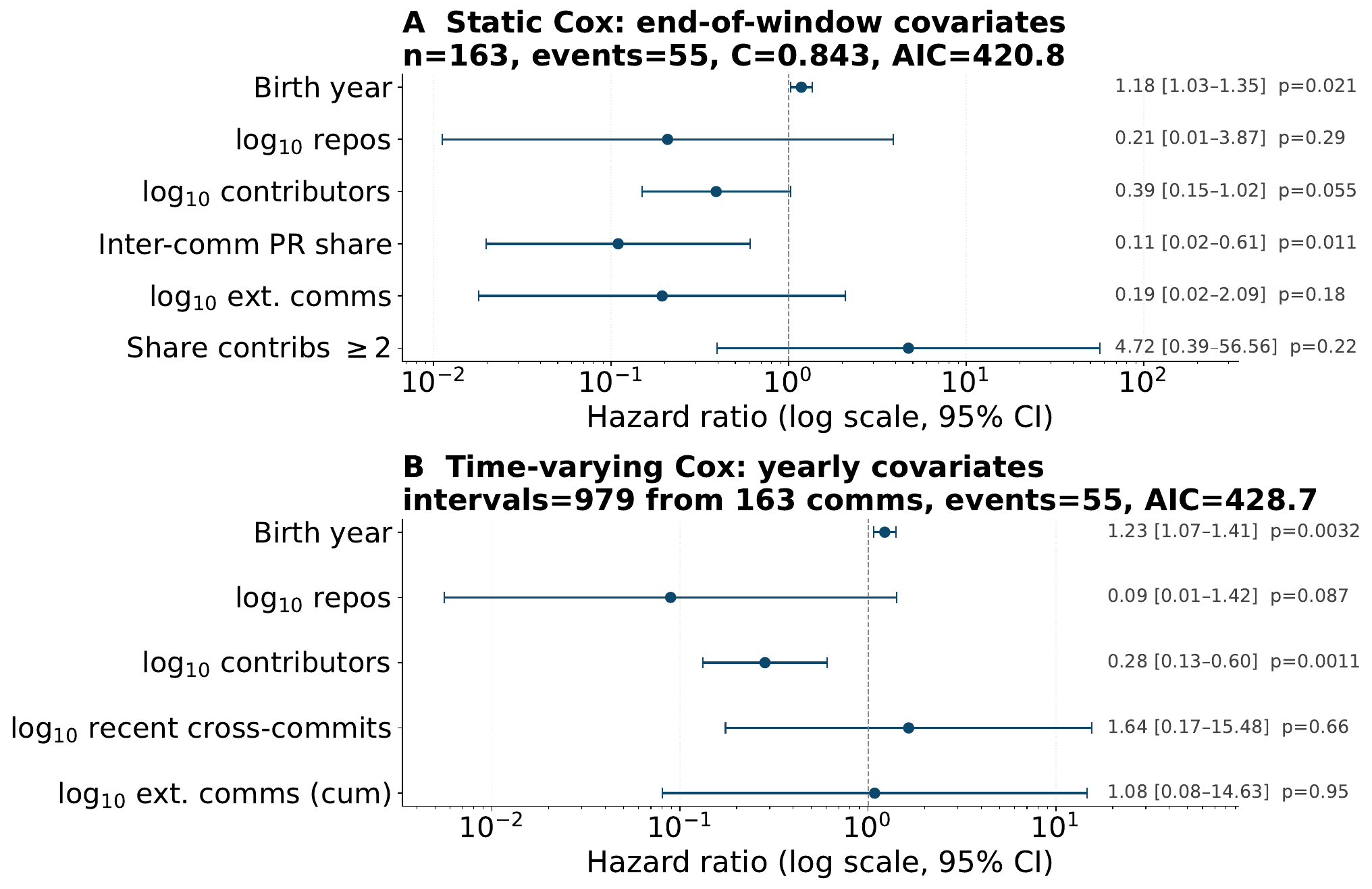}
    \caption{\footnotesize{{\bf Static and time-varying Cox proportional-hazards models of community residualisation.} \textbf{A.}~The static-covariate model, reproduced from Figure~\ref{fig:SFig_hazard_predictors} so that the two specifications can be read against each other [$n=163$, $55$ events, $C=0.843$, AIC $=420.8$]. \textbf{B.}~The time-varying model [$979$ community-year intervals, AIC $=428.7$]. In both panels, hazard ratios on a logarithmic axis with $95\%$ confidence intervals; the dashed vertical line is HR$=1$. Connectivity covariates are recomputed at the end of each year of a community's life from commits dated on or before that year, so the model uses no future information. Both time-varying covariates are incoming commit-flow measures and neither is significant [incoming cross-community commit flow HR $=1.64$, $p=0.66$; incoming community breadth HR $=1.08$, $p=0.95$]; birth year [HR $=1.23$, $p=0.003$] and $\log_{10}$ \#\,contributors [HR $=0.28$, $p=0.001$] remain associated with the hazard. The inter-community pull-request share of Table~\ref{tab:cox_survival} has no counterpart among these covariates and is not tested here. The static-covariate model is also reported on its own in Figure~\ref{fig:SFig_hazard_predictors}.}}
    \label{fig:SFig_time_varying_cox}
\end{figure}

\section{Boundary friction and collaboration depth}\label{sec:SI_friction}

This section supports the boundary-friction analysis in the main text (Figure~\ref{fig:Fig6_depth}) by stratifying the per-PR friction indicators by the author's commit-level breadth $k$.

\subsection{How boundary friction varies with contributor breadth $k$}\label{sec:SI_friction_by_k}

The four boundary-friction indicators reported in the main text (Figure~\ref{fig:Fig6_depth}) compare intra- versus inter-community pull-requests and issues but pool every author together. A natural follow-up question is whether the friction is borne uniformly across contributors or is instead concentrated in a particular layer. We answer this by linking each pull-request to its author's commit-level breadth $k$ [the number of distinct Louvain communities that author has committed to in the corpus] and re-tabulating Figure~\ref{fig:Fig6_depth}A--B by $k$-bin.

\paragraph{Identity reconciliation.} The commit metadata in our primary stream uses \texttt{author\_uniqueID}, which is in $86.9\%$ of rows a numeric GitHub user identifier and in the remainder a login string; the GitHub Archive event stream and the per-PR review-depth table use login strings throughout. To align the two we built an \texttt{actor\_id $\leftrightarrow$ actor\_login} map from the GitHub Archive [one row per event, $490{,}990$ unique \texttt{actor\_id}s, $507{,}395$ unique logins] and re-keyed every commit row by the resulting canonical login. After canonicalisation the $11{,}370$ raw \texttt{author\_uniqueID}s in the commit stream -- the $11{,}371$ distinct identifiers less the \texttt{(no author)} placeholder, and two fewer than the $11{,}372$ contributor nodes of the bipartite graph -- collapse to $11{,}189$ canonical contributors. Of these, eleven are explicit bots [\texttt{[bot]} suffix or known automation accounts: \texttt{dependabot[bot]}, \texttt{dependabot-preview[bot]}, \texttt{renovate[bot]}, \texttt{github-actions[bot]}, \texttt{allcontributors[bot]}, \texttt{deepsource-autofix[bot]}, \texttt{imgbot[bot]}, \texttt{mergify[bot]}, \texttt{pre-commit-ci[bot]}, \texttt{whitesource-bolt-for-github[bot]}, plus \texttt{gitter-badger}] and $688$ are phantom git identities -- numeric identifiers that do not resolve to a GitHub login in our window [these reflect committers whose authorship metadata never matches a recorded GitHub account, e.g.\ legacy or deleted accounts] plus the generic \texttt{root}/\texttt{admin}/\texttt{git}/\texttt{user} placeholders. Both classes are excluded throughout the cross-community contributor analysis because their $k$ counts conflate multiple humans [default git server identities] or reflect automation rather than human boundary-spanning. After filtering, the canonical human cross-community contributor layer at the data-driven boundary $k\geq 7$ comprises nine humans, of which the four with $k\geq 10$ are \texttt{noraj}, \texttt{cclauss}, \texttt{0xflotus}, and \texttt{timgates42}. The two automated $k\geq 10$ identifiers that the unfiltered count would have included are \texttt{dependabot[bot]} [a dependency-update bot that authored $7{,}917$ pull requests in the corpus] and \texttt{gitter-badger}; the two phantom $k\geq 10$ identifiers are the placeholder logins \texttt{root} [$k=20$] and \texttt{148100} [$k=31$]. Removing the bot subset alone subtracts $\sim 24\%$ of the unfiltered $5{,}612$ inter-community merged pull requests [\texttt{dependabot[bot]} contribution] and is the single largest correction induced by the procedure.

\begin{figure}[H]
    \centering
    \includegraphics[width=\textwidth]{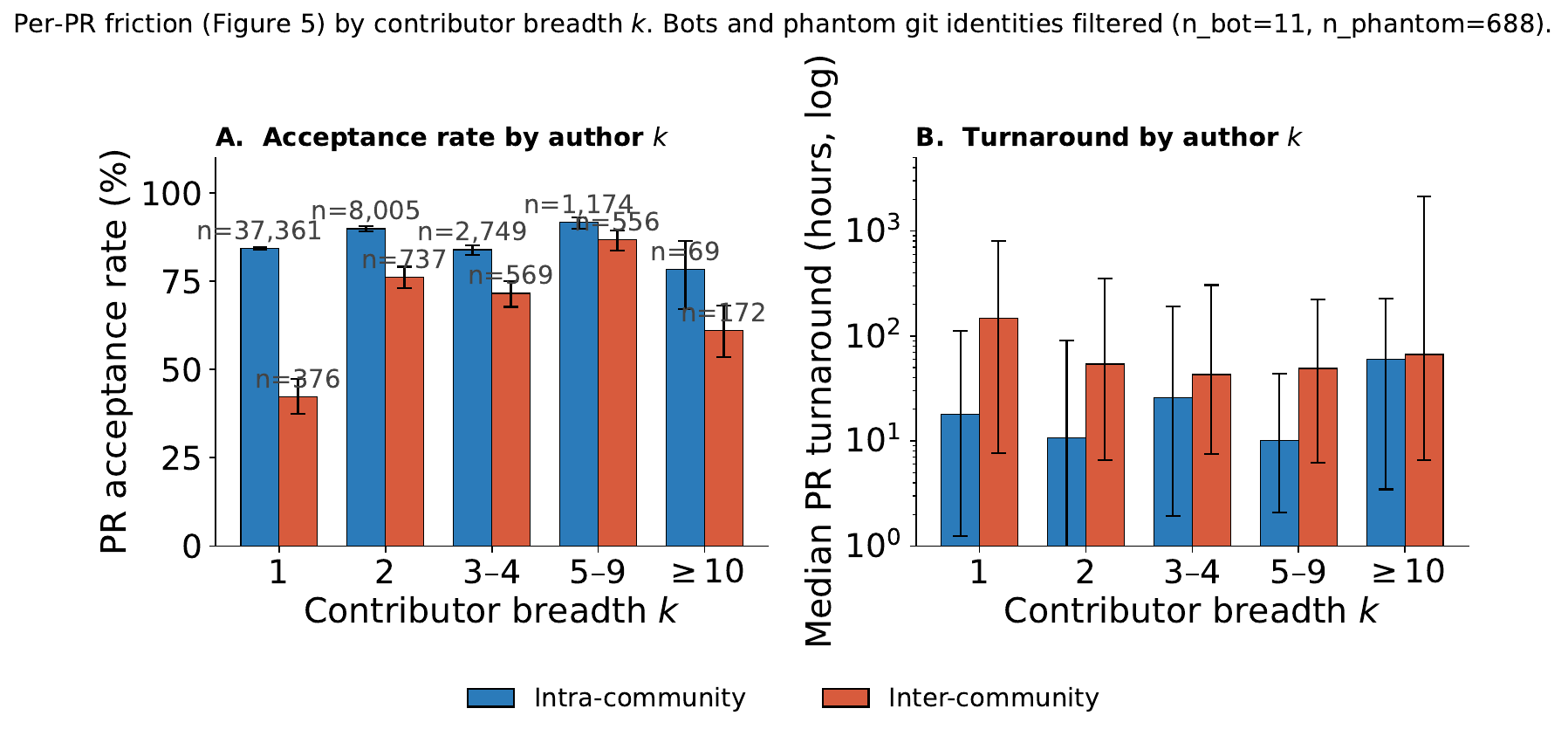}
    \caption{\footnotesize{{\bf Per-PR friction (Figure~\ref{fig:Fig6_depth}A--B) varies sharply with author breadth $k$.} Pull requests are grouped by intra/inter [blue/red] and by their author's commit-level $k$-bin after canonicalising contributor identity [\texttt{author\_uniqueID} $\to$ login via the GitHub Archive \texttt{actor\_id $\leftrightarrow$ actor\_login} map] and removing $11$ explicit bots and $688$ phantom git identities. \textbf{A.}~Acceptance rate [\% of PRs ultimately merged] with Wilson 95\% confidence intervals; per-bin sample sizes annotated above the bars. The intra-community curve is roughly flat at $78\text{--}92\%$, while the inter-community acceptance climbs from $42.3\%$ at $k=1$ to $86.7\%$ at $k=5\text{--}9$ and is $61\%$ at $k\geq 10$ [$n=172$]. \textbf{B.}~Median turnaround [hours from PR open to terminal event] on a logarithmic axis; whiskers indicate the inter-quartile range. Inter-community turnaround at $k=1$ is $\sim$147 hours and falls to $\sim$49 hours at $k=5\text{--}9$ -- a 3-fold compression that narrows the spread across breadth bins; a substantial inter/intra gap nevertheless remains at every $k$, and is widest at $k=5\text{--}9$. These are raw stratifications, not controlled estimates, and the apparent premium at high $k$ does not survive once breadth is measured before the request and the destination community is excluded from it: $k$ is itself partly an outcome of acceptance, since a merged cross-community pull-request lands a commit in the destination community. Supplementary~\ref{sec:SI_friction_controls} sets out the contamination and the controlled analysis that replaces this reading.}}
    \label{fig:SFig_friction_by_k}
\end{figure}

\paragraph{Per-PR friction by author $k$.} Linking the per-PR review-depth table to the canonical $k$ assignment gives $56{,}356$ pull requests with a resolved author, of which $51{,}768$ remain after removing the $11$ bots and $688$ phantom identities. Splitting these by intra/inter and by author's $k$-bin yields the patterns in Figure~\ref{fig:SFig_friction_by_k}.

For \emph{intra-community} pull-requests, acceptance is roughly flat at $78\text{--}92\%$ across all $k$-bins, and median turnaround sits between 10 and 60 hours -- the small variation is consistent with the per-bin sample sizes shrinking from $37{,}361$ at $k=1$ to $69$ at $k\geq 10$.

For \emph{inter-community} pull-requests the friction varies dramatically with the author's $k$. Acceptance rises from $42.3\%$ at $k=1$ [$n=376$] to $76.1\%$ at $k=2$ [$n=737$], $71.5\%$ at $k=3\text{--}4$ [$n=569$], and $86.7\%$ at the $k=5\text{--}9$ bin [$n=556$]. Median turnaround on inter-community pull-requests collapses by a factor of $\sim 3$, from $147.4$ hours at $k=1$ to $48.9$ hours at $k=5\text{--}9$. It does not converge on the intra-community median, however: intra-community turnaround in that bin is $10.1$ hours, so the residual inter/intra gap of $4.8\times$ is the widest of any bin -- the compression closes the gap between broad and narrow crossers, not the gap between crossing and not crossing. The CHANGES\_REQUESTED rate also drops from $2.4\%$ at $k=1$ to $0.7\%$ at $k=5\text{--}9$. The four humans at $k\geq 10$ [$n=172$ inter-community PRs] sit between the $k=5\text{--}9$ bin and $k=2$ contributors on acceptance [$61\%$] and median turnaround [$66$ hours], with a wide upper-quartile envelope that reflects a small sample dominated by a handful of recurring contributor--repository pairs.

\paragraph{Interpretation, and why it does not survive controls.} Read descriptively, this stratification says that the boundary-friction gap reported as a single intra-vs-inter contrast in Figure~\ref{fig:Fig6_depth} is not borne uniformly: it is concentrated among low-$k$ submitters, while high-breadth contributors appear to clear inter-community pull-requests at close to intra-community speed and acceptance. The natural reading of such a gradient is as evidence for a recognition \& repeat-relationship mechanism operating through contributor breadth. \textbf{That reading does not survive scrutiny and should not be drawn.} The breadth measure $k$ counts the communities an author committed to \emph{over the whole observation window}, and a merged cross-community pull-request itself produces a commit in the destination community, raising that author's $k$ by one; the gradient is therefore contaminated by the very outcome it is invoked to explain. Supplementary~\ref{sec:SI_friction_controls} quantifies that contamination, shows that the gradient vanishes once breadth is measured before the pull-request and excludes the destination community, and reports what does survive a full set of controls: a recognition effect specific to the contributor's prior relationship with the destination repository, rather than general to breadth. The tabulation in this subsection is retained as a description of the raw data and should be read only as such.

\subsection{Do the friction gradients survive controls?}\label{sec:SI_friction_controls}

The stratification in Supplementary~\ref{sec:SI_friction_by_k} is descriptive. This subsection asks whether the breadth gradient it displays survives (i) the alternative explanations a reviewer would reasonably raise -- that high-breadth contributors are simply more experienced, more technically skilled, more likely to hold write access, or better at targeting their pull-requests -- and (ii) a measurement problem that turns out to matter more than any of them.

\paragraph{Analysis sample and covariates.} The unit of observation is a closed pull-request. Linking the per-PR review-depth table to the canonical contributor assignment and keeping requests with a resolved non-bot, non-phantom author, a known intra/inter status and a known outcome gives $51{,}709$ pull-requests, of which $2{,}409$ cross a community boundary, authored by $6{,}447$ canonical humans across $354$ repositories. Beyond the fields listed in Methods Section~\ref{sec:gharchive-enrichment}, the GitHub Archive extraction also carries pull-request size (\texttt{additions}, \texttt{deletions}, \texttt{changed\_files}, \texttt{commits}; populated for every \texttt{PullRequestEvent} in every year of the window), the login of the account performing each merge, and \texttt{PushEvent} and \texttt{MemberEvent} records. These allow one covariate per alternative explanation, each computed strictly \emph{before} the focal pull-request opened so that no look-ahead is possible: contributor tenure [days since the author's first event anywhere in the corpus]; cumulative prior pull-requests across the corpus; the author's prior acceptance rate on \emph{other} repositories; pull-request size at submission; prior write access in the target repository [a \texttt{PushEvent}, a merge performed by the author, or a \texttt{MemberEvent} there before the focal PR]; prior reviewing of other authors' pull-requests in that repository; and prior pull-requests to the target repository and to the target community. Repository and year fixed effects absorb heterogeneity in how lenient individual projects are and drift in acceptance over time, and standard errors are clustered on the author throughout. A final specification adds author fixed effects, absorbing every time-invariant author characteristic -- skill, seniority, general reputation -- and identifying the boundary effect from within-author contrasts among the $801$ authors who submit both within- and cross-community pull-requests ($2{,}326$ of the $2{,}409$ inter-community requests).

\paragraph{The breadth measure is endogenous.} A merged cross-community pull-request creates a commit in the destination community and therefore increments its author's $k$. Among the $746$ inter-community pull-requests whose author had never previously committed to the destination community, $67.0\%$ of those whose request was merged subsequently appear as committers there, against $16.9\%$ of those whose request was rejected. Acceptance manufactures breadth at four times the rate of rejection, so regressing acceptance on end-of-window $k$ is in part a regression of acceptance on itself.

To separate the artefact from any genuine effect we recompute breadth as $k^{\mathrm{prior}}_{\mathrm{excl}}$: the number of distinct communities the author had committed to \emph{strictly before} the pull-request opened, \emph{excluding} the destination community. Excluding the destination breaks the mechanical link for the focal observation while leaving the author's breadth elsewhere intact. Applying the two corrections separately identifies which one matters [Table~\ref{tab:SI_breadth_decomposition}]: restricting breadth to what was accumulated before the request leaves the gradient intact, whereas additionally excluding the destination community flattens it.

\begin{table}[h]
\centering
\footnotesize
\caption{Acceptance rate on inter-community pull-requests under three breadth definitions. Column (A) is the measure used in the previous version of this manuscript. Column (B) removes look-ahead only; column (C) additionally excludes the destination community and is the non-circular measure.}
\label{tab:SI_breadth_decomposition}
\setlength{\tabcolsep}{4pt}
\begin{tabular}{lccc}
\toprule
breadth bin & (A) whole window & (B) before the PR & (C) before, destination excl. \\
\midrule
$0$   & --- & $41.4\%$ [$n=232$] & $69.5\%$ [$n=570$] \\
$1$   & $42.3\%$ [$n=376$] & $62.0\%$ [$n=644$] & $68.1\%$ [$n=894$] \\
$2$   & $76.1\%$ [$n=737$] & $78.7\%$ [$n=684$] & $68.8\%$ [$n=336$] \\
$3$--$4$ & $71.7\%$ [$n=568$] & $80.6\%$ [$n=427$] & $85.4\%$ [$n=424$] \\
$5$--$6$ & $87.5\%$ [$n=489$] & $88.5\%$ [$n=287$] & \multirow{2}{*}{$62.7\%$ [$n=185$, $\geq 5$]} \\
$7+$  & $80.6\%$ [$n=67$] / $61.0\%$ [$n=172$] & $61.5\%$ [$n=135$] & \\
\bottomrule
\end{tabular}
\end{table}

\paragraph{Model ladder.} We fit logistic models of acceptance and ordinary-least-squares models of $\log$ integration latency, adding controls in blocks: the boundary indicator alone; plus breadth and its interaction with the boundary indicator; plus experience and technical effort; plus the revealed track record; plus repository fixed effects; plus insider status; plus relationship-specific history; and finally the author-fixed-effects specification. The focal quantity is the interaction between the boundary indicator and breadth -- whether breadth buys \emph{extra} leniency specifically on boundary-crossing requests, over and above whatever it buys on ordinary ones.

Using the whole-window measure (A), the binned inter-community premium relative to $k=1$ is large, highly significant, and survives every control: odds ratios $4.21$, $3.38$, $3.55$, $7.66$ and $2.93$ for $k=2$, $3$--$4$, $5$--$6$, $7$--$9$ and $\geq 10$, all $p<0.02$, with repository fixed effects, insider status, pull-request size and prior repository history in the model. We report this because it is what a reader would expect to see, but we do not treat it as evidence: none of these controls addresses the endogeneity of $k$ itself.

Using the non-circular measure (C), the premium disappears. The binned profile is flat and uniformly non-significant -- odds ratios $0.94$ [$95\%$ CI $0.63$--$1.39$], $1.19$ [$0.70$--$2.00$], $1.61$ [$0.69$--$3.76$] and $0.78$ [$0.38$--$1.61$] for prior breadth $1$, $2$, $3$--$4$ and $\geq 5$, all $p>0.27$. The continuous interaction is null in every specification of the ladder [OR $0.57$, $p=0.21$ with no controls; OR $1.03$, $p=0.93$ with the full control set, matching Table~\ref{tab:SI_friction_controls}], as is the author-fixed-effects linear probability model [$-3.4$ percentage points, $p=0.16$]. Integration latency tells the same story, and in the fully controlled specifications a slightly stronger one. Under the published-breadth measure the interaction is non-significant throughout [$0.11 \leq p \leq 0.99$]. Under the non-circular measure (C) it is non-significant once repository fixed effects, insider status and repeat relationship are added [$0.55 \leq p \leq 0.77$ for M4--M6], but in the intermediate specifications and in the author-fixed-effects model the coefficient is significantly \emph{positive} [$p=0.006$, $0.012$ and $0.015$ for M2, M3 and M7]: greater prior breadth is there associated with \emph{longer}, not shorter, inter-community latency -- the opposite sign to a breadth premium. Consistently, under the non-circular measure the raw inter-community median latency does not decline with prior breadth either [$44$, $67$, $42$, $56$ and $94$ hours across bins $0$, $1$, $2$, $3$--$4$ and $\geq 5$].

\paragraph{What does survive.} Recognition is supported in a relationship-specific form and unsupported in a breadth-general one. In the fully controlled specification with repository and year fixed effects and author-clustered standard errors [Table~\ref{tab:SI_friction_controls}], what predicts a boundary-crossing pull-request being accepted, and accepted quickly, is a prior relationship with \emph{that specific destination}: prior write access in the target repository, prior pull-requests to that repository, and prior pull-requests to that community. All count-valued predictors enter as base-10 logarithms, so a unit increase is a ten-fold increase and the effects below read \emph{per decade}; the latency outcome remains in natural logarithm so that percentage changes are the usual $\exp(\beta)-1$. Acceptance effects are reported as odds ratios [OR$\,=\exp(\beta)$, the multiplicative change in the odds of a merge for a one-unit increase in the predictor]; note that an odds ratio is not a ratio of probabilities, so an OR of $2.26$ against the $68.8\%$ baseline acceptance of inter-community requests by non-insiders corresponds to a predicted acceptance of $83\%$ rather than to a $2.26$-fold change in probability. Descriptively, inter-community pull-requests are merged $92.1\%$ of the time when the author already holds write access in the target repository against $68.8\%$ otherwise, and $87.3\%$ of the time for authors with eleven or more prior pull-requests to that repository against $68.6\%$ for first-time submitters. The latency effect of write access is conditional rather than marginal: raw median latencies by insider status are indistinguishable [$54.1$ against $53.1$ hours], and the reduction emerges only once repository composition and pull-request size are held fixed.

\begin{table}[h]
\centering
\footnotesize
\caption{Fully controlled models of pull-request acceptance and integration latency, non-circular breadth measure, with repository and year fixed effects and standard errors clustered on the author [$n=50{,}396$ pull-requests for the acceptance models, which drop to the repositories retained by the fixed-effects screen; $n=51{,}708$ for the latency models]. Acceptance effects are odds ratios (OR); latency effects are percentage changes in hours.}
\label{tab:SI_friction_controls}
\setlength{\tabcolsep}{3pt}
\begin{tabular}{lcc}
\toprule
predictor & acceptance & integration latency \\
\midrule
Prior write access in the target repository & OR $2.26$ [$p=1.6\times10^{-15}$] & $-66\%$ [$p=5\times10^{-20}$] \\
Prior pull-requests, same repository [per decade] & OR $1.70$ [$p=0.0006$] & n.s. \\
Prior pull-requests, same community [per decade] & n.s. & $-35\%$ [$p=0.04$] \\
Prior acceptance rate on other repositories & OR $1.96$ [$p<10^{-4}$] & $-42\%$ [$p=0.02$] \\
Crosses a community boundary & OR $0.77$--$0.81$ [$p=0.04$--$0.10$] & $+48\%$ [$p=0.003$] \\
Breadth $\times$ boundary [focal] & OR $1.03$ [$p=0.93$] & $-10\%$ [$p=0.77$] \\
\bottomrule
\end{tabular}
\end{table}

\paragraph{Consequences for the paper's claims.} Three points follow. First, the boundary gap itself is unaffected: cross-community pull-requests remain less likely to be merged and slower to integrate after every control. Second, the mechanism that reduces that gap is a repeat relationship with a particular destination, not a general capacity for spanning communities -- which is a better-identified claim than the one it replaces, since it survives repository and author fixed effects. Third, the descriptive carrier-layer results reported in Sections~\ref{sec:results-cross-community} and~\ref{sec:results-depth} are counts and concentration shares rather than gradient estimates, and are therefore untouched: the layer remains numerically thin and remains responsible for a disproportionate share of cross-boundary work. What no longer stands is the claim that membership of that layer, as such, buys leniency at the boundary.

\section{Legacy analyses retained for traceability}\label{sec:SI_legacy}

The three analyses collected here informed earlier drafts and are retained so that
the analytical history of the paper remains inspectable. None of them supports a
claim in the main text. The first two are exploratory clusterings whose boundaries
proved not to be robust; the third uses the \emph{retired} connector/master
labelling based on repository counts ($\geq 5$ and $\geq 10$ repositories), which
has been superseded throughout the paper by the carrier layer defined on
community breadth $k$ (Supplementary~\ref{sec:SI_carrier_threshold}). Figures in
this section should not be read as evidence for the current analysis.

\subsection{Exploratory unsupervised temporal clustering of communities}\label{sec:SI_temporal_clustering}

\begin{figure}[H]
    \centering
    \includegraphics[width=\textwidth]{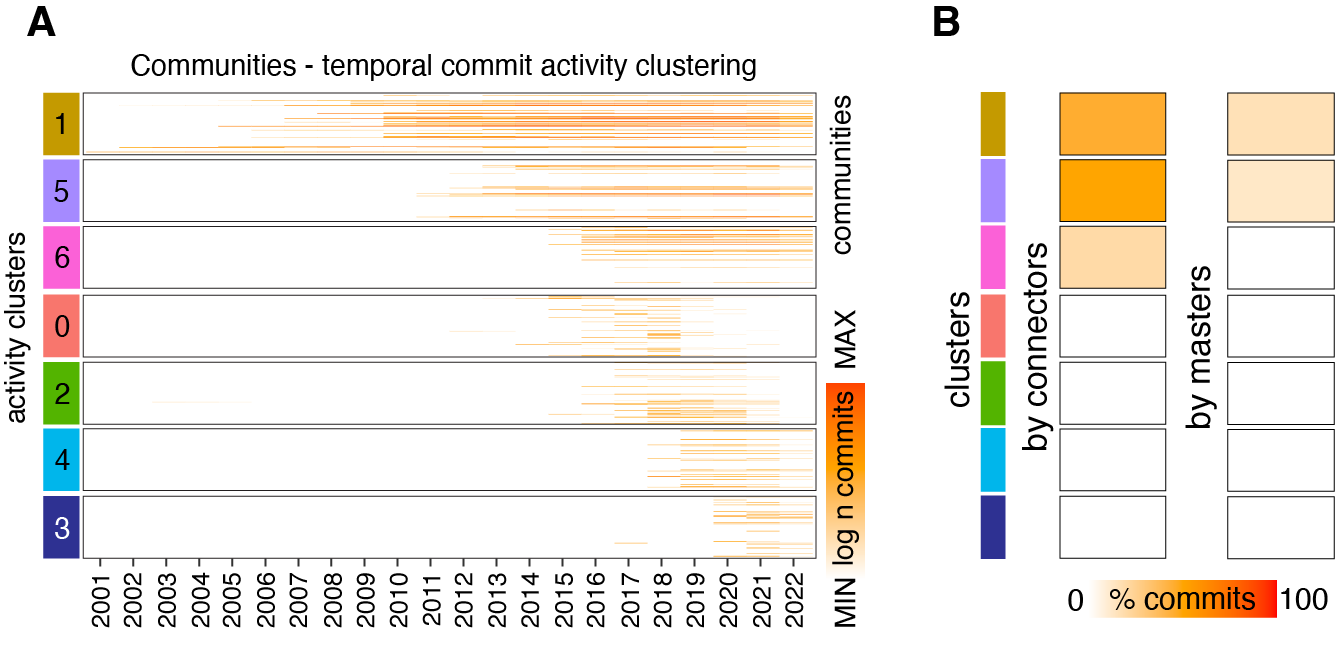}
    \caption{\footnotesize{{\bf Exploratory unsupervised temporal clustering of communities (not used in the main text).} \textbf{A.}~Heatmap stack, one per cluster, detailing log-transformed annual commit activity for each non-singleton Louvain community [rows] over 2001--2022 [columns]; cluster colour-coding shown at the left margin. The pipeline uses Seurat-style normalisation, PCA, UMAP on the leading components, and Louvain-type clustering on the resulting two-dimensional embedding [Methods Section~\ref{sec:methods}]. A minimum pairwise adjusted Rand index of $0.51$ [range $0.51$--$0.94$] across $n_\text{neighbors}\in\{5,15,30\}$ indicates that the cluster boundaries are not robust to local-connectivity choices, and the same temporal heterogeneity is now visualised in the main text as a per-community heatmap sorted by birth year (Figure~\ref{fig:anatomy}A). \textbf{B.}~Distribution of commit shares attributable to contributors with $k\geq 5$ and $k\geq 10$ repositories, by temporal cluster [this figure was produced before the present analysis adopted the data-driven $k\geq 7$ cross-community contributor threshold, so the legacy bins are kept here for visual reference only]: high-$k$ contributors tend to concentrate in sustained-engagement clusters, an observation that resurfaces in the main text under a continuous, contributor-level lens (Figure~\ref{fig:cross_community}C--D).}}
    \label{fig:SFig_temporal_clustering}
\end{figure}

\subsection{Legacy UMAP view of community temporal clustering}\label{sec:SI_legacy_umap}

The figure below accompanied the exploratory clustering above in earlier drafts.

\begin{figure}[H]
    \centering
    \includegraphics[width=\textwidth]{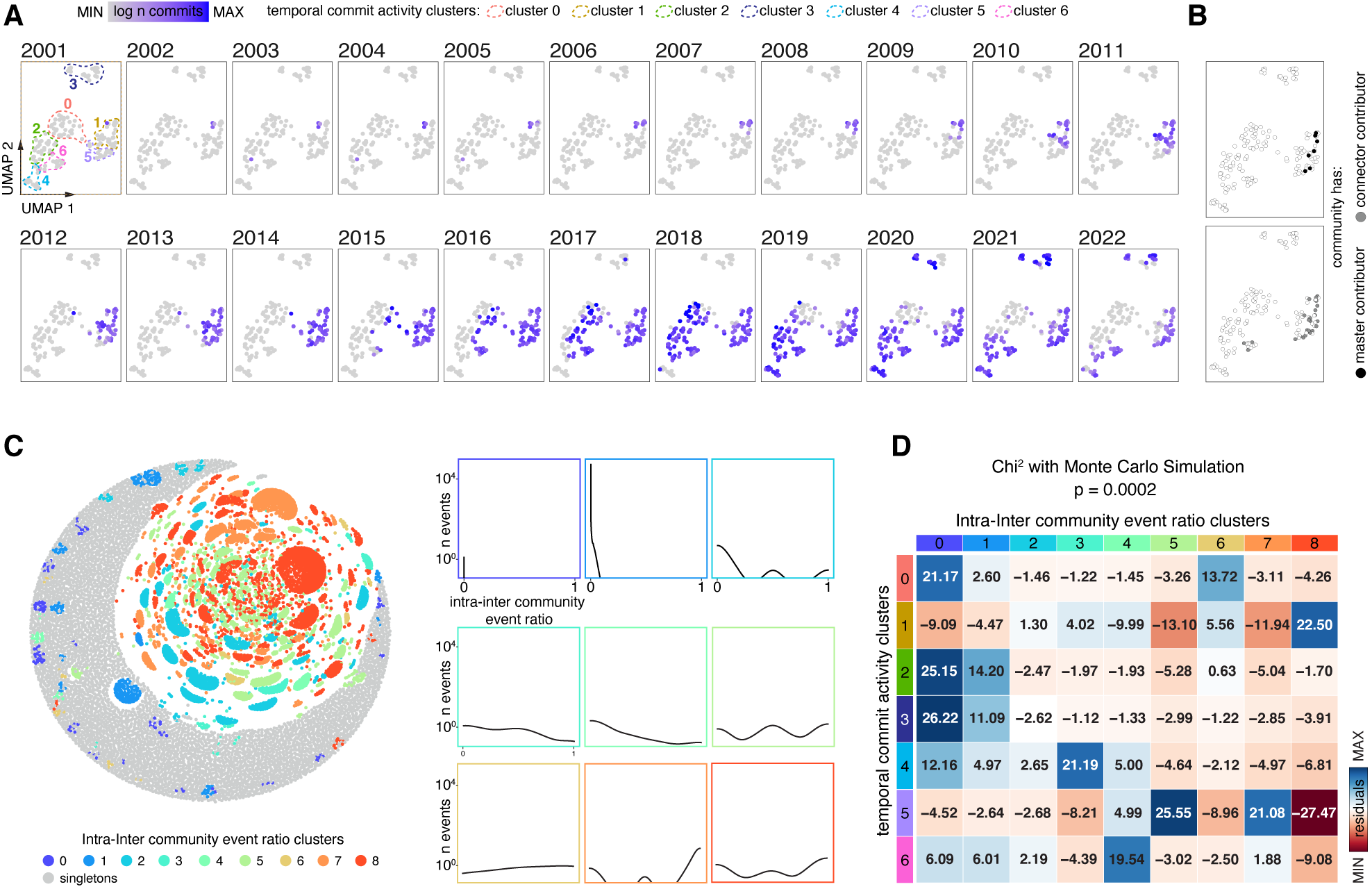}
    \caption[Community temporal clustering and event-ratio clustering]{\footnotesize{{\bf Community Temporal clustering, Event Ratio clustering, Clustering Association. }{\bf A.} Grid of yearly scatter plots illustrating UMAP positioning of communities by their pattern of temporal activity [community temporal clustering]. First scatter on the grid highlights cluster boundaries and labels. Color-coding of UMAP scatter plots indicate the log number of commits made at each year by each community. {\bf B.} Scatter plots with UMAP community temporal activity coordinates highlighting communities that host contributors with $k\geq 5$ [grey, top] and $k\geq 10$ [black, bottom] repositories [the figure pre-dates the data-driven $k\geq 7$ cross-community contributor threshold of the current analysis and is kept here only for the legacy comparison]. {\bf C.} (left) OSS Activity Network color coding contributors as by their event ratio clustering labels. (right) Grid line plots split by event ratio clusters showing the relation between number of events each contributor is associated to and their positioning in the intra-inter community event ratio score. Color-coding is included in left side legend and in line plot outline. {\bf D.} Heatmap showing the Chi-squared residuals for the association test examining the relation between the two types of clustering covered in this figure: temporal commit activity clusters [y-axis] and event ratio clusters [x-axis]. Color-coding from brown (negative residuals) to blue (positive residuals).}}
    \label{fig:SFig4_gitpaper}
\end{figure}

\subsection{Contributor types in the bipartite network}\label{sec:SI_contributor_types}

\begin{figure}[H]
    \centering
    \includegraphics[width=\textwidth]{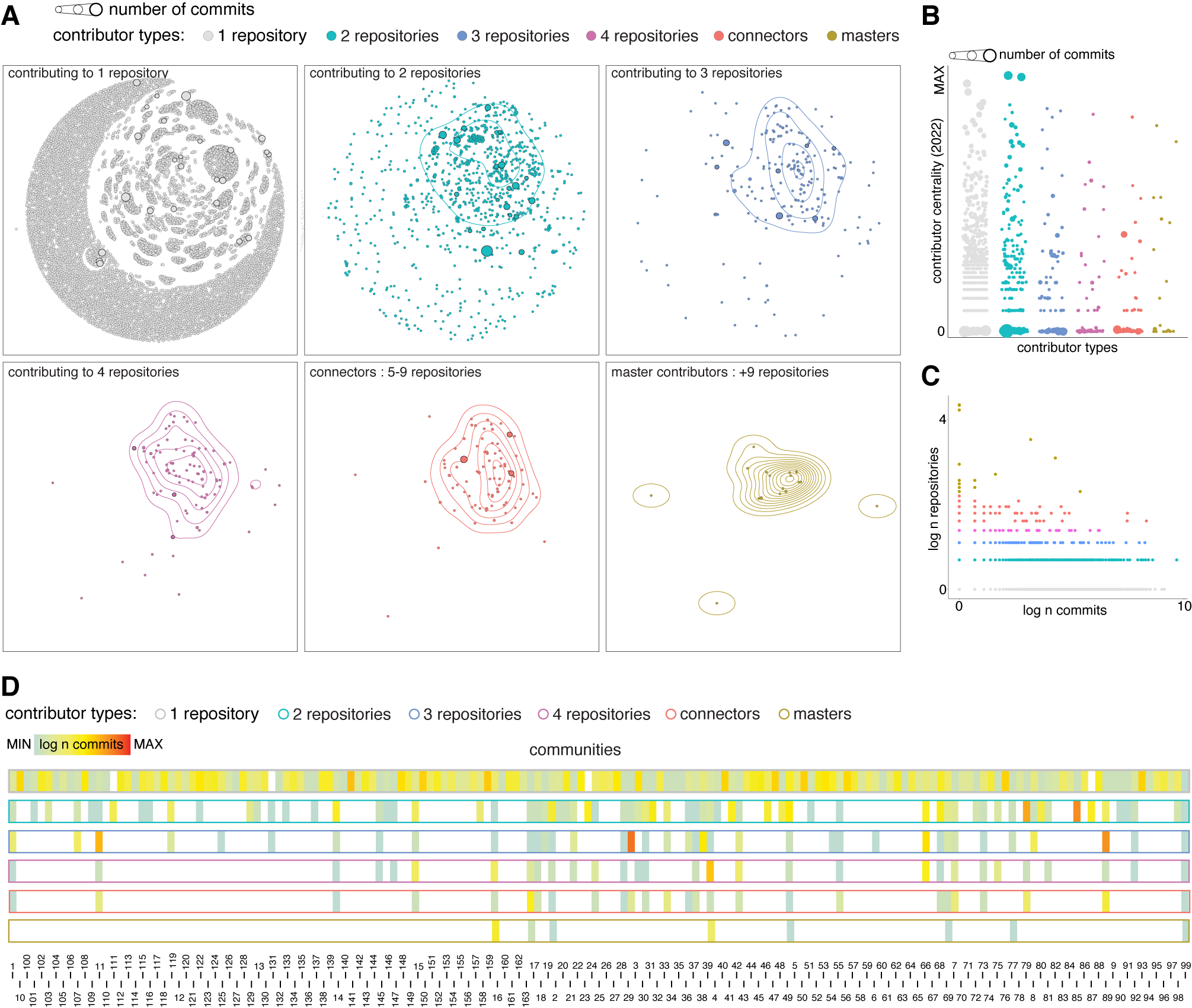}
    \caption[Contributor types in network, activity and community belonging]{\footnotesize{{\bf Contributor types in Network, Activity and Community Belonging. }{\bf A.} Grid of scatter plots depicting the position of contributors, depending on their type, in the OSS network layout. Color-coding signals contributor types by number of repositories touched: $1$ [grey], $2$ [sky blue], $3$ [marine blue], $4$ [purple], $5$--$9$ [orange], $\geq 10$ [beige]; the two upper bins are the legacy split that has since been replaced in the main text by the data-driven $k\geq 7$ cross-community contributor threshold. Point size of contributors representation indicates the number of commits they made through time, outlier contributors in number of commits are highlighted with a black stroke. {\bf B.} Scatter plot displaying contribute centrality in the network [y-axis] depending on contributor type [x-axis] and following contributor type color-coding and size-coding as in A. {\bf C.} Scatter plot illustrating the distribution of contributors by the log-number of repositories they contribute to [y-axis] and the log-number commits they made through time. Color-coding as by contributor type in A.  {\bf D.} Heatmap stack showing the log number of commits made through time on each community by each contributor type [order of rows follows contributor type legend in D and color-coding order, also corresponding to heatmap row outline colors].}}
    \label{fig:SFig3_gitpaper}
\end{figure}






\end{document}